\RequirePackage{fix-cm}
\documentclass[smallextended,natbib]{svjour3}       % onecolumn (second format)
\smartqed  % flush right qed marks, e.g. at end of proof
\usepackage{graphicx}
\usepackage{color}
\usepackage{amssymb}
\DeclareMathAlphabet{\mathpzc}{OT1}{pzc}{m}{it}
\definecolor{myblue}{rgb}{0,0.08,1}

\newcommand{\bA}{{\mathbf A}}
\newcommand{\bB}{{\mathbf B}}
\newcommand{\bE}{{\mathbf E}}
\newcommand{\bJ}{{\mathbf J}}

\newcommand{\bk}{{\mathbf k}}
\newcommand{\bfe}{{\mathbf e}}
\newcommand{\bfj}{{\mathbf j}}
\newcommand{\bn}{{\mathbf n}}

\newcommand{\bq}{{\mathbf q}}
\newcommand{\bv}{{\mathbf v}}

\newcommand{\bx}{{\mathbf x}}

\newcommand{\bnabla}{{\mathbf \nabla}}
\newcommand{\HH}{{\cal H}}

\newcommand{\cd}{\cdot}
\newcommand{\al}{\alpha}

\newcommand{\de}{\delta}
\newcommand{\De}{\Delta}
\newcommand{\ep}{\epsilon}
\newcommand{\ga}{\gamma}

\newcommand{\ka}{\kappa}

\newcommand{\La}{\Lambda}
\newcommand{\la}{\lambda}
\newcommand{\Om}{\Omega}
\newcommand{\om}{\omega}
\newcommand{\si}{\sigma}

\newcommand{\ra}{\rightarrow}

\newcommand{\be}{\begin{equation}}
\newcommand{\ee}{\end{equation}}
\newcommand{\gsim}{\stackrel{>}{\sim}}
\newcommand{\lsim}{\stackrel{<}{\sim}}
\newcommand{\bea}{\begin{eqnarray}}
\newcommand{\eea}{\end{eqnarray}}
\newcommand{\bean}{\begin{eqnarray*}}
\newcommand{\eean}{\end{eqnarray*}}
\newcommand{\dd}{\partial}

\newcommand{\doi}{DOI}

\newcommand{\gr}{$\gamma$-ray}
\begin{document}

\title{Cosmological Magnetic Fields: Their Generation, Evolution and Observation
%\thanks{Grants or other notes
%about the article that should go on the front page should be
%placed here. General acknowledgments should be placed at the end of the article.}
}
\subtitle{}

%\titlerunning{Short form of title}        % if too long for running head

\author{Ruth Durrer        \and
        Andrii Neronov %etc.
}

%\authorrunning{Short form of author list} % if too long for running head

\institute{R. Durrer \at
D\'epartement de Physique Th\'eorique and Center for Astroparticle Physics, \\ Universit\'e de Gen\`eve\\
24 Quai E. Ansermet, 1213 Gen\`eve Switzerland\\
\\
             % Tel.: +123-45-678910\\
              %Fax: +123-45-678910\\
              %\email{fauthor@example.com}           %  \\
%             \emph{Present address:} of F. Author  %  if needed
           \and
           A. Neronov \at
ISDC Data Centre for Astrophysics, Observatoire de Geneve  and Center for Astroparticle Physics, Universit\'e de Gene\`eve\\
16 Chemin d'Ecogia, 1290 Versoix,Switzerland
}             
\date{Received: date / Accepted: date}
% The correct dates will be entered by the editor

\maketitle

\begin{abstract}
We review the possible mechanisms for the generation of cosmological magnetic fields, discuss their evolution in an expanding Universe filled with the cosmic plasma and provide a critical review of the literature on the subject. We put  special emphasis on the prospects for observational tests of the proposed cosmological magnetogenesis scenarios  using radio and gamma-ray astronomy and ultra high energy cosmic rays.  We argue that primordial magnetic fields are observationally testable. They lead to magnetic fields in the intergalactic medium with magnetic field strength and correlation length in a well defined   range.

We also state the unsolved questions in this fascinating open problem of cosmology and propose future observations to address them. 
% \PACS{PACS code1 \and PACS code2 \and more}
% \subclass{MSC code1 \and MSC code2 \and more}
\end{abstract}

%Introduction (s:intro)
%%%%%%%%%%%%%%%%%%%%%%%%%%%%%%%%%%%%%%%%%%%%%%%%%%%%
\section{Introduction}
\label{s:int}
%%%%%%%%%%%%%%%%%%%%%%%%%%%%%%%%%%%%%%%%%%%%%%%%%%%%

Magnetic fields are ubiquitous in the Universe. Wherever we have the means of observing them, they are present:
in our solar system, in stars \citep{Donati09}, in the Milky Way \citep{Wielebinski05} in other low~\citep{Kronberg:1993vk,Fletcher:2011fn,Beck:2012} and high 
redshift~\citep{Kronberg92,Bernet08} galaxies, in galaxy clusters~\citep{Clarke:2000bz,Bonafede10,Feretti12}, in superclusters~\citep{Xu:2005rb}
and even in voids of the Large Scale Structure (LSS) ~\citep{Neronov:2010,Dolag:2010ni,tavecchio-2010,tavecchio-2010.2,Vovk12,Taylor:2011bn,dermer11}. Interestingly, the magnetic field strength in galaxies is typically of the order of a few-to-tens of $\mu$Gauss independent of the galaxy redshift \citep{Kronberg92,Bernet08}.  Also the magnetic fields in clusters are of the order of  $\mu$Gauss \citep{Clarke:2000bz,Bonafede10,Feretti12}.

According to a well accepted paradigm, magnetic fields in astronomical structures of different sizes, from stars (sizes $R\sim 10^{11}$~cm) up to galaxy clusters ($R\sim 10^{24}$~cm) are produced by amplification of pre-existing weaker magnetic fields via different types of dynamo \citep{Parker55,Ruzmaikin:1988,Kulsrud99,Brandenburg:2004jv,Kulsrud08} and via flux-conserving compression during gravitational collapse accompanying structure formation. On short distance scales, magnetic fields dissipate their energy into turbulent and thermal motions of astrophysical plasmas, so that a continuous re-generation of the field is needed on the time scales shorter than the life time of the astronomical object carrying the field. This is the case for e.g. the magnetic fields of the Earth and Sun and other stars and planets. This is also partially true for the galactic magnetic fields, including the field of our own Milky Way galaxy. Weak magnetic fields on the largest distance scales, from $10^{-2}$ to 1~Mpc, from the large scale fields in the galaxies to those in galaxy clusters, might not have enough time to dissipate their energy into plasma motions. Once amplified by dynamo and compression mechanisms, they conserve their strength on time scales comparable to the age of the Universe. 

The dynamo and compression amplification mechanisms can act only if a non-zero magnetic field is present. This "seed" field for the amplification might be tiny, but it has to be generated by a different mechanism, which pre-dates the structure formation epoch or operates at the onset of structure formation. The uncertainty of the strength and of the origin of this initial seed field constitutes the long-standing problem of the origin of cosmic magnetic fields \citep{Kronberg:1993vk,Grasso00,Widrow02,Kulsrud08,Kandus12,Widrow12}. Two broad classes of models for the origin of the seed fields are discussed. One possibility is that the weak seed fields are produced in the early universe, during epochs preceding the structure formation. Another possibility is that the process of generation of the seed fields accompanies the gravitational collapse leading to structure formation. 

The existing data on magnetic fields in galaxies and galaxy clusters cannot provide direct constraints on the properties and origin of the seed fields. This is related to  uncertainties of the details of the dynamo mechanisms operating in galaxies and clusters on the one hand and,  on the other hand, to the numerous saturation effects which drive the galactic and cluster field strengths to fixed values largely independent of the properties of the initial seed fields. 

The only potential opportunity for understanding the nature of the initial seed fields is to search for places in the Universe where these fields might exist in their original form, not distorted by the complicated plasma and magnetohydrodynamics (MHD) processes. 

The only places where such "primordial" magnetic fields might reside is the intergalactic medium (IGM), more precisely, the voids of large scale structure (LSS). If weak magnetic fields were indeed present in the Universe before the onset of structure formation, they did not suffer much amplification because of the absence of the dynamo and compression of the IGM in the voids. Cosmologically produced magnetic fields might passively evolve (be diluted by the expansion of the Universe) still today. Potential measurements of Intergalactic Magnetic Fields (IGMF) using available observational techniques of radio, microwave and \gr\ astronomy might, therefore, provide an important clue on the origin of the seed fields. This idea is the prime motivation for the numerous efforts to detect the IGMF.   

If successful, detection and measurement of the properties of primordial magnetic fields in the voids of LSS will provide an extremely important source of cosmological data. Typical scenarios for generation of magnetic fields in the early universe concentrate on possibilities of field production via charge separation and/or generation of vortical currents at the moments of cosmological phase transitions: the electroweak and the  QCD phase transitions, and the moments of photon decoupling and recombination. Another possibility is the quantum generation of very long wavelength photons during inflation which then are converted into magnetic fields at reheating. In most of the models the moment of cosmological "magnetogenesis" pre-dates the epochs of formation of Cosmic Microwave Background (CMB) signal and the Big Bang Nucleosynthesis (BBN). This means that the detection of relic magnetic fields might provide  observational data on very early physical processes in the hot Universe with temperature in the range above 100~MeV. 
 If they stem from the electroweak phase transition or from inflation, they may even probe physics beyond the standard model. 
 
It is not obvious a-priori that weak magnetic fields which reside in the voids of LSS are primordial. Alternatively, they could be produced at the late stages of evolution of the Universe (at redshifts $z<10$) by outflows from already formed galaxies. These outflows can be galactic winds generated by the star formation activity \citep{Bertone06} and/or relativistic outflows generated by the Active Galactic Nuclei (AGN) \citep{Rees87,Daly90,Ensslin97}.  Both types of outflows are essential elements of the structure formation process. They are responsible for washing out the baryon content of galaxies leading to the "missing baryons" problem \citep{Cen99} and for metal enrichments of the IGM \citep{Aguirre01}. If these outflows are (a) strongly magnetized and (b) able to spread into the voids of the LSS, they can result in "pollution" of the voids with magnetic fields which are much stronger than the relic magnetic fields of primordial origin. The presence of magnetic fields spread by galactic wind into the IGM may prevent measurement of the relic cosmological fields. At the same time, detection and measurement of IGMF spread by winds can constrain the properties of galactic winds and in this way shed light into	 the physics of the "feedback" provided by the winds on the formation and evolution of galaxies. Winds are thought to be responsible for the  regulation of star formation activity of galaxies at different stages of their evolution \citep{Kennicutt12}.

In this review we summarize and critically asses the current knowledge of the weakest magnetic fields in the Universe.  Besides describing the present status of observations of magnetic fields in the IGM, there are two fundamental questions which arise and which we shall address: 
\begin{enumerate}
\item Are IGMF primordial, in the sense that they have been present before the galaxy formation process  took place, with all its complicated non-linear and non-gravitational physics, or have they been formed during galaxy formation e.g. in star formation and AGN activity and then spilled out into the galaxy and into intergalactic space?
\item If they are generated in the early Universe how do they evolve? Are they just decaying with redshift $z$ like $1/(z+1)^2$
as flux conservation would demand and maybe damped on small scales by diffusion (what are these 'small scales') 
or can MHD processes move power from small to larger scales?
\end{enumerate}

To address the first question, we review the mechanisms to generate primordial magnetic fields. 
They fall into three broad classes: inflationary magnetic field generation, the generation of magnetic fields during phase transitions and magnetic fields from second order cosmological perturbation theory. 

Once the fields are produced at a certain "magnetogenesis" epoch in the early universe, they evolve interacting with different types of plasma of charged particles. Such plasma is present in the Universe both before the moment of recombination and after the epoch of reionization. A judgement of the primordial nature of IGMF is not possible without an understanding of the evolution of the field from the event of production until the present. 

To address the second question, we review current understanding of the evolution processes and describe evolutionary tracks of magnetic fields in the $(B, \lambda_B)$ parameter plane, where $B$ and $\lambda_B$ are the most important integral (i.e. distance scale-averaged) characteristics of magnetic field: its strength and  correlation length.  

Finally, we review present observational constraints on the IGMF and discuss possible ways how to distinguish whether the IGMF comes from primordial fields or from  fields produced by galactic outflows.

Our review is organized in the following way. In Section \ref{s:mhd} we summarize the basic equations governing the evolution of magnetic fields and charged particle plasma in the Universe. Next, in Section \ref{s:gen} we review previously proposed mechanisms of generation of magnetic fields in the early universe. 
In Section~\ref{s:evo} we discuss the evolution of primordial magnetic fields. We discuss different 
damping and amplification mechanisms, determine their characteristic scales as function of cosmic time
and we study their effects on different types of primordial magnetic fields. We also investigate the possibility of an inverse cascade.
In Section~\ref{s:obs} we review the observational situation and compare different constraints on IGMF. We also discuss future capabilities of several different observational strategies. In Section \ref{s:constraints_testable} we re-discuss different cosmological magnetogenesis models to single out models which can be tested by observations. We show that relic cosmological fields occupy a distinct region in the  $(B,\lambda_B)$ parameter plane. This region is different from that expected for magnetic fields from galaxy outflows. We argue that this opens a possibility to distinguish between the primordial and galactic outflow produced fields observationally. In each section we also point out the main open problems which still have to be addressed in future research. In Section~\ref{s:con} we conclude.

%Dynamics (s:mhd)

%%%%%%%%%%%%%%%%%%%%%%%%%%%%%%%%%%%%%%%%%%%%%%%%%%%%
\section{Basics of magnetic fields in an expanding Universe}
\label{s:mhd}
%%%%%%%%%%%%%%%%%%%%%%%%%%%%%%%%%%%%%%%%%%%%%%%%%%%%

%%%%%%%%%%%%%%%%%%%%%%%%%%%%%%%%%%%%%%%%%%%%%%%%%%%%
\subsection{Notations and definitions}
\label{s:notations}
%%%%%%%%%%%%%%%%%%%%%%%%%%%%%%%%%%%%%%%%%%%%%%%%%%%%

In this review we consider magnetic fields evolving in a flat expanding universe described by the metric 
\be
\label{metric}
ds^2 = a^2(t)\left[ -dt^2 +\de_{ij}dx^idx^j\right]\,,
\ee
where $a$ is the scale factor which we normalize to unity today such that it is related to the cosmological 
redshift by $1+z=1/a$. Here, $t$ denotes conformal time which is related to physical time $\tau$ by $adt=d\tau$.
The Hubble parameter is
\be
H =\frac{da/d\tau}{a} = \frac{\dot a}{a^2} = \HH/a \,.
\ee
An overdot denotes derivative wrt. conformal time $t$, and $\HH$ is the conformal Hubble parameter. This focus
on conformal variables is useful since electromagnetism is conformally invariant so that, as we shall see, with suitable re-scaling, the equations for a magnetic field in expanding space are identical to those in Minkowski space.

We denote spacetime indices by greek letters and 3d spatial indices by latin letters, 3d vectors are 
denoted in boldface. We use  natural units in which the Planck constant and the speed of light are unity: $\hbar=c=1$. The reduced Planck mass is denoted by $M_P$, such that $8\pi G =1/M_P^2$. We also set $k_b=1$ so that temperature is a measure of energy. More precisely, $1K=0.86\times 10^{-4}$eV.

The electric and magnetic field are defined in the reference frame comoving with the coordinate system in which the space-time metric has the form (\ref{metric}), with the time axis directed along the vectors of the Hubble flow $(u^\mu)=a^{-1}(1,0,0,0)$. As  in \cite{Barrow:2006ch}, we define
\be\label{e2:EB}
E_\mu = F_{\mu\nu}u^\nu\,, \qquad B_\mu = \ep_{\mu\nu\ga}F^{\nu\ga}/2\,,
\ee
such that
\be
 F_{\mu\nu} = u_\mu E_\nu -u_\nu E_\mu + \ep_{\mu\nu\ga}B^\ga\,.
\ee
Here $ \ep_{\mu\nu\ga}= u^\al\ep_{\al\mu\nu\ga}$ is the totally antisymmetric tensor on the 3-space normal to $u^\mu$, while $\ep_{\al\mu\nu\ga}$ is the totally antisymmetric tensor in 4 dimensions with $\ep_{0123} = \sqrt{-g}$. The indices of $\ep_{\mu\nu\ga}$ are are raised and lowered with 
$h_{\al\beta}=g_{\al\beta} +u_\al u_\beta$. Note that both $B^\al$ and $E^\al$ are normal to the four-velocity $u^\mu$.

These definitions are useful also for a generic 4-velocity $u^\mu$ which need not even by hypersurface-orthogonal. However, since in a perturbed Friedmann Universe, both, the deviation of matter / plasma velocities $u^\mu$ from the Hubble flow and the electromagnetic field are small, it suffices to consider the background velocity in the definition of $E_\mu$ and $B_\mu$, so that the electric and magnetic fields are parameterized by the 3d vectors 
$(E_\mu)=a(0,\bE)$, $(B_\mu)=a(0,\bB)$. Maxwell's equations in terms of $\bE$ and $\bB$ are given in Appendix~\ref{a:max}, eqs.~(\ref{a2:maxF1}) to (\ref{a2:maxF4}). They are simply re-scaled versions of the equations in Minkowski space. Here and in the following we denote 3d vectors in boldface.

Homogeneity and isotropy of the Universe implies that statistically the spatial structure of magnetic fields is the same at any location in the Universe. In the view of this, it is often convenient to study the properties of magnetic field in terms of its Fourier components,  
\be\label{e2:fourier}
\bB(\bk,t) = \int d^3x \bB(\bx,t) e^{i\bk\cd\bx} \,.
\ee
Statistical homogeneity and isotropy then imply that expectation values cannot depend on any vector except $\bk$ and on any tensor except $\de_{ij}$ and $\ep_{ijm}$ as well as combinations of these. The spectrum of the magnetic field therefore is of the form
\be\label{e2:specsa}
a^4\langle B_i(\bk,t)B_j^*(\bk',t)\rangle = (2\pi)^3\de(\bk-\bk')\left[(\de_{ij} -\hat k_i\hat k_j)P_{B}(k) -i\ep_{ijm}\hat k_mP_{aB}(k)\right]\,,
\ee
where $\hat\bk = \bk/k$ and $k=|\bk|$. The bracket $\langle~\rangle$ denotes an ensemble average, i.e. an average over many realizations of the stochastic magnetic field. In observations we of course always only measure one realization, but it is usually justfied to assume an 'ergodic hypothesis', namely that spatial average over many independent patches of size $L\gg 2\pi/k$ is a good approximation to the ensemble average, especially, if the size $L$ is larger than the cosmological horizon at the time when the magnetic field was generated.

$P_B$ and $P_{aB}$ are the symmetric and antisymmetric part of the magnetic field spectrum.  The $\de$--function is a consequence of spatial homogeneity and
the tensor structure comes from the homogeneous Maxwell equation $\bnabla\cd\bB=0$.
In terms of a right handed orthonormal system $(\bfe^{(1)},  \bfe^{(2)},  \bk)$ with
$\bfe^{(+)} =(\bfe^{(1)} +i \bfe^{(2)})/2$ and $\bfe^{(-)} =(\bfe^{(1)} -i \bfe^{(2)})/2$, we have
\bea
\bB(\bk,t) ~=~ B_{(+)}(k,t)\bfe^{(+)} + B_{(-)}(k,t)\bfe^{(-)} && \mbox{ and} \\
\label{e2:spec+}
\langle B_{(+)}(\bk)B_{(+)}^*(\bk')\rangle +\langle B_{(-)}(\bk)B_{(-)}^*(\bk') \rangle &=&  (2\pi)^3\de(\bk-\bk')P_B(k)/a^4\\
\label{e2:spec-}
\langle B_{(+)}(\bk)B_{(+)}^*(\bk')\rangle - \langle B_{(-)}(\bk,t)B_{(-)}^*(\bk') \rangle &=&   (2\pi)^3\de(\bk-\bk')P_{aB}(k)/a^4\,.
\label{eq:pb}
\eea 
Again, the angular brackets signify "ansemble averaging".
Eqs.~(\ref{e2:spec+}) and  (\ref{e2:spec-}) are obtained easily with the help of the identities $(\bfe^{(+)})^2 =(\bfe^{(-)})^2=0$ and $\bfe^{(+)}\cd\bfe^{(-)}=1$
together with $\hat\bk\wedge\bfe^{(+)} =-i\bfe^{(+)}$ and $\hat\bk\wedge\bfe^{(-)} =i\bfe^{(-)}$.
Parity transforms $ B^{(+)}$ into  $ B^{(-)}$ and vice versa, hence $P_B$ is even and $P_{aB}$ is odd under parity.
Eqs.~(\ref{e2:spec+}) and  (\ref{e2:spec-})  also imply that $P_B\ge |P_{aB}|$. Equality is reached if one of the helicity modes vanishes completely; such a field is called totally helical.

The energy density of the magnetic field (in Heavyside-Lorentz units see~\cite{Jackson:1962}) is given by
\be
\label{eq:rhob}
a^4\rho_B = a^2\int dk\rho_B(k) =\frac{1}{2\pi^2}\int \frac{dk}{k}k^3P_B(k)\,.
\ee
Here $\rho_B$ is the ensemble average of the magnetic field energy density, which is independent of position. Eq.~(\ref{eq:rhob}) is obtained by noting that
\bea
\rho_B &=& \frac{1}{2}\langle \bB(\bx)\bB(\bx)\rangle = \frac{1}{2(2\pi)^6}\int d^3kd^3k' \langle \bB(\bk,t)\bB^*(\bk',t)\rangle\exp(i\bx\cdot(\bk-\bk'))  \nonumber\\
&=&  \frac{a^{-4}}{(2\pi)^3}\int d^3k P_{B}(k) =\frac{a^{-4}}{2\pi^2}\int \frac{dk}{k}k^3P_B(k) \,. \nonumber
\eea
Note that $\bB(\bx)^2$ has the dimension of energy density, hence $\bB(\bk)\bB(\bk')$ has the dimension of
energy density$\times$length$^6$ and $P(k)$  has the dimension of energy density$\times$length$^3$ 
which is the dimension of the Fourier transform of the energy density as required.
Hence $\rho_B(k)=d\rho_B/dk=k^2P_B(k)/(2\pi^2)$ is the energy density per unit  $k$ interval.
We shall sometimes also employ the energy density "per log interval": $d\rho_B/d\log(k) = k^3P_B(k)/(2\pi^2)$.

It is also convenient to introduce the "characteristic" magnetic field strength at  scale $\lambda=2\pi/k$,
\be
B_\lambda=\left.\sqrt{2\frac{d\rho_B}{d\log(k)}}\right|_{k=2\pi/\lambda} \quad { and } \quad
B = \sqrt{2\rho_B},
\ee
where $\log$ denotes the natural logarithm. We shall systematically use the field strength on scale $\la$ and the scale-averaged field strength   $B$ in the following sections. 

Below we shall also need the power spectrum of the fluid velocity field, we therefore also introduce it here.
\be\label{e4:vfourier}
\bv(\bk,t)=\int e^{i\bk\cd\bx} \bv(\bx,t) d^3x
 \ee
 and
 \be\label{e4:vspecsa}
\langle v_i(\bk,t)v_j^*(\bk',t)\rangle = (2\pi)^3\de(\bk-\bk')\left[\de_{ij}P_{sK} -\hat k_i\hat k_jP_{vK} -i\ep_{ijm}\hat k_mP_{aK}(k)\right]\,.
\label{eq:pk}
\ee
Like the index $B$ for "magnetic", the index $K$ stands for  "kinetic". 
Note that for $P_{sK}=P_{vK}$ the velocity field is divergence free.
The fact that $\langle v_jv_j^*\rangle \ge 0$ implies $P_{sK}\ge P_{vK}$.

 Similarly to the energy of magnetic field, given by Eq. (\ref{eq:rhob}), 
we also introduce the spectral kinetic energy 
\begin{eqnarray}
\rho_K&=& \frac{1}{4\pi^2} \int  k^2\left(3P_{sK}(k)-P_{vK}(k)\right)dk  = \int \rho_K(k)dk \,,
\end{eqnarray}
where  $\rho_K(k)$ is kinetic energy  per unit wave number interval. Note that, 
contrary to $\rho_B$, $\rho_K$ is dimensionless and has to be 
multiplied by the plasma energy density $\rho$ to yield the true kinetic energy. We shall however 
use the customary language and refer to $\rho_K$ as kinetic energy density.

We assign a characteristic correlation length to the stochastic magnetic and velocity fields, defined by
 \begin{eqnarray}
\lambda_B&=&2\pi\rho_B^{-1}\int \rho_B(k)k^{-1}dk \,,\nonumber\\
\lambda_K&=&2\pi\rho_K^{-1}\int \rho_K(k)k^{-1}dk \,.
\end{eqnarray}
This is the characteristic scale, also called 'integral scale' of the magnetic field and the velocity field. We shall sometimes also call it the 'correlation scale' even though this is not strictly correct in a statistical sense\footnote{In statistical mechanics correlations decay exponentially on scales larger than the correlation scale while our correlations usually decay like a power law. Therefore, even though, most of the magnetic/kinetic field energy is concentrated on scales close to $\la_B$ respectively $\la_K$, this is not true for all its cumulants.}. 

The magnetic field and the velocity field are said to be in equipartition if $\rho_B/\rho \simeq\langle v_A^2\rangle/2
= \rho_K =\langle v^2\rangle/2$. Here $v_A$ is the Alfv\'en speed defined by $v_A^2 = B^2/(2\rho)$.
If $P_B(k) \simeq a^4\rho P_K/2$ we speak of detailed equipartition or equipartition on all scales.

%%%%%%%%%%%%%%%%%%%%%%%%%%%%%%%%
\subsection{Helicity}
%%%%%%%%%%%%%%%%%%%%%%%%%%%%%%%
 
The magnetic helicity is the volume integral
\be
H(V) = \int_V\bA\cdot \bB dv \,,
\ee
over a volume through the boundary of which no magnetic field lines cross. The 3d vector $\bA$ is the magnetic vector potential. The above volume   can also be infinite  if the magnetic field decays sufficiently rapidly at infinity.
The helicity is gauge independent, since under a gauge transformation, $\bA \ra \bA +\bnabla \al$
\be
H(V) \ra H(V) +  \int_V\bnabla\al\cdot \bB dv =) \ra H(V) + \oint_{\dd V}\al \bB\cdot\bn ds = H(V) \,.
\ee
Here $\bn$ is the normal to the boundary $\dd V$ and we have assumed that $\bB\cdot\bn=0$.
Magnetic helicity has a simple topological interpretation in terms of linking and twist of isolated flux tubes, see~\cite{Brandenburg:2004jv}.
The helicity density of the magnetic field is then given by $\mathpzc{h} = \bA\cd\bB$.

Using a gauge in which $\bA$ is transverse, $\bk\cd\bA=0$,
we have in Fourier space $k\bA =i\hat\bk\wedge\bB$ which yields
\be
\mathpzc{h} =  \int \frac{dk}{k}\frac{d\mathpzc{h}}{d\log(k)} =\frac{1}{2\pi^2}\int \frac{dk}{k}k^3P_{aB}(k)\,.
\ee  
Hence $k^3P_{aB}(k)/(2\pi^2)$ is the helicity density per log-k interval.

In a Friedmann-Lema\^\i tre universe, the electromagnetic Lagrangian, 
$$L =\frac{1}{4}\sqrt{-g}F_{\mu\nu}F^{\mu\nu} =  \frac{1}{4}\eta^{\mu\al}\eta^{\nu\beta}F_{\mu\nu}
F_{\al\beta}$$
is independent of the scale factor $a(t)$. Hence a freely propagating electromagnetic field $F_{\mu\al}$ is independent of $a$.
This is simply a manifestation of conformal invariance of electromagnetism in 4-dimensions.
This implies that $B_i \propto 1/a$ and $B^i \propto 1/a^3$ such that $\bB^2\propto a^{-4}$. We have taken out this
trivial conformal scaling in the power spectra in Eqs.~(\ref{e2:specsa},\ref{e2:spec+}) and (\ref{e2:spec-}).

Below we see that this scaling remains true when interactions with the cosmic plasma are relevant
 in the special but cosmologically
most relevant interacting case of the magnetic hydrodynamic (MHD) limit due to flux conservation.

If we write $\bB = B^i\bfe_i$ for the orthonormal basis $\bfe_i =a^{-1}\dd_i$, the scaling of the components is $B^i \propto a^{-2}$.

%%%%%%%%%%%%%%%%%%%%%%%%%%%%%%%%%%%%%%%%%%%%%%%%%%%%%%
\subsection{Co-evolution of the magnetic field and the cosmic plasma}
\label{s:radiation}
%%%%%%%%%%%%%%%%%%%%%%%%%%%%%%%%%%%%%%%%%%%%%%%%%%%%%%

Dynamical equations for the evolution of the interacting matter and electromagnetic fields in the expanding Universe are derived starting from the law of conservation of stress-energy tensor, $T^{\mu\nu}_{;\nu}=0$, for the stress-energy tensor consisting of the electromagnetic and plasma (fluid) contributions. In the simplest case of an ideal fluid, its stress-energy tensor is
\be
T^{\mu\nu}_P=(\rho+p) u^\mu u^\nu-pg^{\mu\nu}
\ee
where $\rho$ and $p$ are the energy density and the pressure of the fluid and $u^\mu$ is its four-velocity. We assume that the different components (photons, electron/positrons gas etc.) of the dominant relativistic particles are sufficiently strongly coupled so that we can consider them as one fluid. (For $T\lsim 1$MeV this means that we neglect the neutrinos in our qualitative considerations). The stress-energy tensor of electromagnetic field is, see Appendix~\ref{a:max}
\bea
&& T_{\mu\nu}^{\rm (em)} =  F_{\mu\lambda}{F^\lambda}_\nu-\frac{1}{4}g_{\mu\nu}F^{\lambda\sigma}F_{\lambda\sigma}\\
&& =\frac{1}{2}(E^2 \!\!+ \!B^2)u_\mu u_\nu \!+\! \frac{1}{2}(E^2\!\! +\! B^2)h_{\mu\nu}\!
- \!E_\mu E_\nu\! -\! B_\mu B_\nu \!+\! P_\mu u_\nu \!+\! u_\mu P_\nu \, .
\eea 
where $P_\mu$ is the Poynting vector.
Following \cite{Brandenburg96}, we introduce the following conveniently rescaled quantities,
\bea 
\tilde\rho = a^4\rho,~ \tilde p = a^4p , ~\tilde B^i =a^2 B^i, ~ \tilde E^i =a^2 E^i, ~\tilde J^i=a^3J^i\,.
\eea
As the present value of the scale factor is unity, this implies that the tilde-quantities correspond to their values scaled to today. In the rest of the review, when ever we indicate $\tilde\bB$ or $\tilde\rho$ we mean the value of the magnetic field or of the energy density scaled to today.
 
Using  Maxwell's equation, see Appendix~\ref{a:max}, we can write  the conservation equations, $(T^{\mu\nu}_P+T^{\mu\nu}_{EM})_{;\nu}=0$ in the form
\begin{eqnarray} \label{e2:ene}
\frac{\partial }{\partial t}[\tilde\rho(1+4v^2/3)] +\frac{4}{3}\bnabla(\tilde\rho\bv) &=&
    -\tilde\bJ\cdot\tilde\bE \\
\label{e2:impu}
\frac{4}{3}\left(\tilde\rho\frac{\partial \bv }{\partial t}+\bv\frac{\partial \tilde\rho}{\partial t} +\tilde\rho (\bv\cdot\bnabla)\bv +\tilde\rho\bv(\bnabla\cdot\bv)\right) &=& -\bnabla\tilde p +\tilde\bJ\wedge\tilde\bB \,. \qquad
\end{eqnarray}
The derivatives are wrt. conformal time $t$ and comoving coordinates $\bx$.
Here we have neglected terms which are of third order in the perturbed quantities like 
$\bv$, $\tilde B$, $\dd_t\tilde\rho$ etc. Even though we consider a relativistic fluid with $p=\rho/3$, peculiar (bulk) velocities are small.  Nevertheless, we want to keep quadratic terms in order to be able to describe non-linearities which can provoke modifications in the spectrum like an inverse cascade.

In the early Universe, conductivity is very high, for relativistic electrons we typically 
have, see~\cite{Enqvist:1994mb,Arnold00,Arnold03} and Appendix~\ref{a:visc}.
\be \label{e2:cond}
\si \simeq \frac{T}{\al\log(\al^{-1})} \,.
\ee
where $\alpha$ is the fine structure constant.
It therefore makes sense to work in the ideal MHD limit where 
\be
\tilde \bE = -\bv\wedge\tilde \bB \quad \mbox{and }\quad \tilde \bJ = \bnabla\wedge \tilde \bB
\ee
to lowest order.  The first equation is simply the condition that the Lorentz force on charged particles vanish. The second equation follows from Amp\`ere's law using 
$\bE\ll \bB$, see Appendix~\ref{a:max}. In this limit $\tilde \bE$ is already of quadratic order and we can consistently neglect the 3rd order term $\tilde\bJ\cdot\tilde\bE$ in Eq.~(\ref{e2:ene}). In this approximation, up to first order 
$\dd_t\tilde\rho = -(4/3)\tilde\rho\bnabla\cd\bv$. 

However, viscosity  can become significant and we want to take it into account. We also take into account the damping of the magnetic field due to 
Ohmic losses. Rescaling also  shear viscosity, $\tilde\nu =\nu/a$  and the conductivity 
$\tilde\si=a\si$, including dissipation in Eqs.~(\ref{e2:ene},\ref{e2:impu}) they 
become, see~\cite{Banerjee}
\begin{eqnarray}   \label{e4:conserv}
&&\frac{\partial \tilde\rho}{\partial t}+\bnabla\left((\tilde p+\tilde\rho)\bv\right)=0 \,, \\
&&\frac{\partial \bv}{\partial t}+(\bv\cdot\bnabla)\bv+\frac{\bv}{(\tilde\rho+\tilde p)}\frac{\partial \tilde p}{\partial t}+\frac{\bnabla \tilde p}{(\tilde\rho+\tilde p)}+\frac{\tilde	\bB\wedge(\bnabla\wedge\tilde\bB)}{(\tilde p+\tilde\rho)}=  \nonumber\\  && \qquad\qquad \tilde\nu\left( \bnabla^2\bv+\frac{1}{3}\bnabla(\bnabla\cdot\bv)\right) \,, \label{e4:euler} \\  
&&\frac{\partial\tilde\bB}{\partial t}- \bnabla\wedge(\bv\wedge\tilde\bB)  =
    \frac{1}{\tilde\si}\bnabla^2\tilde\bB\,. \label{e4:induc}
\end{eqnarray}
In addition, we consider a radiation dominated equation of state $\tilde p=\tilde\rho/3$.
With respect to ~\cite{Banerjee} we have neglected 'heat losses' as in the radiation dominated era the photons are part of the plasma and their energy density is included in $\rho$. 

The first equation is just the continuity equation for the cosmic plasma. The second equation is the Euler equation with the dissipation term on the right hand side. 
The rescaled shear viscosity is of the order of the comoving mean free path of the plasma,
$\tilde\nu \sim \lambda_{\rm mfp}/5$, see Appendix~\ref{a:visc}. 

On small scales $\la \ll \lambda_{\rm mfp}/5$, 
we have to replace diffusion damping by damping due to free streaming. This can be done by replacing the dissipation term by $-\tilde\alpha\bv$, where $\tilde\alpha\propto \la_{\rm mfp}^{-1}$ (see~\citet{Banerjee:2004df}).
A more rigorous treatment would require to solve the Boltzmann equation. Nevertheless, qualitatively 
we expect fluctuations to be damped by diffusion and by free streaming on scales $\la \lsim \la_{\rm mfp}$.

The third equation
is the magnetic induction equation, with the Ohmic dissipation term on the right hand side.

If the dissipation terms  are subdominant to the non-linear terms, 
MHD turbulence develops. This is controlled by the Reynolds numbers,
\bea
{\rm R}_k(k) &=&  \frac{v_k}{k\tilde\nu}   \qquad \mbox{(kinetic Reynolds number),}  \label{eq:reynkin} \\
{\rm R}_m(k) &=&  \frac{v_k\tilde\si}{k}   \qquad \mbox{(magnetic Reynolds number),}\\
 {\rm P}_m = \frac{{\rm R}_m}{{\rm R}_k} &=& \tilde\si\tilde\nu   \qquad \mbox{(Prandl number).}
 \label{eq:reynolds}
\eea
Here $k$ is some comoving wave number and $v_k=\sqrt{\langle|\bf v|^2\rangle} = \sqrt{P_K(k)k^3/(2\pi^2)}$.
In Appendix~\ref{a:visc} we compute these numbers for $k=k_B =2\pi/\la_B$ rsp. $k=k_K$ as  functions of the temperature and show that for $T<100$~GeV the Prandl number is much larger than one, so that we may neglect
magnetic diffusion with respect to the kinetic one which is much faster.

On scales where the Reynolds numbers are large, the quadratic terms
$(\bv\cd\bnabla)\bv \sim kv^2$ and $\bnabla\wedge(\bv\wedge\tilde\bB)\sim kv\tilde B$ dominate
over the damping terms $\sim \tilde\nu k^2v$ (or $k^2\tilde B/\si$ for small Prandl number) and turbulence develops.

On small scales, $k>k_d \simeq v_k/\nu$, the damping term dominates and the velocity field is damped. 
Due to the coupling to the magnetic field, this damping is only a power law, but nevertheless very rapid (on the timescale $t_d \sim k_d^{-1}$. Once the velocity field is essentially damped away, the quadratic term in the induction equation drops and the magnetic field remains frozen. Later on, when the viscosity scale becomes smaller, the magnetic field re-generates a velocity field. 

If (which is not the the case in the situations we are interested in) $k>k_d \simeq v\si> v/\nu$, the magnetic field is damped and non-magnetized fluid turbulence remains.

If the fluid is incompressible, $\dd_t\tilde\rho=0$, Eq.~(\ref{e4:conserv}) implies 
$\bnabla\cd\bv=0$ and the fluid motion is purely vortical. It has been argued, see e.g.~\citet{Jedamzik98,Banerjee:2004df}, that this is the case for cosmological magnetic fields. However, even though we know that cosmological density fluctuations are small on large scales, this need not be the case on small scales. Especially, since the energy density $\rho$ also contains kinetic energy, we expect its fluctuations to be at least of order $v^2$. This means that compressible terms in the MHD equations are, in general, as important as the incompressible ones. We shall see in Section~\ref{s:evo}, that this is  relevant for the evolution of the magnetic field spectrum.

\citep{Boyarsky12a} have shown that an additional effective 
degree of freedom (chiral asymmetry) should be added to these equations. 
Its origin is a subtle quantum effect -- the chiral anomaly -- that couples 
the change in the number of left and right-chiral particles with the 
change of the helicity of the magnetic field. Taking into account this 
degree of freedom and its interaction with electromagnetic fields 
significantly changes the evolution in the case of the strong helical 
magnetic fields at temperatures above a few MeV.

Let us summarize the situation as follows:
From equation (\ref{e4:euler}) we see that magnetic field
sources the velocity field. Thus, the process which leads to the production of magnetic fields simultaneously sets the plasma in motion. The nonlinear form of the Euler equation assures that the plasma motions are turbulent, so that the process of generation of magnetic fields is inevitably accompanied by the excitation of plasma turbulence. Furthermore,
the turbulent velocity field couples back to the magnetic field via the term $\bnabla\wedge (\bv\wedge \tilde\bB)$ which leads to turbulence also in the magnetic field. Thus, a consistent description of  the co-evolution of magnetic field and plasma in the radiation dominated Universe has to be described in the "language" of  MHD turbulence (see e.g. the books by \citet{Biskamp:2003,Tsytovich:1977}).

%Generation (s:gen)
%%%%%%%%%%%%%%%%%%%%%%%%%%%%%%%%%%%%%%%%%%%%%%%%%%%%
\section{Generation of primordial magnetic fields}
\label{s:gen}
%%%%%%%%%%%%%%%%%%%%%%%%%%%%%%%%%%%%%%%%%%%%%%%%%%%%

%%%%%%%%%%%%%%%%%%%%%%%%%%%%%%%%%%%%%%%%%%%%%%%%%%%%%%%
\subsection{Inflationary magnetic field production}\label{s:infla}
%%%%%%%%%%%%%%%%%%%%%%%%%%%%%%%%%%%%%%%%%%%%%%%%%%%%%%%

The electromagnetic field is conformally coupled and does not 'feel' the expansion of the Universe.
Therefore, in order to generate magnetic fields during inflation, one has either to couple the electromagnetic 
field to the inflaton or to introduce another coupling which breaks conformal invariance, e.g. a term 
$L_{\rm int} \propto R_{\mu\nu\al\beta}F^{\mu\nu}F^{\al\beta}$ or even break gauge invariance, like 
$R_{\mu\nu}A^\mu A^\nu$ for example.
These possibilities have first been investigated by~\cite{Turner:1988aa} and by \cite{Ratra:1991bn} and later been revisited by many authors~ \citep{Martin:2007ue,Subramanian:2009fu,Kunze:2009bs,Kandus:2010nw,Motta:2012rn,Jain:2012ga} to cite a few recent accounts. Another possibility is that during inflation gauge symmetry is broken and the gauge fields become massive, which also breaks conformal symmetry \citep{Enqvist:2004yy}. In Section~\ref{s:constraints_testable}  we discuss possible observational signatures of inflationary magnetic fields. Below we indicate some of the constraints which are summarized in 
Fig.~\ref{fig:exclusion_inflation}.

%%%%%%%%%%%%%%%%%%%%%%%%%%%%%%%%%%%%%%%%%%%%%%%%%%%%%%%
\subsubsection{Standard inflaton coupling}\label{ssec:infF}
%%%%%%%%%%%%%%%%%%%%%%%%%%%%%%%%%%%%%%%%%%%%%%%%%%%%%%%
We consider the Langangian
\be
L =  \sqrt{-g}\left[\frac{1}{2\ka^2}R + \frac{1}{2}\nabla_\mu\phi\nabla^\mu\phi+V(\phi) +\frac{f(\phi)}{4}F_{\mu\nu}F^{\mu\nu}\right]\,.
\ee
Adopting Coulomb gauge $A_0(\bx,t)=0$, $\partial_j A^j(\bx,t)=0$ and following the notation of 
\cite{Subramanian:2009fu}, Maxwell's equations, $[f^2F^{\mu \nu}]_{,\nu}=0$, lead to an evolution equation for the space components $A_i(\bx,t)$. In a cosmological background it
reads~\citep{Subramanian:2009fu}
\be
\ddot A_i+2\frac{\dot f}{f}\dot A_i-\De A_i=0~,
\label{e2:A}
\ee
where $\De$ is the comoving spatial Laplacian. For a Fourier mode $\bk$, we simply have $\De = -k^2$.
The time evolution of the vector potential depends on the coupling function $f(\varphi)$. One may adopt, 
at least for a short time, a simple power law in conformal time~\citep{Martin:2007ue}:
\be
\label{f}
f(t)=f_1\left(\frac{t}{t_1}\right)^\gamma\;. 
\ee 
For example for power law inflation with an exponential potential this corresponds to a coupling of the 
form $f\propto \exp(-\al \phi/M)$.
For this coupling the damping term is simply $2\ga/t^2$ and in Fourier space eq.~(\ref{e2:A}) can be solved in terms of Bessel functions. Setting for the electromagnetic potential in Fourier space $\bA^{(\pm)}(\bk,t)= {\tilde A}(k,t)\bfe^{(\pm)}/a$
we obtain
\be
{\tilde A}(\bk,t)=\sqrt{\frac{x}{k}}\Big[C_1(\ga)J_{\gamma-1/2}(x)+C_2(\ga)J_{-\gamma+1/2}(x)\Big]\; ,
\label{e2:Asolution}
\ee
where $x\equiv |kt|=-kt$, $J_\nu$ denotes the Bessel function of order $\nu$, and $C_1,~C_2$ are $\ga$ dependent coefficients which are fixed as usual by imposing vacuum initial condition on sub-horizon scales, $-kt\rightarrow\infty$\citep{Subramanian:2009fu}.  Note that during inflation conformal time $t$ is negative.
For the symmetric magnetic and electric field spectra and their correlator we 
obtain~\citep{Martin:2007ue,Subramanian:2009fu}
\bea
P_B &=& 4\pi \frac{k^2}{f^2} |{\tilde A}(k,t)|^2\; , \qquad
P_E ~=~ 4\pi\left|\left(\frac{{\tilde A}(k,t)}{f}\right)'\right|^2\; \mbox{ and}\\
P_{EB} &=& 4\pi \frac{k}{f} \left(\frac{{\tilde A}(k,t)}{f}\right)'{\tilde A}^*(k,t)  \,.
\eea
In this case no anti-symmetric part is generated, $P_{aB}=0$.

On super-horizon scales, $x\ll 1$, we can approximate the Bessel functions by power laws so that these spectra become
\bea
P_B(k,t) &= & \frac{4\pi k}{f^2} \left\{\begin{array}{ll} |c_1|^2x^{2\ga} & \mbox{if } \ga<1/2 \\
|c_2|^2x^{2-2\ga} & \mbox{if } \ga>1/2 \,,\end{array} \right.\label{PB}\\
P_E(k,t)  &=& \frac{4\pi k}{f^2} \left\{\begin{array}{ll} \frac{4|c_1|^2}{(\ga+1/2)^2}x^{2\ga+2} & \mbox{if } \!\ga
   \!<-1/2 ,\\
(1\!-\!2\ga)^2|c_2|^2x^{-2\ga} & \mbox{if } \ga\! >\!-1/2 \end{array} \right. \label{PE}\\
P_{EB} (k,t) &=& \frac{4\pi k}{f^2}   
\left\{\begin{array}{ll} \frac{-2|c_1|^2}{\ga+1/2}x^{2\ga+1} & \mbox{if } \ga < \!-1/2 \\
 (2\ga-1)c^*_1c_2&
\mbox{if } -1/2\!<\ga\!<\!1/2 \nonumber\\
(2\ga-1)|c_2|^2x^{1-2\ga} &\mbox{if } \ga\!>\!1/2\,. \end{array} \right.\\
\label{PEandB}
\eea  
The coefficients $c_i$ are $\ga$-dependent but of order unity. We want to discuss the dependence of these spectra 
on $\ga$. First of all, in order to avoid an infrared singularity in this simple model we must require $-2\le\ga\le 2$. At the boundary the divergence is logarithmic and can be removed in a way which depends only very weekly on the cutoff. Also, when $\ga<0$ the magnetic energy density dominates while for $\ga>0$ the electric energy density dominates. For $\ga =-2$ the magnetic power spectrum is scale invariant and we obtain
\be
\frac{d\rho_B}{d\log k} \simeq \frac{2}{\pi}\frac{|c_1|^2}{f_1^2t_1^4a^4} < \rho_\phi \sim M^2_P/(a^2t^2)\,.
\ee
The condition $\frac{d\rho_B}{d\log k}<\rho_\phi$ is required such that we can neglect the effects of the magnetic energy density on inflationary expansion (backreaction). The same condition has to be satisfied independently by $\rho_E$. Normalizing $e^2$ such that $f=f_1=1$ after inflation, we obtain in all cases
the ratio
\be\label{e3:ampli}
\frac{\rho_B}{\rho_{\rm rad}} \simeq  \frac{1}{(t_1a_1M_P)^2} \simeq \left(\frac{H_{\rm inf}}{M_P}\right)^2 \,.
\ee
With a suitable choice of $f_1^2t_1^4$ it is then easy to obtain magnetic fields of the order of e.g. $10^{-9}$Gauss$/a^2$
on all cosmologically relevant scales in the scale invariant case, $\ga\simeq -2$, while maintaining the condition $f_1t_1a<M_P^{-1}$ during all of inflation, in order to prevent back reaction. After inflation, the conductivity of the cosmic plasma is very high and the 
electric field is rapidly damped. The inflaton is frozen and the function $f(\phi)\ra1$. This scenario has  one serious problem: For $\ga \sim -2$ $f$ is a rapidly growing function during inflation. On the other hand, the electron
field does not couple to $\bA$ but to the canonically normalized electromagnetic potential, $\sqrt{f}\bA$. 
The charge of the electron is therefore $e/\sqrt{f}$ rapidly decreasing. To arrive at $e^2 =1/137$ at the 
end of inflation, $e^2/f$ must have been much larger than 1 during most of inflation. The electron field becomes 
strongly coupled and we cannot trust our perturbative quantum field theory calculation anymore. This problem 
has been noted first by~\cite{Demozzi:2009fu}. We cannot solve it by simply changing $A_\mu$ to $\sqrt{f}A_\mu$
since such a coupling violates gauge invariance. This is actually the only way to save this model, to violate
gauge invariance. The consequences of this, e.g. the generation of electron-positron pairs due to this coupling to the inflaton have not yet been studied.

This problem is avoided if $\ga>0$, hence $f$ is decreasing. But then, since the magnetic field power spectrum is a power law with spectral index $n_s =1+2\gamma$ (if $\gamma<1/2$) and $n_s=3-2\gamma$ (if $\gamma>1/2$) 
the magnetic field spectrum is very blue. Let us denote  by $k_{\max}$ the smallest scale on which $B$ is still generated
during inflation, i.e. the scale that exits the horizon briefly before the end of inflation. If on this scale, the magnetic field energy
density is a fraction $\ep$ of the radiation density after inflation, on some other scale $k_1$ we then have
\be\label{e2:Bblue}
\frac{1}{2}B^2(k) = \left.\frac{d\rho_B(k)}{d\log k}\right|_{k=k_1} = \ep\rho_{\rm rad}
\left\{
\begin{array}{ll}
\left(k_1/k_{\max}\right)^{4+2\ga}&,\gamma<1/2\\
\left(k_1/k_{\max}\right)^{6-2\ga}&,\gamma>1/2.
\end{array}
\right.
\ee
This magnetic field spectrum for the case $\ga=0$ is shown in figure~\ref{fig:exclusion_inflation} for two different values of the inflation scale.

Knowledge of the initial spectrum of the magnetic field allows to make predictions for the expected "relic" magnetic field which might survive until the present epoch. We shall in the following term 'naive evolution of the power spectrum' and evolution where $P_B(k)$ does not change on large scales and is simply damped away beyond a certain 
small damping scale $k_{\rm damp}(t)$ which may depend on time. This 'naive evolution' assumes that beyond the damping scale, the magnetic field just scales like $B\propto 1/a^2$ which is required by flux conservation. We shall show later, in Section~\ref{s:evo}, that evolution is usually more complicated.
Assuming naive evolution, Eq.~(\ref{e2:Bblue}) is nearly time independent since both, $\rho_{\rm rad}$ and $\rho_B$ scale as $a^{-4}$ (apart from the 
changes in the number of relativistic degrees of freedom which we do not expect to account for more that one or two orders of magnitude). The present radiation density  is given by
\be\label{e2:radGauss}
      \rho_{\rm rad}(t_0) \simeq 2\times 10^{-15}({\rm eV})^4 \simeq  
     4.66\times 10^{-34}{\rm g/cm}^3 \simeq \frac{\left(3\times 10^{-6} {\rm G}\right)^2}{8\pi} \,.
\ee
Inserting this in eq.~(\ref{e2:Bblue}) we obtain
\be\label{e2:Bb}
\left.\frac{d\tilde\rho_B(k)}{d\log k}\right|_{k=k_1} = \ep\left(\frac{k_1}{k_{\max}}\right)^{4+\ga}\!\! \frac{\left(3\times 10^{-6} {\rm G}\right)^2}{8\pi} \,.
\ee

If inflation happens at high energy with 
$H_{\rm inf} \sim E_{\rm inf}^2/M_P$ with $E_{\rm inf} \simeq 10^{15}$GeV, this yields 
$$
k_{\max} \sim 1/t_{\rm end}  \sim \frac{H_{\rm inf}}{1+z_{\rm end}}
\sim \frac{E_{\rm inf}T_0}{M_P} \simeq 10^{-3}{\rm cm}^{-1}\left(\frac{E_{\rm inf}}{10^{15}{\rm GeV}}\right)\,.
$$
where $t_{\rm end}$ is the comoving itme at the end of Inflation and  $T_0$ is the present temperature of CMB, $T_0\simeq 2.3\times 10^{-4}$~eV. On a scale of say $k_1^{-1} \sim 1$Mpc $\simeq 3\times 10^{24}$cm, eq.~(\ref{e2:Bb}) then yields a tiny left over field of
\be\label{e2:Bb1}
\left.\frac{d\tilde\rho_B(k)}{d\log k}\right|_{k_1=1\mbox{ Mpc}} \sim \frac{\ep}{8\pi}\left(3\times 10^{-49}{\rm Gauss}
 \left(\frac{10^{15}{\rm GeV}}{E_{\rm inf}}\right)^2\right)^2\,.
\ee
Here we have set $\ga \sim 0$ to obtain the most optimistic value, which is still devastatingly small.

Lowering the inflation scale helps somewhat but even when setting it to the electroweak scale, 
$E_{\rm inf}\sim$200~GeV, we obtain only fields of $10^{-23}$Gauss on Mpc scales, assuming $\epsilon\sim 1$. However, the natural normalization of the magnetic field energy spectrum is $\rho_B\sim k_{\rm max}^4\sim H_{\rm inf}^4\sim$, while the energy density of the Universe scales as $\rho\sim M_P^2 H^2$. This means that typical model calculation results in $\epsilon\sim H^2_{\rm inf}/M_P^2\ll 1$ for $H_{\rm inf}\ll M_P$. Thus, in fact, lowering the energy scale of inflation generically results in weakening of magnetic fields. 

Very generically we shall see that if we want to generate magnetic fields early and if we want
to have reasonably large fields also on large scales, in order for the small scale fields 
not to over-close the Universe their spectrum should not be very blue. We must have either
$n_s\sim -3$, or an evolution which raises the magnetic field power on large scales by some plasma processes. The latter is called an 'inverse cascade'.

Furthermore, from inflation we expect $\ep =\rho_B/\rho_{\rm rad} \simeq H_{\rm inf}^2/M_P^2$. This value is indicated by the thick solid line in Fig.~\ref{fig:exclusion_inflation}.To obtain much larger amplitudes, like e.g. equipartition, $\ep \simeq 1$ as assumed in the dashed line annotated by $E_{\rm inf} =200$GeV, we need in addition a dynamo mechanism e.g. during reheating which rapidly amplifies the magnetic field to equipartition.
This possibility is not excluded but also not confirmed by any detailed study.

%%%%%%%%%%%%%%%%%%%%%%%%%%%%%%%%%%%%%%%%%%%%%%%%%%%%%%
\subsubsection{Coupling to curvature}\label{ssec:infR}
%%%%%%%%%%%%%%%%%%%%%%%%%%%%%%%%%%%%%%%%%%%%%%%%%%%%%%

We want to discuss biefly also another possibility, namely that the electromagnetic field is coupled to curvature. We consider the Langrangian 
\be
L = \sqrt{-g}\left[\frac{R}{2\ka^2} + \frac{1}{4}\left(F^{\mu\nu}F_{\mu\nu} +\frac{\al}{m^2}R^{\mu\nu\al\beta}F_{\mu\nu}F_{\al\beta}\right) \right]\,.
\ee
Varying the action with respect to $A^\mu$ we find
\be
\dd^\mu\left( F_{\mu\nu} +\frac{\al}{m^2}{R_{\mu\nu}}^{\al\beta}F_{\al\beta}\right) = 0 \,.
\ee
In a Friedmann Universe we have
\be
{R^{0i}}_{0j} = \frac{\dot\HH}{a^2}\de^i_j \quad \mbox{ and } \quad {R^{ij}}_{\ell m} = 
\frac{\HH^2}{a^2}\left(\de^i_\ell\de^j_m-\de^i_m\de^j_\ell\right)
\quad \mbox{ with }\ee 
\be   \HH^2/a^2 =\rho/(3M_P^2) \quad \mbox{ and } \quad\dot\HH/a^2 =(\rho+3P)/(6M_P^2)\, . 
\ee
During perfect de Sitter expansion, 
${R_{\mu\nu}}^{\al\beta}$ is constant and the curvature term does not affect the equations of motion. This is also
true for the other possible curvature terms, $RF_{\mu\nu}F^{\mu\nu}$ and $R_{\mu\nu}F^{\mu\al}{F^\nu}_\al$.
Therefore, in this case magnetic field production is suppressed by the slow roll parameters.

We now assume $p=w\rho$ with $-1\lsim w$. This is the case of power law inflation, where  the scale factor
and the energy density behave like 
\bea
a \propto t^{\frac{2}{1+3w}} & \mbox{ and } & \rho \propto t^{\frac{-6(1+w)}{1+3w}} , \mbox{ so that} \\
 (\dot\HH/a^2) =(1+3w)\HH^2/a^2\,, && (\dot\HH/a^2)^. = -6(1+w)\HH^2/(a^2t)\,.
\eea
Inserting this in the equation of motion for $\tilde\bA$ in Coulomb gauge we find
\be\label{e2:ARall}
(1-\frac{2\al}{m^2}(1+3w)\HH^2)\ddot{\tilde A} + \frac{12\al}{m^2}(1+w)\HH^2\frac{1}{t}\dot{\tilde A} +k^2(1-\frac{2\al}{m^2}\HH^2)\tilde A =0\,.
\ee
Typically, the relevant mass scale is the electron mass, $m\sim m_e$ and the curvature terms dominate
in the early universe when $\rho >M_P^2m^2$. Terms of this form do actually occur in 1 loop vacuum polarization calculations, see~\cite{Drummond:1978hh}.  This is the situation we want to consider.
We therefore neglect the standard term and obtain
\be\label{e2:AR}
\ddot{\tilde A}_i - \frac{6(1+w)}{(1+3w)}\frac{1}{t}\dot{\tilde A}_i +k^2\frac{1}{(1+3w)}\tilde A_i =0\,.
\ee
On large scales, $|kt|\ll1$ there is an uninteresting constant mode and a mode behaving like 
$$ \tilde A_i \propto t^{1+\frac{6(1+w)}{(1+3w)}} \propto a^{(7+9w)/2} \,.$$
For $w>-7/9$ this is a growing mode. The general solution is again given in terms of Bessel functions
with coefficients which are determined by the initial conditions. At early times, $|kt|\gg 1$, we may 
neglect the non-standard first derivative term and start from the Minkowski vacuum.  Using gain the variable $x=-kt$
we find in terms of Hankel functions of the second kind (see \citealt{Abramowitz:1972})
\be
\tilde A_i =\frac{c}{\sqrt{k}}x^\nu H^{(2)}_\nu(x)\,, \qquad \nu =\frac{7+9w}{2(1+3w)}\,.
\ee
 On large scales, $x\ll 1$, the magnetic field spectrum is given by 
\be\label{e2:specR}
P_B(k,t) = k^2\tilde A^2 \simeq |c|^2k\left\{\begin{array}{ll}
 x^{2\nu} & \mbox{ if } \nu<0 \, \qquad n_s=\frac{8+12w}{1+3w}\\
  1 & \mbox{ if } \nu> 0 \, \qquad n_s=1 \end{array} \right.
\ee
If $\nu<0$, i.e. $-7/9<w<-1/3$ we can obtain a red spectrum. Actually, we must ask that $w< -4/7$ 
in order to avoid an infrared divergence, i.e., to obtain $n_s>-3$.

On the other hand, the scalar spectral index $n$ of CMB fluctuations in power law inflation is then given 
by (see e.g.~\citealt{Durrer:2008aa})
\be
n-1 = \frac{6(1+w)}{1+3w} \quad \mbox{ such that } \quad 
   w = -\frac{1-(n-1)/6}{1-(n-1)/2} \simeq -1 + (1-n)/3 \, .
\ee
With present data~\cite{Komatsu:2010fb}, which requires $1-n\le 0.05$ we cannot  reach $w>-7/9$.
However, this argument is not entirely solid since $w$ varies slowly during inflation. We know that it has 
been very close to -1 when the CMB scales of order several 100Mpc exited the Hubble scale, but it may 
have been larger later, when e.g. the scale of 1Mpc, relevant for primordial magnetic fields exits the horizon.
Hence a running spectral index with $n \sim 0.96$ at 100Mpc and $n\sim 2$ at 1Mpc such 
that $w\sim -0.6$
at the time when 1Mpc exits the horizon might be marginally possible. Even though simple running is also
strongly constrained by $dn/d\log k = -0.022 \pm 0.02$ at the pivot scale of about 100Mpc~\cite{Komatsu:2010fb}.

However, the CMB results are certainly not compatible with power law inflation at constant $w$ and $w>-7/9$.

The maximal amplitude is again such that
\be
\frac{\rho_B}{\rho_{\rm rad}} \simeq \left(\frac{H_{\rm inf}}{M_P}\right)^2 \,,
\ee
and in $n_s\neq -3$, a dynamo mechanism after inflation is needed to obtain fields with observable amplitude, see figure~\ref{fig:exclusion_inflation}.
 
%%%%%%%%%%%%%%%%%%%%%%%%%%%%%%%%%%%%%%%%%%%%%%%%%%%%%
\subsubsection{Helical inflaton coupling}
%%%%%%%%%%%%%%%%%%%%%%%%%%%%%%%%%%%%%%%%%%%%%%%%%%%%%

We can also add a term $\De L = \sqrt{-g}f(\phi)\tilde F F$ to the Lagrangian, where
$$ \tilde F^{\mu\nu} =\frac{1}{2}{\ep^{\mu\nu}}_{\al\beta}F^{\al \beta}$$
is the Hodge dual of the 2-form $F=\frac{1}{2}F_{\al\beta}dx^\al\wedge dx^\beta$. In terms of electric and magnetic components we have $\tilde F F = -4\bB\cdot\bE$.
Along the same lines as above one can now derive the equation of motion for the gauge potential.
In Coulomb gauge, writing the  Fourier component of the vector potential in the helicity basis,
$\tilde\bA(\bk) = \tilde A_{(+)}(\bk) \bfe^{(+)} (\bk) + \tilde A_{(-)}(\bk) \bfe^{(-)}(\bk) $ we obtain
\be\label{2e:Ah}
\ddot {\tilde A}_{(\pm)} +[k^2\pm  k\dot f]\tilde A_{(\pm)}  =0 \,.
\ee
It is interesting to compare this equation with (\ref{e2:A}). The main difference is the factor $k$ which replaces 
here one time derivative. This comes from the fact that the mixed term $\bB\cdot \bE$ has one time and one spatial derivative. Another very important difference is of course the different sign for the two helicities. There is always one helicity which will be enhanced and the other which will be supressed.

The new term  $k\dot f\tilde A_{(\pm)}$ is much smaller than the $k^2\tilde A_{(\pm)}$ on sub-horizon scales and much smaller than the term $\ddot {\tilde A}_{(\pm)}$ on super horizon scales. Only at horizon crossing in can be relevant. 

Choosing $f$ such that $\dot f = f_N/t$ with a roughly constant pre-factor $f_N$, Eq.~(\ref{2e:Ah}) can be 
solved exactly in terms of Coulomb wave functions~\citep{Durrer:2010mq},
\be
\tilde A_{(\pm)}(x) =\frac{1}{\sqrt{2k}}\left[G_0(\mp f_N/2,x) +iF_0(\mp f_N/2,x) \right] \,, \qquad x =|kt| \,.
\ee
Here $G_0$ and $F_0$ are the irregular and regular Coulomb wave functions~\citep{Abramowitz:1972},
and the pre-factors can be obtained by requiring vacuum initial conditions.
From the asymptotics of these functions for small $x$ one finds on super Hubble scales
\be\label{e2:sphel}
P_B(k) =  k \frac{\sinh(\pi f_N)}{\pi f_N}\, \qquad P_{aB}(k) =  k \frac{\cosh(\pi f_N) -1}{\pi f_N} \,,  \qquad n_s=n_a =1\,.
\ee
The amplitude of this spectrum can become very large if $f_N\gg 1$, however, then the new interaction Lagrangian dominates over the standard term and it is not clear that the perturbative approach adopted here is still valid.
The spectral energy density  grows like $k^4$ and is dominated by the upper cutoff,
$$ \frac{d\rho}{d\log k} \simeq \frac{k^4}{a^4}\frac{\sinh(\pi f_N)}{4\pi^3 f_N}\,, \qquad
\rho_B(t_{\rm end}) \simeq H_{\rm inf}^4\frac{\sinh(\pi f_N)}{16\pi^3 f_N} \,.$$
Here we use that the Hubble parameter is approximately constant during inflation, $H_{\rm end} \sim H_{\rm inf}$.
$$\frac{\rho_B}{\rho_{\rm{rad}}} \simeq \Om_B (t_{\rm end}) \simeq \frac{\sinh(\pi f_N)}{58\pi^3 f_N} \left(\frac{H_{\rm inf}}{M_P}\right)^2 \sim \left(\frac{H_{\rm inf}}{M_P}\right)^2\, .
$$
From this we first conclude that if $f_N$ is not too large, back reaction is unimportant since $H_{\rm inf} \ll M_P$.
However, since this energy density is dominated by the contribution at the high-$k$ end, $k\sim H_{\rm inf}$,
and since the spectrum is again blue, $n_s =1$, we have to draw the same conclusion as in Section~\ref{ssec:infF}.

In this case, however, helicity conservation requires an inverse cascade which alleviates the constraints
somewhat. An  analysis using the evolution of the spectrum during an inverse cascade as proposed in~\cite{Campanelli:2007tc} is presented in~\cite{Durrer:2010mq}, see also Sections~\ref{s:evo} and \ref{s:constraints_testable} of this work.

In conclusion we retain: inflation usually leads to a blue spectrum of magnetic fields. It can generate a scale-invariant spectrum  only if either the 
spectrum of scalar inflaton fluctuations is very blue $n \sim 1.8$ below about 1Mpc (curvature coupling) 
or if the charge of the electron becomes very large during inflation (inflaton coupling). This latter conclusion can 
be evaded if gauge invariance is broken during inflation. A helical coupling to the inflation generically leads maximally helical fields with spectral index $n_s =n_a =1$. If simply scaled to today, such a blue spectrum from the early universe has far too little power on Mpc scale to account for the magnetic fields in galaxies, clusters and voids. 

The typical amplitude expected from the quantum generation of magnetic fields during inflation is like the one of gravitational waves given by
\be
\frac{\rho_B}{\rho} \simeq \left( \frac{H_{\rm inf}}{M_P}\right)^2 \,, \quad v_A \sim \frac{H_{\rm inf}}{M_P}\,.
\ee

This fraction can in principle be amplified by dynamo action after inflation, e.g. during reheating to near equipartition. 
A possibility which has been proposed by~\cite{Sigl97} for fields generated at first order phase transitions, but which may as well be realized at reheating.

%%%%%%%%%%%%%%%%%%%%%%%%%%%%%%%%%%%%%%%%%%%%%%%%%%%%%
\subsection{Magnetic fields from cosmological phase transitions}\label{s3:trans}
%%%%%%%%%%%%%%%%%%%%%%%%%%%%%%%%%%%%%%%%%%%%%%%%%%%%%

Let us now investigate another possibility, namely that magnetic fields are generated during a 
phase transition. 

Even if the electroweak phase transition is very weak, of second order or only a cross-over, magnetic fields with correlation length at the phase transition of the order of $a\lambda_*\sim 1/T \sim 1/m_W$
can form (As before, $\la$ denotes comoving scales hence the physical correlation length is $a\la_*$). \cite{vachaspati91} and \cite{enquist93} (see also~\cite{grasso97}) have estimated that these fields have an amplitude of the order of $B\sim m_W^2$. However, the above correlation scale is smaller than the mean free path of particles in the plasma and is of the order of the inter-particle distance in the plasma, so that one can hardly speak about a persistent magnetic field on time scales larger than $1/T$ in this case. Furthermore, the Ohmic dissipation time on this distance scales, $\tau_{\rm Ohmic}\sim (a\lambda_*)^2\sigma\sim T^{-1}$ is many orders of magnitude shorter than the Hubble time, $H^{-1}\sim M_P/T^2$ at the electroweak phase transition see Appendix~\ref{a:visc}. Therefore, even if such fields are generated, they are rapidly damped away. 

Magnetic fields with correlation length significantly larger than $T^{-1}$ can be generated in the scenario of \citet{Joyce:1997uy}. There the correlation scale is enhanced by the left-right asymmetry in the 
leptonic sector. The scale of magnetic fields then becomes 
of the order of $a\lambda_* \sim \mu^{-1}\sim  (T/\mu)T^{-1}$, where $\mu \lesssim T$ is  
the chemical potential for right-handed leptons. The scenario of \citet{Joyce:1997uy} operates in the temperature range much above the electroweak scale, $T\gtrsim 80$~TeV. 

At the QCD phase transition, the situation is somewhat different. For $100$~GeV$>T>1$~MeV, damping by viscosity  
(see~\citealt{Caprini:2009yp} and Appendix~\ref{a:visc}) is dominated by 
the neutrinos with the mean free path  $a\la_{\rm mfp} \simeq (3G_F^2T^5)^{-1} \sim \left[100\mbox{ GeV}\right]^{-1}\left(100\mbox{ GeV}/T\right)^5$,  while the magnetic diffusivity is of the order $\al(T)/T$, which assures that Ohmic dissipation still damps away the fields on small scales $a\lambda_*\sim 1/T\ll a\lambda_{\rm mfp}$.  Here $G_F\simeq 1/(292$GeV$)^2$ is the Fermi scale and $\al(T)$ is the fine structure constant at energy $T$. Magnetic fields on scales shorter than $a\la_{\rm mfp}(T)$ but larger than the Ohmic dissipation scale are frozen in. 

As long as the electrons are relativistic, both, the correlation length and the magnetic diffusivity scale grow like the scale factor. Hence the relation $\mu\ll a\la_*$ is maintained. However, as we shall see in Sections~\ref{s:evo} and~\ref{s:constraints_testable}, also these fields subsequently decay.

In this section we discuss in some detail the magnetic field spectrum which may result from a first order phase transition. A realistic value for the correlation length, somewhere in-between the extreme values $a\la_*\sim 1/T$ and $\la_*\simeq\ell_H$, where $\ell_H=1/\HH$ denotes the comoving Hubble scale, is still a matter of some debate. For a second order phase transition we do expect it to be of the order of $1/T$.
However, if the transition is first order, we expect a correlation scale which is of the order of the size of the largest bubbles at coalescence, which are of the order of $\lambda_*\sim 0.01\ell_H$. This result has been obtained with numerical simulations, see~\cite{Kamio,Huber:2008hg}.

 A first order phase transition proceeds via bubble nucleation which is  a very violent event likely to lead to turbulence in the cosmic plasma. In a highly conducting cosmic plasma, turbulence is usually MHD turbulence, and
 a turbulent flow generates both, eddies and magnetic fields in the plasma. A detailed account of the fascinating field of MHD turbulence can be found e.g. in~\cite{Biskamp:2003}. In this section we shall not
 enter into any details of MHD turbulence but just discuss some generic aspects which will already allow us to make very strong statements.

First, we just note that the two known transitions of the standard model, the electroweak transition and the QCD 
transition are both not first order. In fact, they are not even true phase transition but just crossovers~(see 
e.g.~\cite{Kajantie:1995kf,Kajantie1996:aa,Csikor:1997ff} for the electroweak transition, if the Higgs mass is 
$m_H\gsim 80$GeV and~\cite{Roberge:1986mm,deForcrand:2002yi},  for the QCD transition at vanishing 
chemical potential).

However, many modifications of the standard models predict a first order electroweak transition, 
see e.g.~\cite{Grojean:2004xa,Huber:2006ma}. It also has been suggested, that the QCD transition can be first order
if the neutrinos have a sufficiently large, but cosmologically allowed chemical potential~\citep{Schwarz:2009ii}. Such a potential is even required if dark matter is to be a sterile neutrino, see~\cite{Boyarsky:2009ix}. With this in mind, we summarize that taking into account present experimental constraints, it is still possible for both, the electroweak and the QCD phase transitions in cosmology to be of first order. In this case they lead to the generation of the magnetic fields which we now study.

As has been discussed by~\cite{Shaposhnikov:1987tw} and \cite{Turok:1990in}, if the electroweak phase transition is first order, it can also explain the baryon asymmetry of the 
Universe. Interestingly, the electromagnetic part of the Chern-Simons number which determines the net baryon number generated at the transition is simply the helicity. This relates the helicity of the magnetic field generated at the transition to the baryon number as worked out in \cite{Vachaspati:2001nb}.

Let us now consider a first order phase transition where correlation lengths can diverge. This divergence is of course obtained in a static, thermodynamical context where all the modes are in thermal equilibrium.
In cosmology the fact that the Universe is expanding  leads to an effective (comoving) maximal 
length scale $\la_{\max} =t$ over which correlations can extend. In other words,  arbitrary correlations $\xi$
which are generated in cosmology {\em after inflation} satisfy
\be\label{e2:cfini}
\xi(\bx,\bx',t) = 0  \mbox{ if } \quad |\bx-\bx'|=r > t\;.
\ee
Here $\bx$ and $\bx'$ are comoving coordinates, and if the process that generates the correlations is statistically
homogeneous and isotropic, the correlation function $\xi$ is a function of $r$ and $t$ only. 

The power spectrum of such causal correlations, which is the Fourier transform of the correlation function, is therefore analytic for $kt\lsim 1$. For the case of magnetic fields this implies that both, $(\de_{ij} -\hat k_i\hat k_j)P_B(k,t)$ and
$\hat k_mP_{aB}(k,t)$ are analytic at small $k$. Hence, $P_B\sim k^{n_s}$, $P_{aB}\sim k^{n_a}$ where  $n_s\ge 2$ is an even integer and $n_a\ge 1$ is an odd
integer. The fact that   $P_B \ge |P_{aB}|$ even requires $n_a\ge 3$.  For more details about this conditions which are simply a consequence of  causality together with the fact that $\bB$ is divergence free, see~\cite{Durrer:2003ja}.

After the phase transition, the 
magnetic field spectrum  has roughly the following form:
\bea\label{e3:PBtransi}
P_B &\simeq& 2\pi^2B_*^2k_*^{-3} \left\{\begin{array}{ll}
   \left(\frac{k}{k_*}\right)^2  & \mbox{ for }~~ k<k_* \\
    \left(\frac{k}{k_*}\right)^{-\al } & \mbox{ for }~~ k<k_*<k<k_d(t) \\
    0 & \mbox{ for }~~ k_d(t)< k \end{array} \right. \\
P_{aB} &\simeq& \beta 2\pi^2B_*^2k_*^{-3} \left\{\begin{array}{ll}
    \left(\frac{k}{k_*}\right)^3  & \mbox{ for }~~ k<k_* \\
    \left(\frac{k}{k_*}\right)^{-\al' } & \mbox{ for }~~ k<k_*<k<k_d(t) \\
    0 & \mbox{ for }~~ k_d(t)< k \end{array} \right.    
\eea
Here $k_*$ is the correlation scale, $k_*=2\pi/\la_*$, with $\la_*<t_*$, and $t_*$ is the (conformal) time of the phase
transition and $\beta $ denotes the helicity fraction. Typically, $\la_*\sim t_*/100$ is of the size of the largest bubbles which form during the phase transition 
before coalescence. This is a typical number found in numerical simulations by~\cite{Huber:2008hg}, but it depends  sensitively on the strength of the phase transition~\citep{Caprini:2007xq,Espinosa:2010hh}.

In numerical simulations it has been found that helical magnetic fields become totally helical  soon after the phase transition (see~\citealt{Banerjee:2004df,Campanelli:2007tc}), so that soon after the phase transition either $\beta =1$ or helicity vanishes, $\beta =0$.

The spectral index $\al $ which is attained in the so called inertial range is not really certain. We shall assume, that the turbulence is fully developed and we obtain a Kolmogorov spectrum~\citep{LL6} with $\al=\alpha' =11/3$, see Section~\ref{s:evo} for more details. 

The energy density in the magnetic field, as given in Eq.~(\ref{eq:rhob}), is dominated by its value at the 
correlation scale $\la_B \simeq 2\pi/k_*$,
\be
a^4\rho_B = \frac{1}{2\pi^2}\int\frac{dk}{k}k^3P_B \simeq  \frac{1}{2}\tilde B_*^2 \,.
\label{e:rhob}
\ee
For the density parameter we then obtain with (\ref{e2:radGauss})
\be
 \epsilon=\frac{\rho_B}{\rho_{\rm rad}}  \simeq \left(\frac{2}{g_{\rm eff}(t_*)}\right)^{1/3}\left(\frac{B_*}{3\times 10^{-6}{\rm Gauss}}\right)^2 \,.
\ee 
Here we have taken into account the change in the relativistic number of degrees of freedom. We also have assumed 
that today all neutrinos are massive, i.e., $m_\nu> T_0\sim 2.3\times 10^{-4}$eV for all types of neutrinos so that $g_0 =2$.
Assuming adiabatic expansion one requires a constant entropy, $ g_{\rm eff}T^3a^3=g_0T_0^3$
and therefore $a^2\rho_{\rm rad} \propto g_{\rm eff}T^4a^2$ behaves like $g_{\rm eff}^{1/3}$.

In the radiation dominated era, the relation between conformal time and temperature is given by
\bea
\lambda_*\simeq t_* &\simeq&  \left(\frac{\sqrt{3}M_P}{a\sqrt{8\pi g_{\rm eff}}T^2}\right)^{-1} \simeq \frac{3\times 10^5{\rm sec}}{g_{\rm eff}^{1/6}}\left(\frac{100{\rm GeV}}{T_*}\right)  \nonumber \\  \label{e2:tT}
&=& \frac{3\times 10^{-3}{\rm pc}}{g_{\rm eff}^{1/6}}\left(\frac{100{\rm GeV}}{T_*}\right)\,.
\eea

Similarly to the case of the inflation-generated magnetic fields, one can estimate the strength of the relic fields surviving until the present  on large scales. If there is no inverse cascade and the magnetic fields evolve passively, the field strength at $k_1=1$Mpc$^{-1}
\simeq 10^{-14}$sec$^{-1}$ is of the order of
\bea
\left. \tilde B\right|_{k_1=1\mbox{ Mpc} }&=& \left(k_1^3P_B(k_1)/(4\pi^2)\right)^{-1/2} \simeq  3\times 10^{-6}\mbox{ Gauss}\  \epsilon^{1/2}(k_1/k_*)^{5/2} \nonumber \\  &\simeq& 10^{-29}{\rm Gauss}\sqrt{\frac{\epsilon}{g_{\rm eff}(t_*)^{5/6}}}\left(\frac{100{\rm GeV}}{T_*}\right)^{5/2} .
\eea
Hence passively evolving magnetic fields from the electroweak phase transition can at best amount to about
$10^{-29}$~Gauss on Mpc scales while those from the QCD phase transition at $T_* \sim 100$~MeV can amount to about $10^{-23}$~Gauss.

As we shall discuss in Section~\ref{s:obs}, magnetic fields generate a spectrum of anisotropic stresses which
induces a cosmological background of gravitational waves. The spectrum of this background peaks at
frequency $\nu_* =k_*/(2\pi) \sim 100/t_*$. Interestingly, for the electroweak symmetry breaking scale, $T_* \sim 100$~GeV, this corresponds to mili-Hertz frequencies which are in the optimal sensitivity range of the planned space antenna~eLISA (european Laser Interferometric Space Antenna)~\citep{Binetruy:2012ze}.

All the above mechanisms are related to some non-equilibrium processes 
(i.e. the phase transitions or to the relaxation of the initial 
conditions). It was recently demonstrated \citep{Boyarsky12} that 
already in the Standard Model,  long-range magnetic fields can be 
spontaneously generated as part of the equilibrium state in the 
presence of matter-antimatter asymmetry.  At finite  baryon 
or lepton at finite density, quantum corrections due to 
parity-violating weak interactions induce a Chern-Simons term in 
the free energy of the electromagnetic field.  This result is based on a subtle quantum effect that 
appears in the Standard Model at second order perturbation theory (at two 
loops).  This effect can be relevant in both the symmetric phase and the Higgs phase.

%%%%%%%%%%%%%%%%%%%%%%%%%%%%%%%%%%%%%%%%%%%%%%%%%%%%
\subsection{Results from second order perturbation theory}
%%%%%%%%%%%%%%%%%%%%%%%%%%%%%%%%%%%%%%%%%%%%%%%%%%%%

Within first order cosmological perturbation theory and within the strong coupling limit of electrons and protons, no
magnetic fields form due to the inhomogeneities of the matter distribution of the Universe. For this to happen we
need a current with non-vanishing vorticity. For such a current, $\bJ$, Amp\`ere's law gives
\be
\De \bB = -\frac{4\pi}{c}\bnabla\wedge \bJ \,.
\ee
Within linear cosmological perturbation theory $\bJ =e(n_p-n_e)\bv$, so even if we go to 2nd order in the strong coupling limit so that electrons and protons are not perfectly coupled and $n_p\neq n_e$, since $\bv$ is a scalar perturbation, hence a gradient, we have also to go
to second order in the inhomogeneities to obtain 
$$\bnabla\wedge \bJ =e\bnabla(n_p-n_e)\wedge \bv \neq 0\,.$$

Clearly, such second order perturbations are very small on cosmological scales. 

The full system of perturbation equations to second order taking into account the imperfect coupling of
protons and electrons has been derived and studied numerically in several 
papers by~\cite{Ichiki:2007hu,Maeda:2008dv,Fenu:2010kh,Maeda:2011uq}.
Even though the details of the results do not quite agree, they all obtain very small magnetic fields, $\tilde B\lsim 10^{-24}$Gauss, on the scales, $k\lsim 10h/$Mpc where they can calculate the field reliably.
However, the spectrum is raising towards smaller scales and it is not clear whether higher resolution simulations which go up to say $10h/$kpc might not give more promising results. (Even though Ref.~\cite{Ichiki:2007hu} claim to have a result until $k=10^{9}/$Mpc, this is just an interpolation of the result found at (1 - 10)$h/$Mpc, which has been
refuted later by~\cite{Fenu:2010kh}.) 

Even though these results are not fully under control yet, it seems therefore unlikely that straight forward second order perturbations without any initial seed fields can lead to sufficient magnetic fields on galactic and inter galactic scales. 

This situation can change if first order vector perturbations are present. then, the fact that electrons and protons are not perfectly coupled can lead to a different vorticity in each of these fluid at first order.
The generation of a magnetic field by this mechanism was first discussed by \cite{Harrison:1973zz}.

However, in standard inflation such perturbations are not generated, and even if they are generated they decay during the radiation dominated era (see, e.g.~\citealt{Durrer:2008aa}). In order to have genuine, non-decaying vector perturbations, one either has to source them continually, e.g. with topological defects, or one has to modify gravity as, e.g. in the Aether theory of
vector-tensor gravity. The perturbative generation of magnetic fields has been studied for both these cases.
\cite{Hollenstein:2007kg} have shown that vorticity conservation prevents the transfer of vorticity by purely gravitational interactions which would be needed for the Harrison mechanism to work. Therefore, the vector perturbations of topological defects cannot help. 
Also within the aether theory, only very small magnetic fields of order $B\sim 10^{-22}$G can be generated 
(see~\citealt{Saga:2013glg}).

%Evolution (s:evo)

%%%%%%%%%%%%%%%%%%%%%%%%%%%%%%%%%%%%%%%%%%%%%%%%%%%%%%
\section{Cosmological evolution of magnetic fields}
\label{s:evo}
%%%%%%%%%%%%%%%%%%%%%%%%%%%%%%%%%%%%%%%%%%%%%%%%%%%%%%
    
The generation of magnetic fields of strength $\tilde B_*$ by a process operating at the comoving time $t_*$ with comoving correlation scale $\lambda_*$ , sets up the initial conditions for the subsequent evolution of the coupled magnetic field -- primordial plasma system from the moment of magnetogenesis up to the end of the radiation dominated era and the moment of decoupling / recombination. The evolution continues also after recombination in a system where the charge density of the plasma is strongly reduced because most of the electrons and protons/nuclei have combined to neutral atoms \citep{Sethi05}. Plasma effects on the evolution of magnetic fields increase again at the latest stages of evolution, when the density of primordial plasma grows again after the re-ionization at redshift $z\sim 10$. 

The qualitative picture of the evolution of magnetic fields and the primordial plasma is governed by the MHD equations introduced in Section~\ref{s:mhd}. In general, the nonlinear MHD equations are difficult (if not impossible) to solve, both analytically and numerically. Some general properties of the solutions can, nevertheless, be established based on relatively straightforward order-of-magnitude estimates of the importance of the different terms in the MHD  equations see  \cite{Jedamzik98,Subramanian98,Banerjee:2004df,Barrow:2006ch,Jedamzik11,Kahniashvili:2012uj,Saveliev2012}.  In the following subsections we summarize these general properties.

%%%%%%%%%%%%%%%%%%%%%%%%%%%%%%%%%%%%%%%%%%%%%%%%%%
\subsection{Initial conditions for the evolution}
%%%%%%%%%%%%%%%%%%%%%%%%%%%%%%%%%%%%%%%%%%%%%%%%%%

\begin{figure}
\begin{center}
\includegraphics[width=\linewidth]{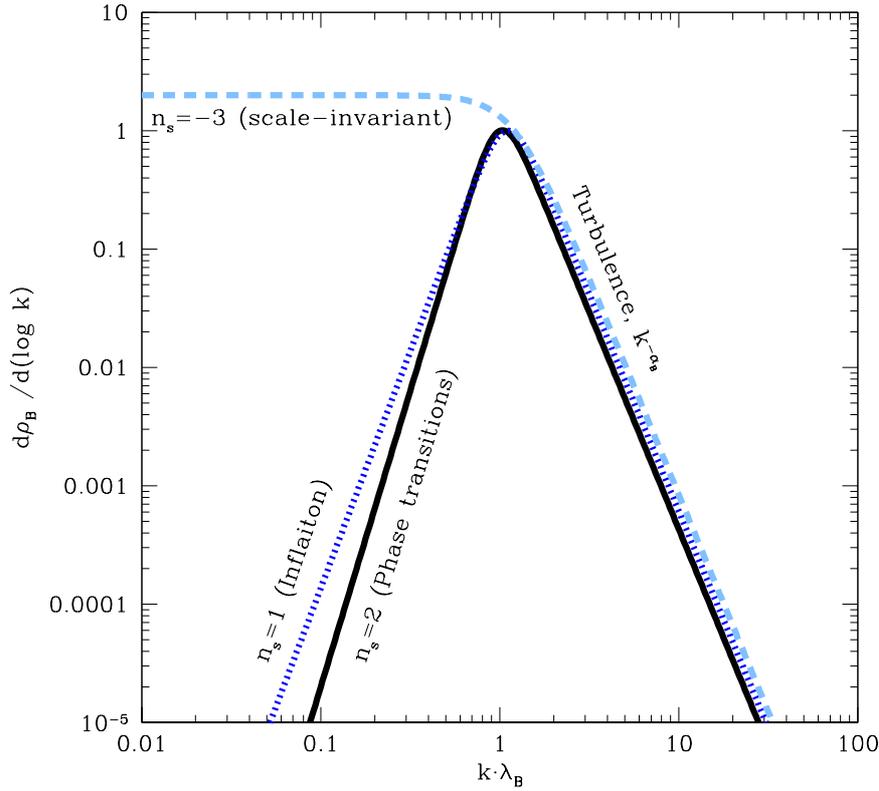}
\end{center}
\caption{Possible spectral energy distributions of cosmological magnetic fields. At small $k$, the spectra of the fields generated at phase transitions in a causal way  follow powerlaw with the slope $n_s=2$. Inflationary mechanisms typically result in the slope $n_s=1$. Inflation could in principle generate a scale-invariant spectrum with $n_s=-3$. At large $k$ all the spectra follow a universal slope formed by turbulence. }
\label{f4:powerspectra}
\end{figure}

The discussion of  mechanisms of generation of magnetic fields in Section~\ref{s:gen} suggests that the magnetogenesis results in the production of a turbulent plasma and magnetic fields characterized by the power spectra $P_B, P_{aB}, P_{sK}, P_{vK}, P_{aK}$ (see Eqs. (\ref{e2:specsa}), (\ref{eq:pk}) for definitions),  in different intervals of wave numbers. At the initial time, the large wavelength (small wavenumber) tail of the power spectra of the magnetic field and plasma on scales larger than the characteristic scale of magnetogenesis $\lambda_*$ are given by
\begin{eqnarray}
P_B(k,t_*)=P_{*B} k^{n_s} \,,\qquad
P_K(k,t_*)= P_{*K} k^{n_k} \,.
\end{eqnarray}
where $P_K$ scaling applies to $P_{sK}$ and/or $P_{vK}$. The largest power on the longest scales is achieved for the minimal possible values of $n_s$, $n_k$. Formally, for t magnetic fields generated at the inflationary epoch, the requirement that the magnetic field energy is not infrared divergent restricts $n_s$ to $n_s> -3$. However, nearly all self-consistent mechanisms of  field generation during inflation proposed so far satisfy stronger constraint, namely $n_s\gsim 1$. As we have argued in Section~\ref{s3:trans}, see Eq.~(\ref{e3:PBtransi}), for causal field generation, e.g. a phase transition, we require $n_s=2$. Different possible spectra of the field at small $k$ are shown in Fig.  \ref{f4:powerspectra}. 

%%%%%%%%%%%%%%%%%%%%%%%%%%%%%%%%%%%%%%%%%%%%%%
%%%%%%%%%%%%%%%%%%%%%%%%%%%%%%%%%%%%%%%%%%%%%%

For the velocity spectrum, the situation is more complicated. First of all, there is no cosmic plasma present during inflation hence turbulent motions develop only after inflation and this in
a causal way. This implies that the velocity power spectrum defined in Eq.~(\ref{e4:vspecsa}) 
is always the Fourier transform of a function of compact support and therefore analytic on large scales. 
The leading term in the Taylor expansion of the kinetic power spectrum $P_K$ at small $k$ is determined by the nature of fluid motions. As mentioned above, considerations of hydrodynamic turbulence often adopt the assumption of incompressibility of fluid motions, which imposes a divergence-free velocity field,
\begin{equation}
\bnabla\cdot\bv=0 \,.
\label{eq:divv0}
\end{equation}
This condition is identical to the divergence-free condition satisfied by the magnetic field, 
and as mentioned above in this case $P_{sK}=P_{vK}\equiv P_{K}$ and causality requires the same asymptotics for the power spectrum at small $k$ as for the magnetic field, 
\begin{equation}
P_K(k)\sim k^2, \ \ \ k\rightarrow 0\ \ \mbox{(incompressible fluid/plasma).}
\end{equation}
Assuming the validity of condition (\ref{eq:divv0}) significantly simplifies the MHD equations and facilitates numerical modeling of turbulence. This is why this condition is commonly adopted in  turbulence modeling (see e.g. \citet{Biskamp:2003}). In particular, it was adopted for the study  of cosmological magnetic fields in the papers by~\citet{Banerjee:2004df} and \citet{Caprini:2009yp}. 

However, there is no particular reason why the process of generation of magnetic fields in the early Universe would excite only incompressible fluid motions. Indeed, it is clear from the system of Eqs. (\ref{e4:conserv} -- \ref{e4:induc}) that the term $\bB\wedge (\bnabla\wedge\bB)$ in the Euler equation provides a source term for both, compressible and incompressible modes. 
Furthermore, the argument that the Mach number $M=v/c_s = \sqrt{3}v\ll 1$ does not suffice, since the additional necessary condition (see \citealt{Biskamp:2003}), that time derivatives $\dd_t$ are much smaller than the term $\bv\cdot \bnabla$ is not satisfied in our situation.  For our relativistic plasma, we expect time derivatives which are of the same order as spatial derivatives.
  
Thus, in a generic situation, the power spectrum of the kinetic energy of the plasma motions is not restricted to have $P_{sK}=P_{vK}$. While analyticity requires $P_{vK} \propto k^2$ for small $k$, $P_{sK}$ has no non-analytic pre-factor  and is in general white noise for small $k$. Then, the asymptotic of $P_K\simeq P_{sK}$ is white noise,
\begin{equation}
P_K(k)\sim k^0, \ \ \ k\rightarrow 0\ \ \mbox{ (compressible fluid/plasma).}
\end{equation}
The evolution of magnetic fields in compressible plasmas is called Burgers 
turbulence~\citep{Tsytovich:1977}. In cosmological settings it has been considered by \citet{Brandenburg96,Jedamzik11,Kahniashvili:2012uj}. 

On scales $k>k_*$ the initial power spectrum of both the magnetic field and the kinetic energy of plasma is suppressed. The detailed shape of the initial spectra in this  regime is usually  irrelevant because turbulence establishes a "universal" slope of the power spectra on wave numbers $k>k_B$ with $k_B\lsim k_*$ which is independent of the initial shape as we discuss in the next subsection. 

%%%%%%%%%%%%%%%%%%%%%%%%%%%%%%%%%%%%%%%%%%%%
\subsection{The regime of freely decaying turbulence}
%%%%%%%%%%%%%%%%%%%%%%%%%%%%%%%%%%%%%%%%%%%%

Both the rescaled magnetic and kinetic energy densities, $\tilde \rho_B,\tilde \rho_K$  and the correlation lengths $\lambda_B,\lambda_K$ defined in Section~\ref{s:mhd} evolve with  time. This evolution can in principle be obtained by solving the system of equations (\ref{e4:conserv}--\ref{e4:induc}), for given 
initial conditions, ${\tilde \bB}(k, t=t_*), \bv(k, t=t_*)$ and $\tilde\rho(k,t=t_*)$. However, the nonlinearity of the evolution equations renders the analytical or numerical solution very complicated, for any realistic set of initial conditions e.g. after the electroweak or QCD phase transitions. 
Because of this difficulty, a common approach is to derive a qualitative picture of the evolution based on an order-of-magnitude analysis of the relative importance of the different terms in 
Eqs.~(\ref{e4:conserv}--\ref{e4:induc}). 

The main process determining the evolution of the magnetic fields and the plasma at intermediate wavenumbers is the establishment of MHD turbulence. This process operates in the regime in which the dissipation terms on the right hand side of the Euler equation~(\ref{e4:euler})  and  on the right hand side of the induction equation~(\ref{e4:induc}) can be neglected. As discussed in Section~\ref{s:mhd} these are scales for which both the magnetic and kinetic Reynolds numbers are large. In this situation, the "mode coupling" terms containing spatial derivatives in the Euler and induction equations continuously generate larger $k$ modes at the expense of the lower $k$ modes and in this way transfer power from large to smaller scales. This process is extensively studied in various contexts of hydrodynamic and MHD turbulence, both analytically and numerically, see e.g. \citet{Biskamp:2003,Tsytovich:1977,LL6} and more. The result of these studies is that the power spectrum of the  magnetic field and kinetic energy evolve to  power law spectra given by
\begin{equation}
P_B\propto k^{-\alpha_B}, \quad \ P_K \propto k^{-\alpha_{K}} \,,  \ \  \alpha_{\bullet}>0
\end{equation}
independently of the details of initial conditions. The most commonly known result is the Kolmogorov turbulence spectrum with 
\begin{equation}
\alpha_K={-11/3}
\end{equation}
encountered in  incompressible hydrodynamic turbulence. The same slope for the magnetic field power spectrum, typically $\alpha_B=\alpha_K$ is found in numerical models of MHD turbulence. The order of magnitude argument that leads to this spectrum is originally due to \citet{Kolmogorov:1941}. We follow the argumentation of~\citet{LL6}:

We assume that turbulence is full developed on the scales under consideration. Energy is transferred from large to smaller scales and dissipated at some dissipation scale $k_d$. On scales $k<k_d$ on which turbulence is developed, the energy transferred to smaller scales per unit time, let us call it $\varepsilon$, can only depend on the mean velocity on this scale, $v_\la =(P_K(k)k^3)^{1/2}$ and on the scale itself, $\la=2\pi/k$. It must have the units of $[\varepsilon] =[v^2/t] = [v^3/\la] = [v^3k]$. Setting $\varepsilon \simeq v_\la^3k$, this implies
\be
P_v(k) \simeq \varepsilon^{2/3}k^{-11/3} \propto k^{-\al_K} \quad \mbox{with } \alpha_K={-11/3}\,.
\ee
Note that this scaling does not depend on the value of $\varepsilon$. The only hypothesis used is that $\varepsilon$ is independent of $k$ which is necessary for the situation to be stationary. Even though this is not evident from the above 'derivation', the Kolmogorov spectrum has also been observed (numerically) to hold in relativistic 
plasmas~\citep{Mueller:2006up}. The part of the magnetic field spectra formed by the free turbulence decay is shown in Fig. \ref{f4:powerspectra}. It is important that the large $k$ behaviour of the spectrum processed by the turbulence is largely independent of the initial spectrum of the field at the moment of generation. 

If the velocity field and the magnetic field are in equipartition on all scales, we expect also 
$\al_B=\al_K =11/3$.

\citet{Goldreich:1994zz}  use a slightly different argument and obtain a spectrum 
for the magnetic field with $\alpha_B=10/3$ while~\citet{Iroshnikov:1964} and \citet{Kraichnan:1965} propose $\alpha_B=7/2$. The latter values are obtained considering the collision of Alfv\'en-waves on a strong background field and are probably not relevant here. Also, simulations are in good agreement with the Kolmogorov slope~\citep{Muller:2005zz}.

The values cited above are all quite close and the precise values of $\alpha_B, \alpha_K$ are, in fact, not important for the general understanding of the process and of the time evolution of the power spectra $P_B, P_K$. It is just important that $\al_\bullet < -3$ such that the turbulent energy is concentrated on scales around $k_B$ and not at the damping scale. The evolution of the correlations scale and of the energy density  can be understood, at least  qualitatively, in the following way. 

Consider a moment of time $t>t_*$.  A generic property of the MHD turbulence is that it transfers energy from large to small scales. On the scale $k_d$, the dissipative terms can no longer be neglected and the turbulent energy is lost into heating up the plasma (see below).  On very large scales, turbulence did not have enough time to fully develop and the initial spectral slope is maintained. As time goes on the largest scale on which turbulence is developed, the 
integral scale or correlation scale $\lambda_B, ~ \lambda_K$ grows and correspondingly  $k_B=2\pi/\la_B$ and $k_K=2\pi/\la_K$ decrease. Beyond  $k_B$ the initial spectrum has been processes e.g. into a Kolmogorov slope. On $k<k_B$, $k<k_K$ the spectrum still has its original slope. 

We expect that the time dependence of these scales follows a power law,
 \begin{eqnarray}
\lambda_B&\sim& t^{\kappa_B}\,, \qquad 
\lambda_K \sim t^{\kappa_K} \,,
\end{eqnarray}
with the indices $\kappa_{\bullet}>0$ (where $\bullet$ is either $B$ or $K$) which we now derive.  The rescaled energy density  in magnetic field and in turbulent motions of the plasma decreases with time due to dissipation
\begin{eqnarray}
\tilde\rho_B&\sim& t^{-\zeta_B} \,, \qquad 
\tilde\rho_K  \sim t^{-\zeta_K} \,,
\end{eqnarray}
with $\zeta_{\bullet}>0$. 

Suppose that at  a given moment of time the rescaled energy density of plasma motions is $\tilde \rho_K(t)$. This energy is associated to a characteristic velocity via the relation
\begin{equation}
\tilde \rho_K=\frac{v_K^2}{2} \,.
\end{equation}
In the same way, a characteristic velocity can be associated to the energy density of magnetic field. This is the Alfv\'en velocity given by
\begin{equation}
\frac{\rho_B}{\rho}=\frac{ v_A^2}{2} \,.
\end{equation}
These velocities characterize the speed of the spread of changes in the configuration of the velocity field and the magnetic field. Typically the size of  regions over which the fields can change in a coherent manner on a time scale $t$ are $\la_{\bullet}\sim v_{\bullet}t$. Unless the magnetic fields are generated in an a-causal way, e.g. during Inflation, the integral scale of the magnetic field and of the plasma motions at the time $t\gg t_*$ cannot exceed $v_{\bullet}t$  and, in general, this scale provides a reasonable estimate of the integral scales (also called "largest processed eddy" scales):
\begin{eqnarray}\label{laBvAt}
\lambda_B&\sim v_At \,,\qquad
\lambda_K&\sim v_Kt \,.
\end{eqnarray}

Coupling between plasma motions and the magnetic field usually establishes equipartition between plasma kinetic energy and magnetic field on a certain distance scale. In this case the conditions
\begin{equation}
\tilde \rho_B\sim\tilde \rho_K\sim \overline\rho\,, \ \ \ v_K\sim v_A\sim \overline v\,, \ \ \ \lambda_K\sim \lambda_B\sim \overline \lambda
\end{equation}
are satisfied.  We then expect already for reasons of dimensionality that  
\be 
\overline\lambda\sim \overline v t.
\label{eq:lam}
\ee
 where $\overline v$ is given by $\overline v^2\sim 2\overline\rho \sim P_{K,B}/\overline \lambda^3 \sim \left(\overline{\lambda}\right)^{-(3+n)}$, where $n=n_s$ or $n=n_k$, depending on whether matter or magnetic field power spectrum initially dominates on the scale $\overline \lambda$.  Here $n$ characterizes the slope of the (dominant) unprocessed part of the power spectrum. Substituting this expression in Eq.~(\ref{eq:lam}), we arrive at the relation
\be\label{e4:laBn}
\overline\lambda\sim t^{\frac{2}{5+n}}=
\left\{
\begin{array}{ll}
 t^{2/5},& \mbox{ when } n_k=0,\ n_s=2  \quad \Rightarrow \quad n=0\,, \\
 t^{2/7},& \mbox{ when } n_k=2,\  n_s=2  \quad \Rightarrow \quad n=2\,.
\end{array}
\right.
\ee
The first case corresponds to compressible turbulence (Burgers turbulence), when the power spectrum of plasma motions is $P_K\sim k^0$ at small $k$ and the power of  plasma motions dominates over the magnetic field power $P_B\sim k^2$ on large scales $k\rightarrow 0$. The second case corresponds to  incompressible turbulence in which $P_K\sim P_B\sim k^2$ at small scales. 

The energy of plasma motions and of the energy of the magnetic field evolve in equipartition according to 
\be
\tilde \rho_B\sim\tilde \rho_K\sim \frac{\overline v^2}{2} \sim t^{-\frac{2(3+n)}{5+n}}=
\left\{
\begin{array}{ll}
 t^{-6/5},& \mbox{ when } n_k=0,\ n_s=2 \,, \\
 t^{-10/7},& \mbox{ when } n_k=2,\  n_s=2 \,.
\end{array}
\right.
\ee

The characteristic magnetic field strength at the scale $\overline\lambda$ evolves as \citep{Banerjee:2004df,Campanelli:2007tc,Jedamzik11}
\be
\tilde B =\sqrt{2 \tilde \rho_B}\sim t^{-\frac{(3+n)}{5+n}}=
\left\{
\begin{array}{ll}
 t^{-3/5}\sim \left(\overline\lambda\right)^{3/2},& \mbox{ when } n_k=0,\ n_s=2 \,,\\
 t^{-5/7}\sim \left(\overline\lambda\right)^{5/2},& \mbox{ when } n_k=2,\  n_s=2\,.
\end{array}
\right.
\ee

More generally, for an arbitrary $n$, the dependence of the magnetic field strength on the correlation length
\begin{equation}
\label{eq:n}
\tilde B_\lambda\sim \lambda^{-\frac{3+n}2}
\end{equation}
holds with $n=\min(n_s,n_k)$.  Here we assume that on the scales where turbulence is developed equipartition between the magnetic field energy and the kinetic energy is established.  

In a hypothetical case of the scale-invariant magnetic field generated during inflation with spectrum $n_s\simeq -3$,  the above equation implies that magnetic field strength does not change in the course of cosmological evolution, while the correlation length grows as $\lambda_B\propto t$. In this case therefore, the correlation scale is always  the same fraction of the horizon scale. A case quite similar to scaling topological defects, see~\cite{Durrer:2001cg}. This scaling has also been argued for 
by~\citet{Christensson:1999tp}, without however, realizing that it is only valid for scale invariant magnetic field spectra.

We now consider the case of a maximally helical magnetic field,  in a cosmic plasma that respects helicity conservation~\citep{Biskamp99,Biskamp:2003}, which we expect in the regime where $T<m_e$  (see below). The helicity density is of the order of $\mathpzc{h} =\langle \tilde \bA\cdot\tilde \bB\rangle \sim \la_B\tilde \rho_B$ and its conservation implies the relation 
\begin{equation}
\tilde B \sim\sqrt{\tilde\rho_B}\sim \lambda_B^{-1/2}\,.
\end{equation}
This is equivalent to the substitution $n=-2$ in the Eq. (\ref{eq:n}). The time evolution of $B, \la_B$ then follows the law \citep{Biskamp99}
\be 
 \la_B \sim t^{2/3}\,\ \  \tilde B_\lambda\sim t^{-1/3}\,.
\ee

It has been shown recently by \citet{Boyarsky12a}, that the chiral 
anomaly of the Standard Model plays an important role for $T > m_e$ even 
though chirality flipping reactions are in thermal equilibrium. Even in 
the homogeneous approximation (neglecting turbulent flows) it was shown 
that an inverse cascade develops solely due to this effect. In the 
process of this inverse cascade the magnetic helicity is approximately 
conserved (changes very slowly). Therefore it is an important future project to include the 
effect of \citet{Boyarsky12a} consistently in the standard MHD analysis.

%It has been shown recently by~\citet{Boyarsky:2011uy}, that the chiral anomaly of the standard model, which violates parity, is %relevant for $T\gsim m_e$. Helicity conservation therefore cannot be invoked and
%the above growth law for the correlation scale $\la_B$ might be modifed at $T>1$MeV even if the field is fully helical.

%%%%%%%%%%%%%%%%%%%%%%%%%%%%%%%%%%
\begin{figure}[h!]
\begin{center}
\includegraphics[width=5.869cm]{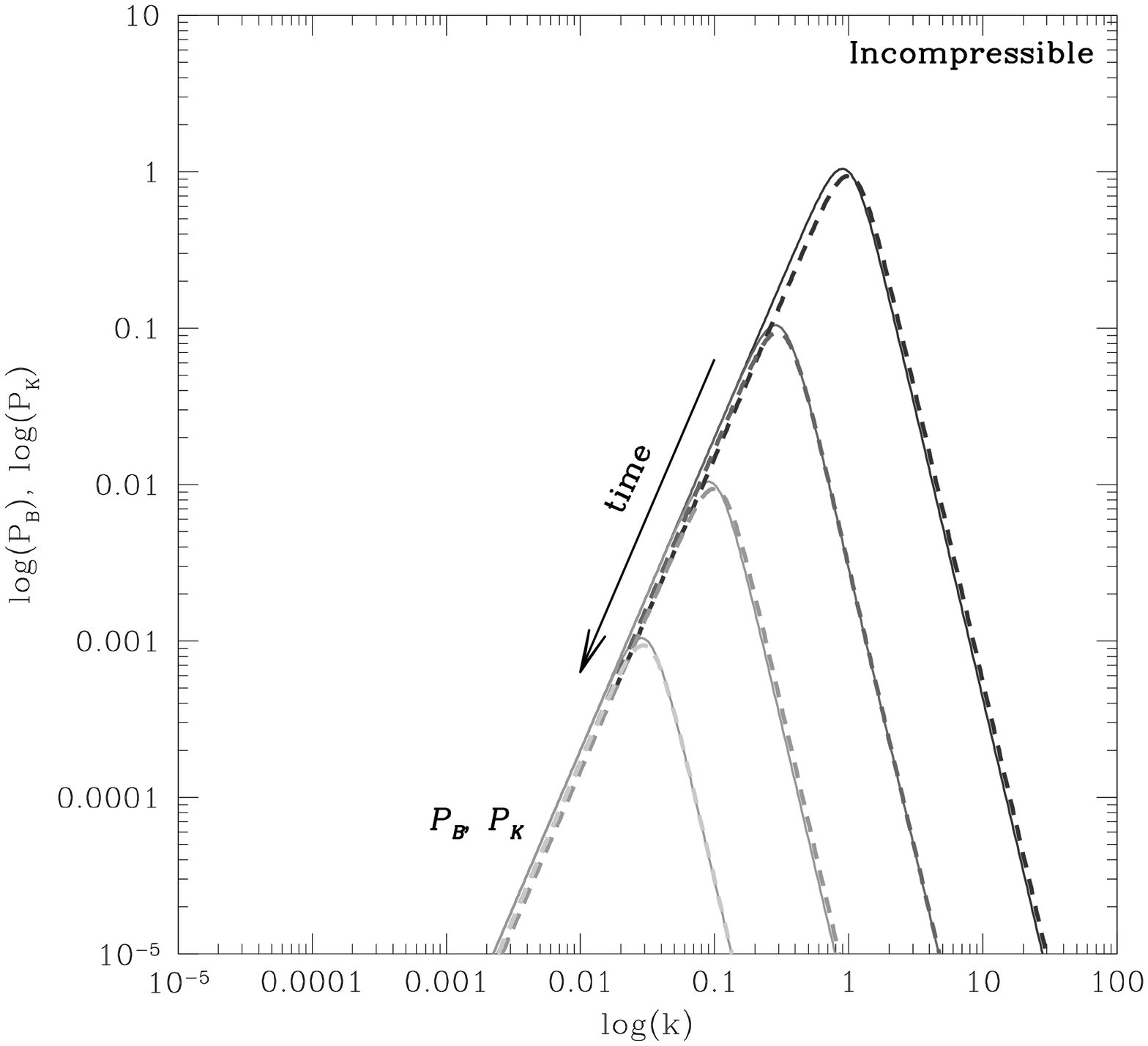} \includegraphics[width=5.869cm]{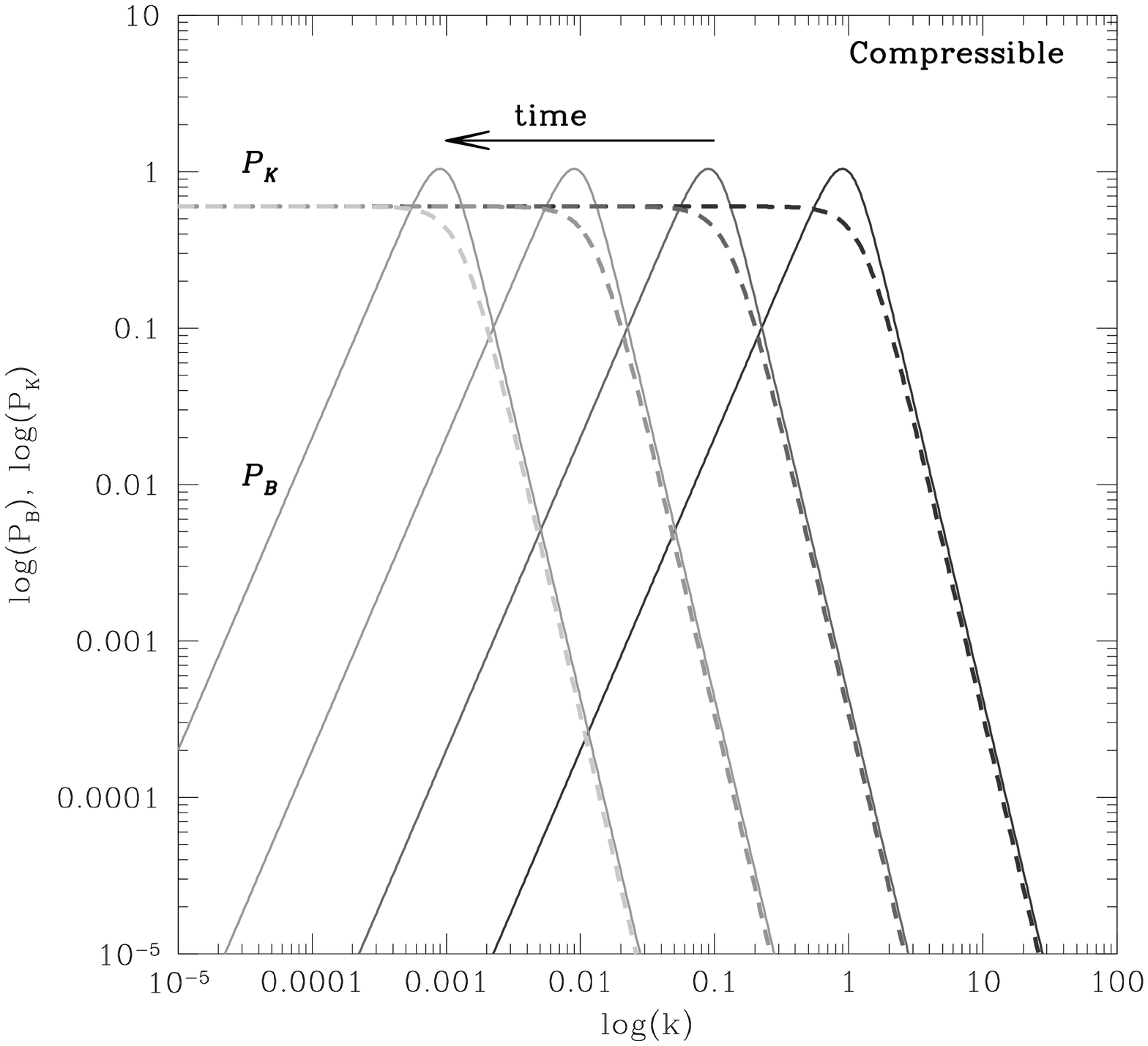} \\  \includegraphics[width=5.99cm]{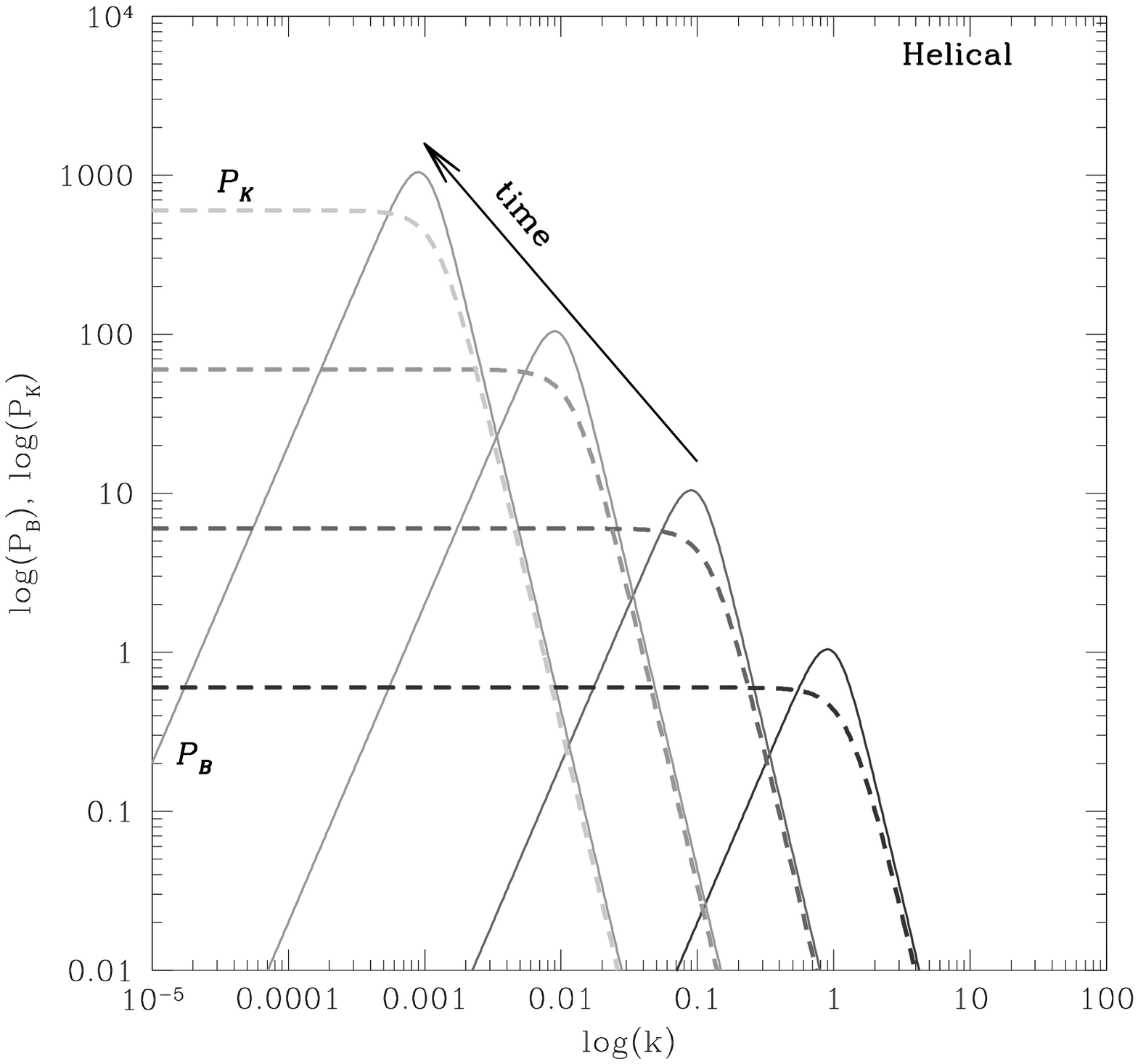}
\end{center}
\caption{The evolution of the magnetic field spectrum. Top left: incompressible flow, top right:
compressible flow, bottom: a fully helical field. The spectra evolve towards smaller $k$. Dashed lines show the kinetic energy spectrum. }
\label{f4:Bevol}
\end{figure}
%%%%%%%%%%%%%%%%%%%%%%%%%%%%%%%%%%

The evolution of causal magnetic field spectra for incompressible fluid motions, compressible fluids and for the fully helical case are shown in Fig.~\ref{f4:Bevol} in the simple case where all components are in equipartition. In the incompressible case, even though the correlation scale is growing, there is no 'inverse cascade' in the sense that no power is transferred from small to larger scales.
The growth of $\la_B$ is simply due to the loss of small scale power by dissipation. On large scales, $\la>\la_B$ the power spectrum is not affected. This is sometimes called 'passive growth'.
This is different for the compressible and the helical cases. There, the power spectrum grows on scales 
$\la>\la_B$ and decreases on scales $\la<\la_B$ which is what we call an inverse cascade. 

Realistic situations might well be more complicated. For example if the field is not maximally helical, or if the 
scalar velocity mode is much smaller than the vector mode, $|\bnabla\cdot \bv| \ll |\bnabla \wedge\bv|$, the field first decays like in the incompressible case until it becomes maximally helical, or until the vector amplitude is on the level of the scalar amplitude and only then it evolves according to the scheme shown in Fig.~\ref{f4:Bevol}.

\begin{figure}
\begin{center}
\includegraphics[width=\linewidth]{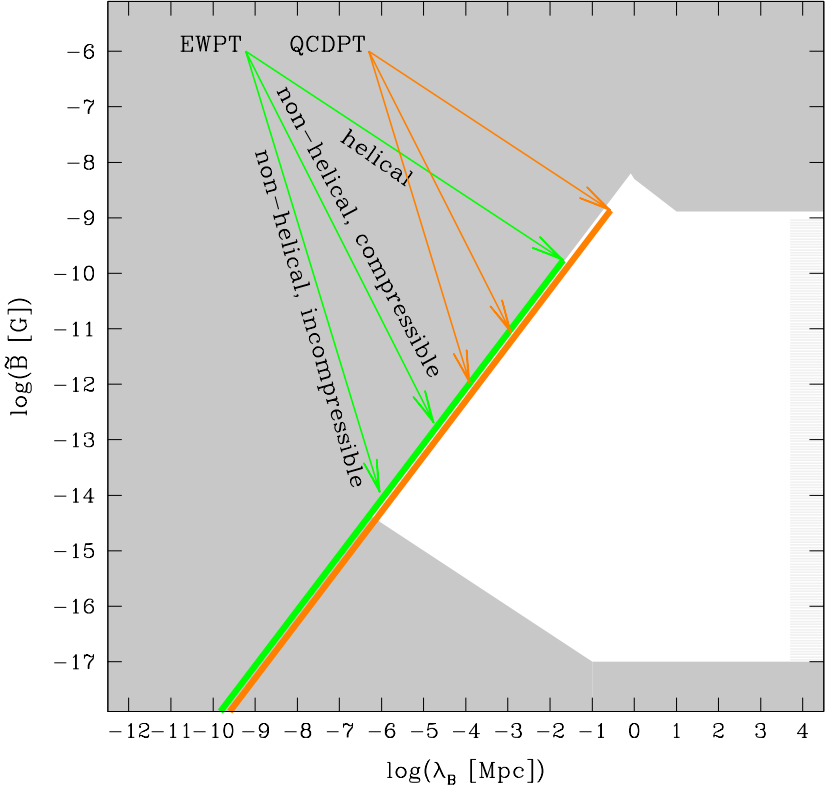}
\end{center}
\caption{The evolution of the magnetic field amplitude and integral scale for helical fields and for non helical compressible and incompressible flow. Both, the electroweak and QCD phase transitions are indicated. The line on which the tracks end is given by the relation $v_A = \la_B/t_{\rm rec}$, with $t_{\rm rec} \sim 200$Mpc.}
\label{f4:evolP}
\end{figure}

The evolutionary tracks of the correlation length and average strength of magnetic field in the $\tilde B,\lambda_B$ diagram are shown in Fig. \ref{f4:evolP}. The locus of the natural termination points of the tracks is the line corresponding to the largest processed eddy size at the end of radiation dominated epoch / recombination when the temperature drops to $T_{\rm rec}\sim 0.3$~eV
\be
\label{e4:limit0}
\tilde B\sim 10\frac{T_{\rm rec}T_0^3 \lambda}{M_P} \simeq 10^{-8}\left(\frac{\lambda}{1\mbox{ Mpc}}\right)\mbox{ G}.
\ee
The factor $\sim 10$ accounts for the numerical coefficients obtained when re-expressing $t$ and $B$ in  Eq.~(\ref{laBvAt}) through $T$ and $M_P$, using the Friedman equation and the identity 
$$\frac{1(\mbox{Gauss})^2}{8\pi} = 1.9\times 10^{-40}\mbox{GeV}^4 = 3.4\times 10^{12}K^4\, . $$  
This linear relation between the magnetic field amplitude and the correlation scale, which is simply a 
consequence of Eq.~(\ref{laBvAt}), has first been pointed out by~\cite{Banerjee:2004df}.

%%%%%%%%%%%%%%%%%%%%%%%%%%%%%%%%%%%%%%%%%%%%%%%%%%%%%%
%%%%%%%%%%%%%%%%%%%%%%%%%%%%%%%%%%%%%%%%%%%%%%%%%%%%%%

%%%%%%%%%%%%%%%%%%%%%%%%%%%%%%%%%%%%%%%%%%%%%%%%%%%%%%
\subsection{Evolution with strong viscous damping}
%%%%%%%%%%%%%%%%%%%%%%%%%%%%%%%%%%%%%%%%%%%%%%%%%%%%%%

In principle, if free turbulent decay proceeds up to the end of  recombination  (when most of the charged plasma disappears and the evolution of the field changes), the integral characteristics of the magnetic field  move with constant velocity along the lines shown in Fig.~\ref{f4:evolP}. However, the periods of free turbulent decay  terminate at least during two regimes where the damping term in Eq.  (\ref{e4:euler}) becomes important at scales up to $\lambda_B$. 

In the cosmological case, for $T < 100$~GeV the Prandl number is very large so that dissipation of magnetic fields proceeds by kinetic diffusion: the magnetic fields generate velocity fields via the source term in the Euler equation and the velocity fields are then dissipated via kinetic viscosity. The kinetic viscosity is provided by the least coupled particle in the plasma and the viscosity coefficient in Eq. (\ref{e4:euler}) is of the order of $\tilde\nu\sim \lambda_{\rm mfp}/5$, where $\lambda_{\rm mfp}$ is the comoving mean free path of the least coupled particle \citep{Weinberg71}, see Appendix~\ref{a:visc} for more details. 

Depending on the temperature, the least coupled particles in the plasma are neutrinos (for the temperatures above $T\sim 1$~MeV) or photons (for the temperatures below MeV down to decoupling at $T\sim 0.3$~eV). At the time of neutrino or photon decoupling, the mean free path of the particles grows beyond the horizon scale. The dissipation scale $\la_d=2\pi/k_d$ is the scale at which the Reynolds number, Eq. (\ref{eq:reynkin}), becomes of order unity
\be
 \lambda_d = \frac{2\pi}{k_d} \simeq \frac{\tilde\nu(T)}{v_K}\sim \frac{\lambda_{\rm mfp}\sqrt{\rho}}{5B} \,.
\ee
In the last estimate we have substituted $v_K\sim v_A\sim B/\sqrt{\rho}$. The growth of 
$\lambda_{\rm mfp}$ at the moments of decoupling of neutrinos and photons leads to the growth of the viscous damping scale, up to  values comparable to the integral scale $\lambda_d\sim \lambda_{B} \simeq\la_K$ and the fluid enters the so called dissipative regime. 

The magnetic field and the plasma no longer evolve according to free turbulent decay, the plasma motions are damped by viscosity already at the integral scale $\lambda_K$. This regime was studied in detail by \cite{Banerjee:2004df}, who found that damping at the integral scale suppresses plasma motions and removes the coupling of the plasma to the magnetic field (setting $v\rightarrow 0$ in the induction equation  (\ref{e4:induc})). Once the coupling between the plasma motions and magnetic field is removed, the rescaled magnetic field temporarily stops evolution, $\partial \tilde B/\partial t\simeq 0$, so that $\tilde B$ and $\lambda_B$ do not change. To the contrary, the kinetic integral scale $\lambda_K$ continues to grow and $\rho_K$ continues to decrease because the turbulent kinetic energy of the plasma is efficiently dissipated into heating up the plasma. 
This happens, as long as $\la_d \simeq \lambda_{\rm mfp}/v_A  \gsim \la_K \gg \lambda_{\rm mfp}$. Therefore, this regime is relevant only
if $v_A\ll 1$.

Once the mean free path of the least coupled particle becomes significantly larger than the typical scale of the system (integral scales $\lambda_K,\lambda_B$), the evolution changes once more. The least coupled particles are now too weakly coupled to the fluid to provide a true viscosity. This situation is called "free streaming". In this regime the viscous damping term in the Euler equation  (\ref{e4:euler}) has to be replaced by a friction term of the form $\al v$ \citep{Banerjee:2004df}:
\be
\tilde\nu\nabla^2\bf v\rightarrow \tilde \alpha \bf v \,.
\ee
The coefficient $\tilde\alpha \propto \lambda_{\rm mfp}^{-1}$ is an ordinary friction or 'drag force' term, analogous
to Silk damping of baryon fluctuations during decoupling~\citep{Durrer:2008aa}. The proportionality factor is of order 1 as long as the electrons are relativistic and becomes $\rho_\ga/\rho_b$ after electron--positron annihilation.

The growth of $\lambda_{\rm mfp}$ leads to the decrease of $\tilde\alpha$ and, as a consequence, to a suppression of this damping term in the Euler equation. This provokes the end of the dissipative regime and the restoration of turbulence soon after the onset of the free-streaming regime. Note, however, that for this to take place, it is important that there are stronger interactions (in our situation electromagnetic forces) with much shorter   mean free path,  $\lambda_{\rm mfp\, 2}\ll \lambda_{\rm mfp}$ so that  the fluid picture still applies, otherwise  the Euler and continuity equations have to be replaced by a Boltzmann equation.  When the next largest $\lambda_{\rm mfp\, 2}$ grows to the value $\lambda_{\rm mfp\, 2}/\la_B^2 \gsim 
\tilde\alpha$, the original Navier-Stokes form of the damping term term is re-installed with $\tilde\nu \simeq \lambda_{\rm mfp\, 2}$ and the coupling to the weakest coupling particle species 
 can be neglected.
   
The kinetic energy of plasma motions is dissipated into heat in the dissipative regime. The energy contained in the magnetic field is constant, up to the trivial dilution due to the overall expansion of the Universe on scales  much larger than the Ohmic dissipation scale given by 
\be 
\lambda_{\rm Ohmic}\sim (\tilde \sigma \HH)^{-1/2}\simeq \sqrt{\alpha\log(\alpha^{-1})}\left(\frac{M_P}{T}\right)^{1/2}T_0^{-1};
\ee 
the rescaled magnetic field is not evolving when the 
velocity field vanishes:  $\partial \tilde B/\partial t\simeq 0$. 
Once turbulence is restored after decoupling of the most weakly coupled particle , the Lorentz force term in the Euler equation (\ref{e4:euler}) 
\be
\tilde\bB \wedge (\nabla\wedge\tilde\bB)
\ee
serves as a source term for plasma motions. The coupling between plasma and magnetic field restores also  equipartition between the magnetic and the kinetic energy, so that a new cycle of free turbulent decay starts. 

The equations determining the correlation length $\overline\lambda\sim \lambda_B\sim \lambda_K$ and the energy density  / velocity scale $\overline v\sim v_K\sim v_A$ in the free turbulence decay regime are largely insensitive to the details of the shapes of the kinetic and magnetic power spectra.  The only property of the spectrum which matters is the total energy density, which determines the average velocity scale $\overline v$ and, as a consequence, the "eddy processing" time scale $t\sim \lambda/\overline v$. Since, the energy density of the magnetic field right after the restoration of the turbulent regime is roughly the same as just before the end of the turbulent regime at transition to the viscous regime, further evolution of the system in the restored turbulence regime proceeds "as if" there was no episode of  viscous damping~\citep{Banerjee:2004df}. In terms of the evolutionary diagram for $\tilde B,\lambda_B$, shown in Fig.~\ref{f4:evolP}, the system always remains on the same track (shown by the lines with arrows). During  free turbulent decay, the system moves along the track. During the viscous damping and free streaming (the dissipative regime) the system halts and remains at the same point of the track $\tilde B=const$, $\lambda_B=const$ until  free turbulent decay is restored. 

This type of evolution is also observed in numerical simulations both for 
incompressible~\citep{Banerjee:2004df} and for compressible~\citep{Kahniashvili:2012uj} MHD and also in the helical case~\citep{Banerjee:2004df,Kahniashvili:2012uj}.

%%%%%%%%%%%%%%%%%%%%%%%%%%%%%
\subsection{Example: evolution of the field generated at electroweak phase transition}
%%%%%%%%%%%%%%%%%%%%%%%%%%%%%

As an example, let us consider magnetic field with initial comoving correlation length $\lambda_*\simeq 0.01 \ell_H$ (here $\ell_H =1/\HH_* =t_*$ is the comoving Hubble scale at time $t_*$) and magnetic energy density $\rho_B\sim 0.1\rho$ produced at the electroweak phase transition at  temperature $T_*\sim 100$~GeV (see Section~\ref{s3:trans}). We assume that the evolution of the magnetic field is governed by the compressible MHD, so that the correlation length evolves with comoving time as\footnote{Even if we have initially $\la_B=\la_i((t-t_i)/\tau)^{-2/5}$ after a few Hubble times this very  turns into $\la_B =\la_*(t/t_*)^{-2/5}$, where $\la_*$ denotes the correlation scale at $t_*$.}
\begin{equation}
\lambda_B=\lambda_*\left(\frac{t}{t_*}\right)^{2/5}\simeq 0.01\frac{M_P}{T_*T_0}\left(\frac{T}{T_*}\right)^{-2/5}
\simeq 10^{14}\left(\frac{T}{100\mbox{ GeV}}\right)^{-2/5}\mbox{ cm} \,.
\end{equation}
In the temperature range $100$GeV$ >T > 1$MeV the dominant contribution to the viscosity is provided by neutrinos with the comoving mean free path 
\be
\lambda_{{\rm mfp},\nu}=\frac{1}{G_F^2T^4T_0}\simeq 1\left(\frac{T}{100 \mbox{ GeV}}\right)^{-4}\mbox{ cm} \,.
\ee
The velocity $\overline v\simeq \sqrt{\rho_B/\rho}$ evolves as 
\be
\overline v\simeq v_A\simeq v_K\simeq 0.3 \left(\frac{t}{t_*}\right)^{-3/5}\simeq 0.3 \left(\frac{T}{100\mbox{ GeV}}\right)^{3/5} \,,
\ee
so that the Reynolds number at the scale $\lambda_B$ evolves as 
\be
{\rm R_k}=\frac{v_A\lambda_B}{\lambda_{{\rm mfp},\nu}}\simeq 10^{12}\left(\frac{T}{100\mbox{ GeV}}\right)^{21/5} \,.
\ee
This means that the free turbulence decay terminates (i.e. ${\rm R_k}\sim 1$)  when the temperature reaches $T\sim 0.1$~GeV. Starting from this moment the rescaled magnetic field 
and $\lambda_B$ stop  evolving. 

Once the mean free path of neutrinos becomes significantly larger than $\lambda_B$, the damping term in the Euler equation changes to $\tilde \nu\nabla^2\bv\rightarrow \tilde \alpha \bv$ with $\tilde \alpha \sim \lambda_{{\rm mfp},\nu}^{-1}$. This means that the Reynolds number (which expresses the relative importance of the damping term compared to the $(\bv\cdot\nabla) \bv$ and/or $\tilde\bB\wedge (\nabla\wedge\tilde\bB$  terms) becomes 
\be
{\rm R_k^{({\rm free})}}=\frac{v_A\lambda_{{\rm mfp},\nu}}{\lambda_B}\simeq 10^{-4}\left(\frac{T}{0.1\mbox{ GeV}}\right)^{-3} \,.
\ee
Thus, at $T\sim 8$MeV, the system returns into to free turbulent decay, corresponding this time 
to large ${\rm R_k^{({\rm free})}}$.  
The fact that in this case the dissipative regime is relatively short comes from the fact that $v_A$ is
rather large and ${\rm R_k^{({\rm free})}} = v_A^2 {\rm R_k}^{-1}$ becomes larger than unity after a damping regime which lasts only about a decade in temperature (and conformal time).

At $T\sim 1$MeV, the neutrinos finally  decouple from the cosmic plasma  
After that time, the main contribution to the viscosity is provided by the photons. Assuming that the electrons are non relativistic, their abundance is suppressed by a factor  $\eta_b = n_b/n_\ga \simeq 2.7\times 10^{-8}\Om_bh^2$ \citep{Durrer:2008aa}. 

The mean free path of photons then is 
\be
\lambda_{{\rm mfp},\gamma}= \frac{1}{a(T)\si_Tn_e} \simeq \frac{1}{\eta_b\sigma_T T^2T_0}\simeq 10^{11} \left(\frac{T}{1\mbox{ MeV}}\right)^{-2}\mbox{ cm}\,,
\ee
and the Reynolds number corresponding to the photon viscosity becomes
\be
{\rm R_k}=\frac{v_A\lambda_B}{\lambda_{{\rm mfp},\gamma}}\simeq 10\left(\frac{T}{1\mbox{ MeV}}\right)^{11/5}\,,
\ee
so that the free turbulent decay terminates again at temperatures around $T\sim 0.3$~MeV because the mean free path of photons becomes comparable to $v_A\lambda_B$. Then the magnetic field correlation length stops evolving for a  while, until the photon mean free path does become significantly larger than $\lambda_B$. Once this is the case, the damping term in the Euler equation takes its free-streaming form  $\tilde \nu\nabla^2\bv\rightarrow \tilde \alpha \bv$, but now with with 
\be
\tilde\alpha =\frac{\rho_\gamma+p_\gamma}{(\rho_{b+e^\pm}+p_{b+e^\pm})\la_{\rm mfp}} \simeq \frac{4\rho_\gamma\sigma_Tn_ea}{3\rho_b}
\ee
where $\rho_{b+e^\pm}$ is the density of the baryon $+$ electron/positron fluid and $\rho_\gamma$ is the photon density. The last equality is valid when $T\lesssim 0.5$~MeV \citep{Subramanian98}.
%$\tilde \alpha \sim \lambda_{{\rm mfp},\gamma}^{-1}$. 

Similarly to the neutrino decoupling regime, the importance of the heat dissipation via photons now decreases and free turbulent decay starts again.  However, this restart of turbulence is somewhat delayed by the large factor $\rho_\ga/\rho_b \sim 10^6(T/{\rm MeV})$. This time the viscosity is provided by the electrons, for which the growing Coulomb collisions cross-section
\be
\sigma_C\simeq \left(\frac{m_e}{T}\right)^2\sigma_T
\ee
assures that their comoving mean free path is shorter than that of photons. We obtain
\be 
\lambda_{\rm mfp,e}=(\si_c n_e a)^{-1} =\frac{1}{\eta_b\sigma_CT^2T_0}\simeq 10^{11}\mbox{ cm}\,,
\ee
which is independent of the temperature.

The natural termination point of this new free turbulence decay period is the moment of 
recombination, $t_{\rm rec}$ when most of the charged plasma disappears.  This is the termination time adopted in Fig.~\ref{f4:evolP}, $\la_B/v_A=t_{\rm rec} \simeq 200$Mpc.
 
A similar discussion of the magnetic field evolution has first been presented by \cite{Banerjee:2004df}.

%%%%%%%%%%%%%%%%%%%%%%%%%%%%%%
\subsection{Late evolution in the matter dominated Universe}
%%%%%%%%%%%%%%%%%%%%%%%%%%%%%%
 
Evolution of the magnetic field modes via MHD interactions with the charged plasma is, in principle, possible also at later stages of evolution of the Universe, i.e. in the matter dominated era. Indeed, some residual charged plasma is still present after  recombination.  Furthermore, the intergalactic medium becomes completely re-ionized again  after reionization of the Universe at the redshift $z_{\rm ri}\simeq 10$. This means that  MHD processes, including turbulent and damped decay can operate also at the late stages of the evolution of the Universe.

Processing of modes via MHD turbulence during the late time can be qualitatively understood in the same terms as  during the radiation dominated era.  The equation
\be
\label{e4:lambdab}
\lambda_B\sim v_A t
\ee
for the size of the largest processed eddies does not depend on details of the evolution of the Universe and is applicable also during the matter and cosmological constant dominated eras. The main qualitative difference with the radiation dominated era comes from the time evolution of the velocity scale $v_A= \sqrt{2\rho_B/\rho}$. 

To understand this difference, it is convenient to consider first the most optimistic case when magnetic field energy density decreases only due to the expansion of the Universe (i.e. no processing via MHD turbulence occurs). During the radiation dominated era, $\rho_B$ and  $\rho=\rho_{\rm rad}$ evolve in the same way so that $v_A\simeq const$. However, during the matter and /or cosmological constant dominated era, $\rho_B$ evolves like the radiation energy density, while $\rho=\rho_{b} \propto a^{-3}$. Here $\rho_b$ is the baryon density. Dark matter is decoupled from the charged plasma and
does not participate in MHD turbulence.
This means that $\rho_B/\rho$ decreases with time and, as a consequence, 
\be
v_A=\sqrt{\frac{2\rho_B}{\rho_{b}}}\propto a^{-1/2}\sim \frac{1}{t}; \ \ \lambda_B\sim 
   \mbox{constant.}
\ee
 Thus, no further growth of the comoving magnetic correlation scale  occurs during the matter dominated era. If we still assume that the magnetic field strength decreases as a result of dissipative processes, $v_A$ decreases even faster with comoving time and the size of the eddies which can be processed via MHD even decreases. 
 
 It is interesting to note that Eq.~(\ref{e4:lambdab}) also applies to magnetic fields which might be generated in the late Universe (e.g. via Galactic winds, see 
 Section~\ref{s:obs}. If relatively strong magnetic fields  generated at rather short distance scales are ejected into the intergalactic medium, they drive turbulence in the intergalactic medium, via the Lorentz force term  $\bB\wedge(\nabla\wedge\bB)$ in the Euler  equation (\ref{e4:euler}). This leads to an processing of the magnetic eddies by MHD turbulence up to the scale given by Eq.~(\ref{e4:lambdab}). The process of free turbulent decay  then moves the magnetic field power to  shorter scales where it is  dissipated into heat. Thus, the strength of the magnetic field is  reduced and its correlation length is increased until the condition
 \be
\label{e4:lambdab0}
\lambda_B\sim v_A t_0\sim \sqrt{\frac{2\rho_B}{\rho_bH_0^2}}
\ee
is satisfied, where $t_0$ is the present age of the Universe. Actually, the time in the above equation should be $t_0-t_i$ where $t_i$ is the injection time, but we assume $t_i \ll t_0$, hence injection at redshift $z_i\ge 1$ so that we may neglect this correction in our order of magnitude estimate. This provides an upper  bound on the fields injected at scale $\la$  into the intergalactic medium by any process (including relic fields produced in the early universe)
\begin{equation}
\label{eq:limit}
\tilde B\lesssim H_0^2 \Omega_b^{1/2}M_P \lambda_B  \simeq  10^{-8}\left[\frac{\la_B}{1\mbox{ Mpc}}\right]\mbox{ G}\,.
\end{equation}
Occasionally, this limit is close to the limit (\ref{e4:limit0}) on the comoving field strength at the moment of recombination. 

The viscosity of non-relativistic plasma \citep{LL10} is $\nu\simeq \overline v_T\lambda_{\rm mfp}$, where $\overline v_T\simeq (T/m_e)^{1/2}$ is the thermal velocity of the plasma electrons.  The Reynolds number is, therefore,  R$_K =5v_A\la_B/(\overline v_T\la_{\rm mfp}) \sim v_A^2t_0/(\overline v_T\la_{\rm mfp})$. In order for these magnetic fields to be in the turbulent regime we therefore have to require also
\be
\overline v_T \la_{\rm mfp} < v_A^2t \,.
\ee
The electron mean free path is dominated by the Coulomb collisions such that
$$
\la_{\rm mfp} \simeq (\si_C n_b)^{-1}  \simeq \frac{10T_{IGM}^2m_p}{\si_T m_e^2 \Omega_bH_0^2M_P^2} \simeq \frac{5\mbox{ pc}}{(1+z)^2}\left(\frac{T_{IGM}}{10^4\mbox{ K}}\right)^{2} \,.
$$
where $T_{IGM}\sim 10^4$~K is the present-day temperature of the IGM, which got re-heated by the reionization at the redshift $z_{\rm ri}\simeq 10$. 
Inserting this in the expression for the Reynolds number we find that the condition R$_K>1$ requires
\be\label{e4:rey0}
\tilde B \gtrsim 10^{-12}\mbox{ G}\left(\frac{z_{\rm ri}}{10}\right)^{-10/6}\,.
\ee
Thus,  fields with correlation length larger than $\lambda_B\sim 0.1$~kpc (see Eq. (\ref{eq:limit})) could have been processed by the turbulence in the IGM. 

Weaker fields with correlation length in the range 5 pc$<\lambda_B<100$~pc might avoid damping via turbulence at the late stages of evolution, because in this distance range IGM plasma velocities are damped by the viscosity due to Coulomb collisions. However, at the onset of reionization, when the temperature of the IGM was much lower than $\sim  10^4$~K, the dampling distance scale might have been shorter, so that fields with shorter correlation length possibly excited turbulence. Detailed understanding of the IGM turbulence excited by the magnetic fields would require modeling of the reionization dynamics.  

At still shorter distance scales, $\lambda_B<\lambda_{\rm mfp}$, the IGM plasma is collisionless, so that its not appropriately described by the MHD equations. Instead, one has to resolve the Botzmann /Vlasov equations for particle distributions \citep{Kulsrud83}. Turbulence could also develop in the collisionless plasma. The generic nature of the relation (\ref{eq:limit}) suggests that it might be also applicable to the collisionless case, although a detailed investigation of the behaviour of the IGM in the presence of relatively strong short scale magnetic fields is needed to verify this. 

In general, magnetic fields which are initially stronger than the  limit~(\ref{eq:limit}) could evolve along the evolutionary tracks outlined above toward this limit. Therefore, the line in the $(B, \lambda_B)$ plane, given by 
Eq.~(\ref{e4:lambdab0}) is the locus of the "termination points" of the evolution tracks for all cosmological fields generated at short distance scales. 
 
Further processing of the fields by the MHD processes occurs during structure formation \citep{Ryu08,Schleicher10,Sur10}. In general, gravitational collapse leading to formation of galaxies and galaxy clusters  amplifies any pre-existing fields via straightforward magnetic flux conserving compression and/or via action of various types of dynamos.  In galaxies, flux conservation yields  an amplification of
\be\label{e4:galflux}
\frac{B_{\rm fin}}{B_{\rm in}} = \left(\frac{\rho_{\rm fin}}{\rho_{\rm in}}\right)^{2/3} \sim 10^4\left(\frac{\rho_{\rm fin}}{10^6\rho_{\rm in}}\right)^{2/3}  \,.
\ee
where $\rho_{\rm in}$, $\rho_{\rm fin}$ are the initial and final average matter densities before and after the gravitational collapse. 

The amplification by dynamo action is much more uncertain, but it may be many orders 
of magnitude larger.
In this case, the final characteristics of the fields in the gravitationally collapsed structures is  largely independent of the initial conditions at the onset of the structure formation, but is given by some dynamo saturation amplitude. The fact that magnetic fields in galaxies are all roughly of the same amplitude hints that this may well be the case. 

The only place where the "relic" initial magnetic fields are preserved is then the intergalactic medium in the voids of the large scale structure. The pre-existing field in the voids is not processed by the MHD effects accompanying structure formation and, therefore, the field in the voids must still satisfy the relation~(\ref{eq:limit}).
Detecting magnetic fields in voids with correlations scale given by this relation would be a strong indication of their primordial nature.

%Observation (s:obs)

%%%%%%%%%%%%%%%%%%%%%%%%%%%%%%%%%%%%%%%%%%%%%%%%%%%%%%
\section{Observational constraints}
\label{s:obs}
%%%%%%%%%%%%%%%%%%%%%%%%%%%%%%%%%%%%%%%%%%%%%%%%%%%%%%

The range of the field strengths $B_\lambda$ and correlation lengths $\lambda_B$, implied by the relation (\ref{eq:limit}) is within the reach of available observational tools based on the methods of radio and $\gamma$-ray astronomy. Therefore, it appears reasonable to explore the possibility of observational detection of the relic fields which might be present in the voids of the LSS. In this Section we summarize the status of the searches of Intergalactic Magnetic Fields (IGMF), including the fields in the voids of the LSS. 

In the absence of positive detections, the discussion of this section is limited to the summary of observational constraints on the strength and correlation length $B_\lambda, \lambda_B$ of IGMF.  It is important to note that relic magnetic fields from the Early Universe are not the only magnetic fields which might populate the voids. Therefore, even a real measurement of of IGMF, which should be possible with future observational facilities, does not necessarily imply the measurement of the relic fields from the Early Universe. We discuss the possibility to distinguish the relic magnetic fields from the fields of different origin at the end of the Section.  

\begin{figure}
\includegraphics[width=\linewidth]{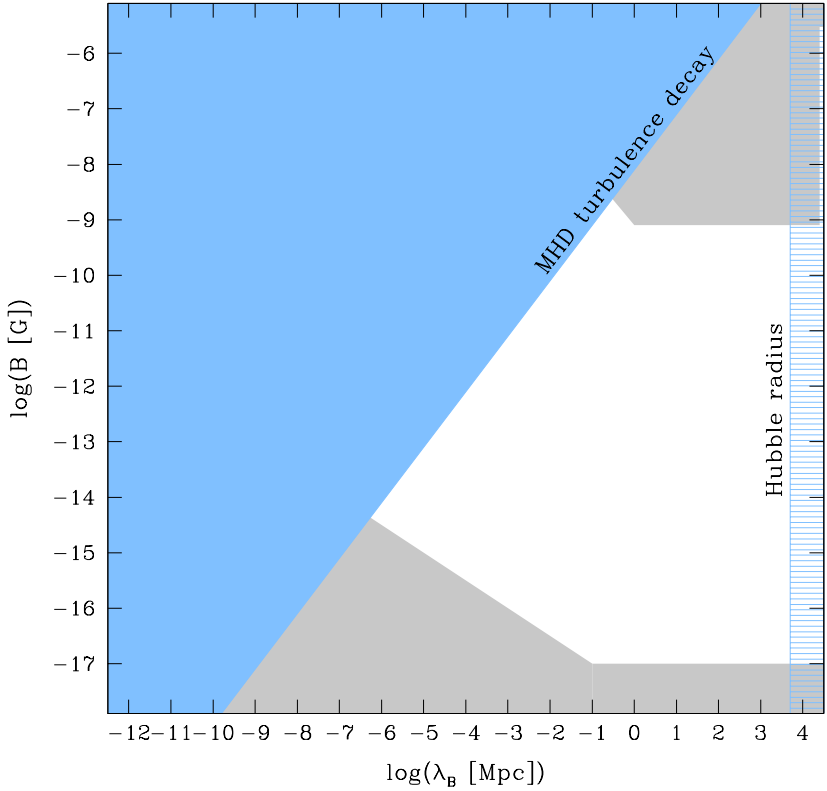}
\caption{Theoretical constraints on the IGMF parameters in the present day Universe.}
\label{f5:theory}
\end{figure}

\begin{figure}
\includegraphics[width=\linewidth]{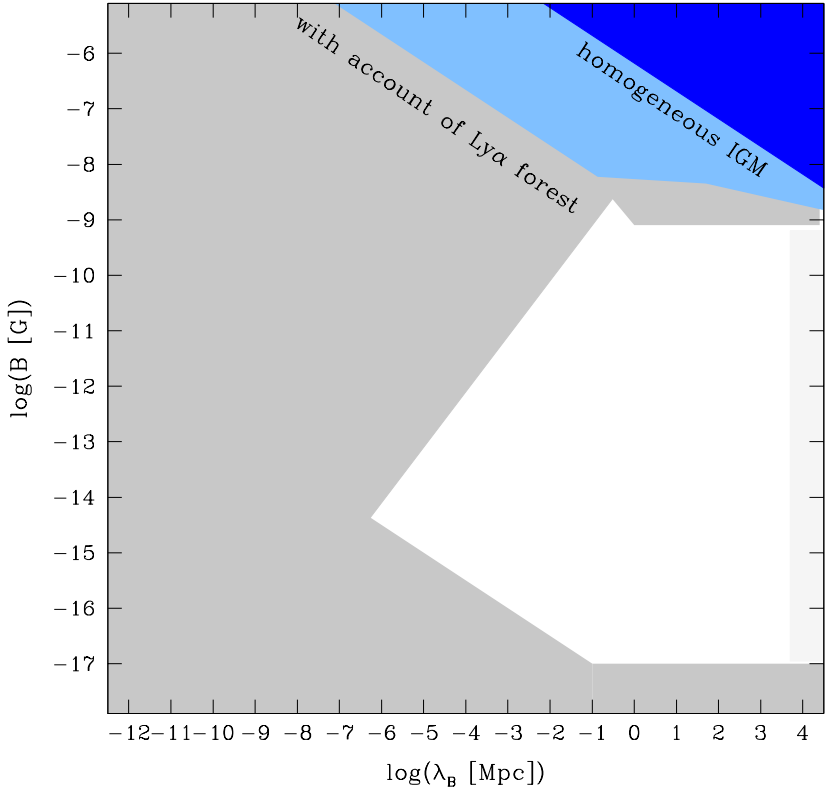}
\caption{Constraints on the IGMF from Faraday rotation measurements.}
\label{f5:faraday}
\end{figure}

Similarly to the previous sections, we present the observational constraints on IGMF in the $(B_\lambda,\lambda_B)$ parameter space, see Fig.~\ref{f4:evolP}. This figure
shows the evolution of the field strength and correlations scale throughout the history of the Universe. On the other hand, it can also be used to show constraints on present day fields, i.e., limits on the allows ranges for the parameters $(B_\lambda,\lambda_B)$ at $z=0$. The white, unshaded area
in Figs.~\ref{f4:evolP} to \ref{f5:CMB} and similar figures shows the allowed range of parameters of IGMF in the present Universe. Each of the constraints (boundaries of the unshaded region) is explained in detail in this section.

%%%%%%%%%%%%%%%%%%%%%%%%%%%%%%%%%%%%%%%%%
%%%%%%%%%%%%%%%%%%%%%%%%%%%%%%%%%%%%%%%%%

%%%%%%%%%%%%%%%%%%%%%%%%%%%%%%%%%%%%%%%%%%%%%%%%%%%%%%
\subsection{General theoretical bounds on $B_\lambda$, $\lambda_B$}
\label{s:mf}
%%%%%%%%%%%%%%%%%%%%%%%%%%%%%%%%%%%%%%%%%%%%%%%%%%%%%%

The straightforward theoretical constraint on the present day strength and correlation length of cosmologically produced IGMF is given by Eq.  (\ref{eq:limit}). Strong magnetic field injected at small distance scales would drive turbulence in the primordial plasma and later in the IGM. Eq.~(\ref{e4:lambdab0}) provides an estimate of the size of the largest eddies which can be processed by the turbulence on a time scale comparable to the age of the Universe. Turbulence removes power from the short-scale modes of magnetic field. This leads to the increase of the field correlation length until the relation (\ref{eq:limit}) is satisfied. The constraint  (\ref{eq:limit}) is shown with the label "MHD turbulent decay" in Fig.~\ref{f5:theory}. 

%%%%%%%%%%%%%%%%%%%%%%%%%%%%%%%%%%%%%%%%%
%%%%%%%%%%%%%%%%%%%%%%%%%%%%%%%%%%%%%%%%%

There is no formal upper limit on the possible correlation length of IGMF. If generated during inflation, it might be even coherent on the scales larger than size of the visible part of the Universe. However we can never observe correlations on scales larger than the present Hubble scale, which is indicated by the line marked "Hubble radius" in Fig.~\ref{f5:theory}. A fields with correlations scale larger than the present Hubble scales would be perceived as a constant field throughout the Universe.
Observable limits on such fields are not so strong, they are of the order of $10^{-9}$Gauss
and mainly come from Faraday rotation in the CMB polarization, see Section~\ref{s:CMB_limits}. A recent  revised discussion on CMB limits for constant magnetic fields, taking into account free streaming neutrinos, can be found in~\cite{Adamek:2011pr}.

%%%%%%%%%%%%%%%%%%%%%%%%%%%%%%%%%%%%%%%%%%%%%%%%%%%%%%
\subsection{Faraday rotation measurements}
\label{s:faraday}
%%%%%%%%%%%%%%%%%%%%%%%%%%%%%%%%%%%%%%%%%%%%%%%%%%%%%%

Upper bounds on the strength of IGMF are imposed by the non-observation of Faraday rotation 
of the polarization plane of linearly polarized radio emission from distant quasars.

Propagation of a linearly polarized electromagnetic wave of wavelength $\lambda$ through plasma causes a rotation of the polarization vector by an angle \citep{Kronberg:1993vk}
\be\label{e5:Faraday}
\Psi=RM\lambda^2
\ee
where the rotation measure ($RM$) is determined by the distance to the source $d(z)$, the strength of the magnetic field component parallel to the line of sight, $B_{||}$ and by the free electron density $n_e$ in the region through which the wave is propagating
\be
\label{eq:rm}
RM=\frac{e^3}{2\pi m_e^2}\int_0^{d(z)} \frac{n_e(z) B_{\|}(z)}{(1+z)^2}dx(z) \,.
\ee
In standard $\La$CDM cosmology, the distance element  $dx(z)$ is related to the redshift by
\be 
dx(z)=\frac{dz}{H_0(1+z)\sqrt{\Omega_{m}(1+z)^3+\Omega_{\Lambda}}} \,.
\ee
$\Omega_{m}$, $\Omega_{\Lambda}$ are the present matter and dark energy fractions. 

Given the distribution of free electrons in the IGM one can derive a measurement (or an upper limit) of $B_{||}$ from the measurements of (or upper limits on) the Faraday rotation of polarized radio emission from distant extragalactic sources of (most of them are distant quasars). 

There are several challenges to the measurement of IGMF from Faraday rotation. First, the distribution of free electrons in the IGM along the lines of sight toward different quasars is uncertain. Depending on the assumptions about this distribution, different bounds on the IGMF have been reported in the past \citep{Kronberg:1993vk,Kronberg82,Blasi99}. 

Furthermore, the effect of Faraday rotation due to the IGMF is small compared to that produced by the magnetic field of the Milky Way. Sensitive constraints on the IGMF can be derived only after a proper characterization and subtraction of the effect of Galactic magnetic field. However, our knowledge of the Galactic magnetic field is rather 
limited~\citep{Han94,Brown07,Pshirkov11,Farrar12,Farrar12a,Oppermann12} and there still are large uncertainties in Galactic field models. This introduces uncertainties in the constraints on the IGMF derived from the Faraday rotation measurements. 

In the simplest approximation, an estimate of the free electron density in the IGM can be obtained from the known mean baryon density $\rho_{b}=\Omega_{b}\rho$, with $\Omega_{b}\simeq 0.02/h^2 \simeq 0.04$~\citep{Komatsu:2010fb}. Here and in what follows we adopt the value $H_0 =h100$km/s/Mpc $=70$km/s/Mpc for the Hubble constant. The IGM is almost completely ionized today, so that 
\be
n_{e0}\simeq \frac{\Omega_b\rho_b}{m_p}
\ee
is a good estimate of the average free electron density in the Universe (and in the voids of LSS) today, implied by the electric neutrality of the Universe. This estimate has been adopted in the early studies of the Faraday rotation constraints on the IGMF \citep{Rees72,Kronberg76,Kronberg:1993vk}. 
 
The redshift dependence of $n_e$ depends on the ionization history of the interstellar medium. After $z\sim 10$ the IGM is mostly ionized so that $n_e$ is about the average baryon density $n_e(z)\sim n_{e0}(1+z)^3$. Assuming that the magnetic field strength at the integral scale decreases only due to the expansion of the Universe, $B_{\|}(z)\sim B_0(1+z)^2$ and substituting this redshift dependence of $B_{\|}$ and $n_e$ into Eq. (\ref{eq:rm}) gives numerically
\be
\label{eq:rm_igmf}
RM_{\rm IGMF}\simeq
%\frac{e^3n_{b0}B_{0\|}}{2\pi m_e^2c^3 H_0(\Delta\lambda)^2}\left[\sqrt{\Omega_{m0}(1+z)^3+\Omega_{\Lambda 0}}-1\right]\simeq 
10\left[\frac{B_0}{10^{-8}\mbox{ G}}\right]\left[\frac{n_{e0}}{10^{-7}\mbox{ cm}^{-3}}\right]\left[\frac{\sqrt{\Omega_{m}(1+z)^3+\Omega_{\Lambda }}-1}{\Om_m}\right]\frac{\mbox{rad}}{\mbox{m}^2}
\ee
for magnetic fields coherent on the Hubble scale $\lambda_B\sim \ell_H =H_0^{-1}$. 
For fields with smaller coherence scale, a suppression factor of $(\la_B/\ell_H)^{1/2}$ is introduced to account for the randomness of the magnetic field direction. Constraints on the IGMF at the level of 
\be
B\lesssim 2\times 10^{-9}\left(\frac{\lambda_{B}}{\ell_H}\right)^{-1/2}\mbox{~G}
\ee
have been derived in this way from the Faraday rotation data by~\citep{Rees72,Kronberg76,Kronberg:1993vk,Blasi99}. (Note that stronger constraints are quoted in the original literature, \citet{Rees72,Kronberg76}, since there $\Omega_{b}\sim 1$ is  assumed.)  This constraint, properly rescaled to $\Omega_{b}\simeq 0.04$, is shown by the dark blue shading in Fig.~\ref{f5:faraday}.

A more elaborate analysis leading to a somewhat more accurate estimate of $n_e$ along the line of sight to distant quasars is based on the account of information obtained from the Ly$\alpha$ forest data \citep{Kronberg82,Blasi99}. \citet{Kronberg82} were first to notice that an excess of Faraday rotation in the signal of distant quasars is observed when the line of sight toward a quasar passes through absorption line systems. They have interpreted this as being due to the presence of magnetic fields in the "clouds" responsible for the absorption lines (these may be either intervening galaxies, or lower density Ly$\alpha$ clouds which have not yet collapsed to form galaxies). 

This idea was further developed by \citet{Oren95} who also made an attempt to subtract the Galactic RM  from the RM measurements of distant quasars, to search for the residual RM due to the IGMF and/or magnetic fields in the intervening clouds / galaxy systems. Their claim is that the contribution of the Galaxy to the RM of distant quasars can be determined with precision better than $\sim 15-20$~rad/m$^2$. This result has been used by \citet{Blasi99}, who has introduced a detailed model of electron density distribution along the line of sight toward quasars, based on the statistics of the density distribution in Ly$\alpha$ clouds. \citet{Blasi99} also assumed that in higher density clouds the magnetic field is amplified by  compression, to that the effect of the density increase in the clouds contributes twice to the Faraday rotation signal (\ref{eq:rm}): once directly through the increased $n_e$ and second time through the increase of $B_\|$. Stronger Faraday rotation signals from the clouds results in stronger constraints on the unamplified IGMF outside the clouds. The constraint derived by \citet{Blasi99} is shown as the light-blue shaded region in Fig.~\ref{f5:faraday}.  

Another application of the correlation of excess rotation measures with intervening structures along the line of sight is developed in \citet{Kronberg08,Bernet08}.  They find strong correlation of the increased RM with MgII absorbing systems with the equivalent width (EW) of the MgII line of $EW>0.3$~\AA. The excess RM introduced by the intervening systems is only $RM\simeq 140$~rad/m$^2$. Measurements of so small additional RM are possible because of the redshift dependence of the additional RM due to the intervening MgII absorption systems.  The typical size of the MgII absorption systems is $\sim 100$~kpc, which indicates that these systems are, most probably, bubbles around the star-forming galaxies produced by galactic winds  \citep{Bordoloi11}. Based on the measurement of the RM and on the estimates of the hydrogen column densities of the MgII absorption halos, \cite{Kronberg08,Bernet08} derive an amplitude  of the magnetic field in these 100~kpc scale galactic halos of $B\sim 10\ \mu$G, under the assumption that the coherence length of the magnetic field is comparable to the halo size. The field correlation length in the halos is most likely significantly shorter than the halo size. Presence of the field reversals would boost the the estimate of the field strength by the square root of the ratio of the halo size to the correlation length (see \citet{bhat13} for further discussion).

The main uncertainty in the measurements of magnetic fields in different components of the LSS, based on the Faraday Rotation technique comes from the uncertainty of the Galactic contribution to the $RM$ signal.
Substituting typical scale height of the Galactic disk $H_{\rm Gal}$ for $d(z)$ and the free electron density in the interstellar medium $n_{e, \rm Gal}$ for $n_e(z)$ in Eq. (\ref{eq:rm}) one finds a Rotation Measure due to the Galactic magnetic field of the order of
\be 
\label{eq:rm_galactic}
RM_{\rm Gal}\simeq 10^2\left[\frac{B_{\rm Gal}}{10^{-6}\mbox{ G}}\right]\left[\frac{n_{e,\rm Gal}}{0.1\mbox{ cm}^{-3}}\right]\left[\frac{H_{\rm Gal}}{1\mbox{ kpc}}\right]\frac{\mbox{rad}}{\mbox{m}^2}
\ee
where $B_{\rm Gal}$ is the magnetic field strength in the interstellar medium. Although "typical" values of all the three quantities, $H_{\rm Gal},B_{\rm Gal},n_{e\rm Gal}$ are difficult to define (e.g. they are widely different for the Galactic disk and halo), the order of magnitude estimate in Eq. (\ref{eq:rm_galactic}) shows that the Faraday rotation accumulated during the propagation of the radio beam through the Galaxy is much larger than the that accumulated during the propagation through the IGM. 

%%%%%%%%%%%%%%%%%%%%%%%%%%%%%%%%%%%%%%%%%%%%%%%%%%%%%%
\subsubsection{Prospects for IGMF measurement with next-generation radio telescopes.}
\label{s:next_generation_radio}
%%%%%%%%%%%%%%%%%%%%%%%%%%%%%%%%%%%%%%%%%%%%%%%%%%%%%%

Recent accumulation of a large data base of Rotation Measures from extragalactic sources \citep{Taylor09,Stil11} has enabled a significant improvement of the knowledge of the global structure of the Galactic magnetic field \citep{Farrar12,Farrar12a,Oppermann12}. This will, in principle, allow a better control of the Galactic RM and, as a consequence, lead to better constraints on the IGMF contribution to the RM. However, an order-of-magnitude improvement in the sensitivity of the Faraday Rotation measurements of the IGMF requires to shrink  the error bars of $RM_{\rm Gal}$ by a factor of 100, which in term requires the  knowledge of the Galactic magnetic field and free electron density with sub-percent precision. Taking into account the remaining large uncertainties in  modeling the free electron distribution \citep{NE2001,Gaensler08}, as well as in the degeneracy of  model parameters of the Galactic magnetic field \citep{Farrar12,Farrar12a,Pshirkov11}, it is not clear whether this precision can indeed be reached. 

The next qualitative improvement of our knowledge of the three-dimensional structure of the Galactic magnetic field and of the free electron distribution in the interstellar medium is expected from the next generation radio telescopes LOFAR ({\tt http://www.lofar.org/}) and SKA ({\tt http://www.skatelescope.org/}), \citep{Beck11}. Qualitatively new survey capabilities of these facilities will further increase the sample of $RM$ measurements for extragalactic sources from $\sim 1$~source/deg$^2$ \citep{Taylor09} up to $\sim 10^3$~sources/deg$^2$ in the case of an SKA all-sky survey with 1~hr exposure per field-of-view \citep{Gaensler06}. This will allow a much more detailed modeling of the Galactic magnetic field, thereby improving the sensitivity for the search of weak extragalactic contributions to the $RM$. 

An improved measurement of the three-dimensional structure of the Galactic magnetic field can be obtained by taking into account not only extragalactic sources, but also sources of linearly polarized radio emission inside the Milky Way. The most important class of polarized Galactic sources are pulsars. Measurements of dispersion and rotation measures of the pulsar emission provide constraints on both the magnetic field and free electron density in the interstellar medium.  Pulsars can be found at different locations inside the Galaxy, so that they in principle allow for a three-dimensional "tomography" of the Galactic magnetic field (provided that sufficiently large number of pulsars can be found in thin distance slices and in different directions \citep{Han06}. Up to now some $\sim 2\times 10^3$ pulsars are known. 554 of them have been used by \citet{Han06} to study the structure of the Galactic magnetic field. SKA will provide a qualitative improvement due to a 10 times larger ($\sim 2\times 10^4$) pulsar detection statistics \citep{Smith09}.

A qualitatively new possibility to distinguish the IGMF contribution to the $RM$ from the Galactic contribution will also arise with some $\sim 10^8$ extragalactic sources in the SKA sky survey. This will be a possibility of a three-dimensional "$RM$ tomography" of the Universe, i.e. study of the gradual accumulation of the $RM$ signal in thin  redshift slices. The most straightforward effect expected from the IGMF contribution to the $RM$ is the characteristic dependence of the signal on $(1+z)$, see Eq. (\ref{eq:rm_igmf}). Detection of such dependence in the $RM(z)$ signal may provide a possibility to "bypass" the uncertainty of the Galactic contribution to the $RM$. 

%%%%%%%%%%%%%%%%%%%%%%%%%%%%%%%%%%%%%%%%%%%%%%%%%%%%%%
\subsection{Limits from CMB observations}
\label{s:CMB_limits}
%%%%%%%%%%%%%%%%%%%%%%%%%%%%%%%%%%%%%%%%%%%%%%%%%%%%%%

Magnetic fields interact with the primordial plasma in the early universe. The presence of sufficiently strong  magnetic field  therefore, affects  the evolution of the plasma. The imprint of  magnetic fields on the state of the plasma can potentially be revealed in the properties of cosmic microwave background (CMB), which encodes information on the state of primordial plasma at the epoch of photon decoupling and recombination, $z_{\rm rec} \sim 1100$. The observed temperature fluctuations and polarization provide the most precise cosmological data set and it is therefore most interesting to study the effect of a primordial magnetic field on these data.

A magnetic field affects the CMB anisotropies and polarization in many ways, 
see~\cite{Barrow97,Durrer:2006pc,Shaw:2009nf,Paoletti11,Shaw12,Paoletti:2012bb}. The results for a constant magnetic field derived in the pioneering paper by \cite{Barrow97} are actually invalid, since there the compensation by neutrino anisotropic stresses (see \citealt{Adamek:2011pr}) which isotropize the Universe are not taken into account. 
\begin{enumerate}
\item\label{pgrav} The energy momentum tensor of the magnetic field perturbs the geometry of the Universe which governs the geodesic motion of CMB photons. This introduces scalar, vector and tensor perturbations in the CMB \citep{Durrer:1999bk}.
\item\label{pfluid} The evolution of the cosmic plasma is affected by the presence of a magnetic field which leads to fast and slow magnetosonic waves \citep{Adams:1996cq} and to Alfv\'en waves~\citep{Subramanian98a,Durrer:1998ya}. The former introduce slight shifts in the acoustic peaks of the CMB \citep{Kahniashvili:2006hy}, while the latter mainly lead to vector perturbations \citep{Lewis:2004ef}.
\item Faraday rotation turns $E$-polarization of the CMB partially into $B$-po\-lari\-za\-tion 
\citep{Seshadri:2000ky,Kahniashvili09}.  Since Faraday rotation is frequency dependent, see 
Section~\ref{s:faraday}, this can be separated from the the effects under points \ref{pgrav}. and \ref{pfluid}. which are 'achromatic'.
Actually magnetic fields generate large vector modes which generate dominantly B-polarization, but with the usual thermal CMB spectrum (see~\citealt{Seshadri:2000ky,Lewis:2004ef}). 
\item If the magnetic field is helical, its parity violation leads to correlations of the temperature anisotropy and  of E-polarization with B-polarization; correlations which are forbidden in a parity invariant universe \citep{Caprini:2003vc}.
\item \label{pSilk}The presence of a magnetic field affects recombination and Silk damping. It therefore alters the damping tail of the CMB anisotropies \citep{Jedamzik:2011cu}.
\item \label{pNt}Non-thermal dissipation of magnetic field energy into the energy of electrons/positrons during the recombination epoch can lead to distortion of blackbody CMB spectrum, mainly by introducing a chemical potential \citep{Jedamzik98}. 
\item Magnetic fields affect the formation of clusters. Their abundance is well determined by  Sunyaev-Zel'dovich (SZ) decrement measurements in the CMB \citep{Shaw12}.
\item Since the energy momentum tensor of the magnetic field and the Lorentz force in the MHD limit are quadratic in the field strength, magnetic fields, even if they are Gaussian, will introduce non-Gaussian CMB anisotropies and polarization.
\end{enumerate} 

Interestingly, all these effects yield limits on magnetic fields of the order of  nG. This is
not so surprising as the fluctuations in the CMB are of the order of $10^{-5}$ and the energy
density in a cosmic magnetic field is
\be
\Om_B  \simeq 10^{-5}\left(\frac{B}{10^{-8}G}\right)^2\Om_\ga \,.
\ee
We therefore expect that magnetic fields of $10^{-9}$G leave an imprint of about 1\% on the
CMB anisotropies and polarization, which is marginally detectable. The only effects on the CMB for which this argument is not valid are the points \ref{pSilk} and \ref{pNt} of the above list.

In the following we explain the above points in more detail.

%%%%%%%%%%%%%%%%%%%%%%%%%%%%%%%%%%%%%%%%%%%%%%%%%%%%%%
\subsubsection{Limits from CMB angular power spectrum}
\label{s:CMB_angular}
%%%%%%%%%%%%%%%%%%%%%%%%%%%%%%%%%%%%%%%%%%%%%%%%%%%%%%
The energy momentum tensor of a stochastic magnetic field is fluctuating from point to point. Its power spectrum is given by the 4-point function of the magnetic field, e.g.
\be\label{e5:TB}
a^4\langle T^{(B)}_{ij}(\bk)T^{*(B)}_{\ell m}(\bk')\rangle = (2\pi)^3\de(\bk-\bk')\int d^3q \langle B_i(\bk-\bq)B_j(\bq)B_\ell(\bq)B_m(\bk-\bq)\rangle
 + \cdots \,.
 \ee
 
In particular, even if the magnetic field distribution is Gaussian, its energy momentum tensor is not. But in this case the power spectrum of the energy momentum tensor can be expressed in terms of the mangetic field power spectrum with the help of Wick's theorem as explained by~\citet{Durrer:1999bk}. For example, for the magnetic energy density spectrum we obtain 
\be
 P_{\rho_B}(k) =\frac{1}{ (2\pi)^3} \int d^3q P_B(q)P_B(|\bk-\bq|) \,.
\ee
Note the difference between this power spectrum which describes {\em fluctuations} in the magnetic energy density and the mean magnetic energy density which is its zero mode given by the $dk/k$-integral of $P_B(k)k^3$.
 
The metric fluctuations from this source of energy and momentum are calculated via the first order Einstein equations and they enter the Boltzmann equation of CMB fluctuations as a source term.
For vector perturbations, which are absent in standard cosmology, they have been implemented in CAMBcode (developed by \citealt{Lewis:2002ah}) by \cite{Lewis:2004ef}. Even though the magnetic 
field is a vector field,  its energy momentum tensor (\ref{e5:TB}) which is quadratic in the field, contains scalar, vector and tensor contributions of similar amplitude. 

For a magnetic field spectrum behaving like $k^{n_s}$ on large  scales, $k<k_B$, the spectrum of the magnetic 
energy momentum tensor is dominated by the upper cutoff and behaves as white noise on large  scales, $k<k_B$ if $n_s>-3/2$. For $n_s<-3/2$ the energy momentum tensor inherits the magnetic field spectral index $n_s$.

A magnetic field with super horizon scale correlations, $\la\gsim \ell_H(z)$, keeps the universe homogeneous but renders it anisotropic, a Bianchi~I model, see \cite{Barrow97,Adamek:2011pr}. At temperatures below 1MeV, where neutrinos free stream, such a global anisotropy is however compensated by the induced neutrino anisotropic stress generated by the gravitational effects of anisotropic relativistic free streaming, see~\cite{Adamek:2011pr}.  For a stochastic magnetic field this leads to a suppression by a factor $(k/\HH)^2$ on super horizon scales \citep{Bonvin:2010nr}.  This effect is very relevant especially for scale invariant magnetic field spectra. However, such spectra which can only come from inflation also generate, in addition to the compensated mode and to the passive mode present in causally generated magnetic fields, e.g, from a phase transition (see \citealt{Shaw:2009nf}), a small constant mode which is not compensated and which can have a scale invariant spectrum~(\cite{Bonvin:2011dt}).

In addition to gravitational effects which dominate on large scales, interactions of electrons with the magnetic field and the CMB affects the CMB anisotropy spectrum also via a modification of the acoustic peaks. The sound speed in the presence of a magnetic field is enhanced, $c_s^2 \ra c_s^2 +(\bk\cdot\bB)^2/\rho$, which affects the position of the acoustic peaks in the CMB anisotropy and polarization. Furthermore, Alfv\'en waves, i.e. vector perturbtions are generated which lead to relatively strong B-polarisation~\citep{Lewis:2004ef} and the Silk damping tail is affected. 

For a given magnetic field spectrum, all these effects can be included in standard n Boltzmann codes like CAMB by \cite{Lewis:2002ah}, which treat the photon+baryon+dark matter system within linear perturbation theory to compute CMB anisotropies and polarization \citep{Shaw12}. The magnetic field contributions just enter via a source term on the right hand side of the linear perturbation equations and by modifying the initial conditions. They lead to a so called 'passive mode' which has the same initial conditions as the inflationary mode. In this mode the magnetic field just intervenes by changing the evolution equation of the baryons due to the Lorentz force. In addition there is a 'compensated mode'  where the gravitational effects of the magnetic field enter. On very large, super horizon scales, these are, however, compensated by initial fluid under densities and neutrino magnetic stresses, at least if the magnetic fields are generated causally. See 
\cite{Shaw:2009nf,Bonvin:2010nr,Adamek:2011pr}. If the magnetic field is generated during inflation, the compensation mechanism after neutrino decoupling is still active, but an additional 'passive mode' due to the matching condition at the end of inflation is introduced \citep{Bonvin:2011dt}. The relation of its amplitude to the late time magnetic field strength depends on the details of reheating. 

%%%%%%%%%%%%%%%%%%%%%%%%%%%%%%%%%%%%%%%%%
\begin{figure}
\includegraphics[width=\linewidth]{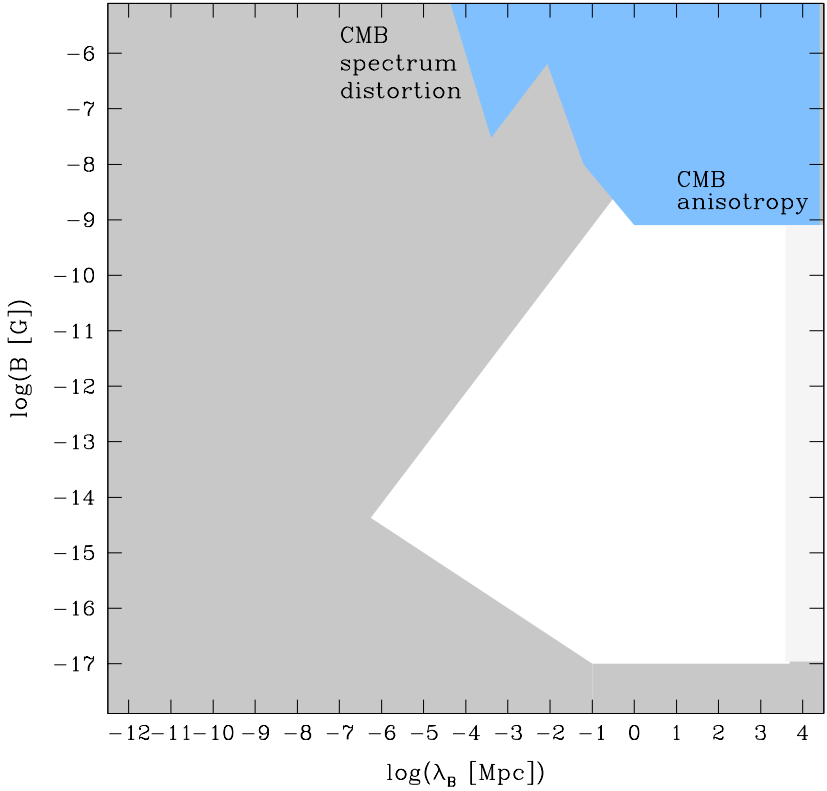}
\caption{Constraints on the cosmologically produced IGMF from CMB anisotropy measurements.}
\label{f5:CMB}
\end{figure}
%%%%%%%%%%%%%%%%%%%%%%%%%%%%%%%%%%%%%%%%%

The non-observation of the large angular scale anisotropies of the CMB have led to an upper limit 
$B\le 4\times 10^{-9}$~G for the fields with correlation scale of the order of the CMB scale, 
$\lambda_B\gsim 10h^{-1}$Mpc  \citep{Barrow97}. Limits depending on  the power law of the magnetic field spectrum have been derived in Refs.  \citep{Durrer:1999bk,Giovannini09,Paoletti11,Yamazaki12}. The envelope of the upper bounds on $(B,\lambda_B)$ for the range of the power law indices $-3\le n_s\le 2$ \citep{Paoletti11} is the lower boundary of the light blue shaded region of  $(B_\lambda,\lambda_B)$ parameter space in Fig. \ref{f5:CMB}. 

Similar limits can also be derived from the absence of non-Gaussianities in the observed CMB anisotropies.
Non-Gaussianities from magnetic fields are mainly of the local type.
The limits on the CMB bi-spectrum, usually parameterized in terms of $f_{nl}$, therefore constrain a possible contribution to CMB anisotropies and polarization from magnetic fields. These constraints can be used to limit the magnetic field amplitude for a given spectrum~\citep{Seshadri:2009sy,Caprini:2009vk,Trivedi:2010gi}. Recently also the tri-spectrum has been calculated \citep{Trivedi:2011vt}.

\subsubsection{Spectral distortions}
\label{s:CMB_spectrum}
Non-thermal dissipation of magnetic field energy into the energy distribution of electrons before recombination  can lead to distortions of the blackbody CMB spectrum~\citep{Jedamzik98}.   This distortion generates a non-zero chemical potential $\mu$, which has been calculated by~\citet{Jedamzik00} in the form of a double integral. In order to obtain bounds on $B,\lambda_B$, we consider two  limiting cases in which the integral can be performed analytically, namely, the cases when the magnetic field correlation length is much smaller or much larger than the characteristic damping length scale:
\begin{equation}
\label{damping}
\lambda_D = \frac{2 \pi}{z_\mu^{3/2}}\sqrt\frac{\overline{t}_0}{15 n_e^0 \sigma_{T}} \approx 400\mbox{ pc},
\end{equation}
where time constant is $\overline{t}_0=2.4\times 10^{19}$ s, $z_\mu$ is the characteristic redshift of freeze-out from double-Compton scattering $z_\mu=2.5\times 10^{6}$,    $n_e^0$ is electron density and $ \sigma_{T}$ is Thomson  cross section. 

Taking into account the constraint  $|\mu|<9 \times 10^{-5}$ at 95 \% confidence level from COBE FIRAS data~ \citep{Fixsen96}, one can derive analytical  limits on magnetic fields in the two limiting cases. 
In the case  $\lambda_B \ll \lambda_D$ one has~ \citep{Jedamzik00}:
\begin{equation}
\label{lim_low}
B < 3.2 \times 10^{-8} G \frac{1}{\sqrt{K}} \left(  \frac{\lambda_B}{400 ~\mbox{pc}}\right)^{-(n_s+3)/2}~,
\end{equation}
where  $K=1.4 \Gamma (n_s/2+5/2) \Gamma(3n_s/5 +9/5)2^{-(n_s+5)/2}(6/5)(n_s+3)$ is a constant of order unity, $K=0.8$ resp. $2.1$ for $n_s=-2$ resp. $+1$. In the  opposite regime, $\lambda_B \gg \lambda_D$, the constraint becomes
\begin{equation}
\label{lim_high}
B < 3.2 \times 10^{-8} G \frac{1}{\sqrt{K_2}} \left(  \frac{\lambda_B}{400 ~\mbox{pc}}\right)~,
\end{equation}
where $K_2$ is another constant of order unity. The strongest limit on the field strength is obtained for the fields with correlation length $\lambda_B\sim\lambda_D$. For such  fields no convenient analytical approximation can be found and instead a numerical integration of the expression given in 
Ref.~\citep{Jedamzik00} has to be performed. In Fig. \ref{f5:CMB} we show the bound on $B,\lambda_B$ implied by the analysis of distortions of the CMB spectrum as extrapolations of the analytical approximations given by Eq. (\ref{lim_low}), (\ref{lim_high}) for the entire ranges $\lambda_B<\lambda_D$ and $\lambda_B>\lambda_D$. Note that the dependence of the limits on $B,\lambda_B$  on the power-law index $n_s$  practically disappears in the case  $\lambda_B\gg \lambda_D$ (only weak dependence remains in the constant $K_2$). For $\lambda<\lambda_D$, we consider the softest power spectrum slope $n_s=2$ corresponding to the fields causally produced at the phase transitions in the Early Universe (see Section \ref{s:gen}).
 
 Apart from producing a non-zero chemical potential, transfer of the magnetic field energy to electrons/ positrons can result in non-zero  Compton parameter $y$. Taking into account restrictions $y<1.5 \times 10^{-5}$ from COBE FIRAS, one finds a limit  $B<3  \times 10^{-8}$ G at $\lambda \sim 0.3-0.6 $ Mpc~ \citep{Jedamzik00}. Note, that limits Eq.~(\ref{lim_low}) and  Eq.~(\ref{lim_high}) constrain magnetic fields created at $z>z_\mu=2.5\times 10^6$, while the limit following from restrictions on $y$ applies for fields created before  $z>z_{el}\simeq 2 \times 10^4$ where elastic Thompson scattering drops out of thermal equilibrium, i.e. $t_{el}>t$, see~\cite{Durrer:2008aa}.

%%%%%%%%%%%%%%%%%%%%%%%%%%%%%%%%%%%%%%%%%%%%%%%%%%%%%%
\subsubsection{Limits from CMB polarization}
\label{s:CMB_polarization}
%%%%%%%%%%%%%%%%%%%%%%%%%%%%%%%%%%%%%%%%%%%%%%%%%%%%%%

Since Thompson scattering is anisotropic, temperature anisotropies in the CMB (more precisely the quadrupole) induce a small net polarization of CMB photons. Depending on the polarization pattern this is called E-polarization (gradient field on the CMB sky) or B-polarization (rotational pattern on the CMB sky). For details see~\cite{Durrer:2008aa}. Scalar perturbations from inflation only generate E-polarization. Magnetic fields lead to
strong vector perturbations which generate significant B-polarization~\citep{Seshadri:2000ky,Lewis:2004ef}.

Magnetic fields which the CMB photons encounter on their way from the last scattering surface into our antennas Faraday rotate this polarization pattern and thereby generate also B-polarization. Due to its dependence on the wavelength given in Eq.~(\ref{e5:Faraday}), this effect can be easily distinguished from achromatic gravitational effects, e.g. primordial gravitational waves which also generate B-polarization.  Constraints on $B,\lambda_B$ stemming from non-observation of this effect were first discussed by~\cite{Kosowsky96} and subsequently updated  using the 5-years data of WMAP by~\cite{Hinshaw09}. The limits coming from non-observation of Faraday rotation in CMB signal are, at present, weaker than the limits imposed by the rotation measures of distant blazars or limits from  the CMB angular power spectrum~\citep{Kahniashvili09}. 

%%%%%%%%%%%%%%%%%%%%%%%%%%%%%%%%%%%%%%%%%%%%%%%%%%%%%%
\subsubsection{Limits from magnetic field effects on LSS}
\label{s:SZ}
%%%%%%%%%%%%%%%%%%%%%%%%%%%%%%%%%%%%%%%%%%%%%%%%%%%%%%

Cosmological magnetic fields present at the onset of  LSS formation may affect the properties and the statistics of the matter perturbations. In this way, strong enough field might affect the abundance  galaxy clusters and, therefore, modify the Sunyaev-Zeldovich effect on CMB angular power spectrum~\citep{Tashiro11}. Using the data of the South Pole Telescope, \citet{Shaw12} derived a limit $B\lesssim 4$~nG at $\lambda_B=1$~Mpc scale. In a similar way, the presence of strong magnetic fields can distort the statistics of column densities of the Ly$\alpha$ clouds. Non-observation of this effect also constrains the field strength. Combining the CMB angular power spectrum, Sunyaev-Zeldovich effect data, and Ly$\alpha$ data, a joind constraint $B\lesssim 1$~nG at Mpc distance scale has be derived by \cite{Shaw12}. 

Strong magnetic fields provide an additional contribution to the pressure and thereby enhance the Jeans mass for gravitational collapse. The influence of magnetic fields can be strong enough to partially suppress the collapse of small structures. This leads to the decrease of the amount of ionizing radiation produced by the star formation in small galaxies. On the other hand, the Lorentz force induces additional perturbations in the baryons which are transferred to dark matter and enhance collapse of small scale structure above the magnetic Jeans mass \citep{Sethi05,Kim:1994zh,Subramanian98}. This can also influence the onset of reionization of the Universe, see \cite{Sethi05} and \cite{Tashiro:2005ua}. Based on this idea,  \citet{Schleicher11} derive a scale-independent upper limit $B\lesssim 3$~nG on cosmological magnetic fields present at the redshift $z\gsim 7$. This limit is comparable to the limits imposed by the CMB data. 

The effects of magnetic fields on structure formation and the  thermal and ionization evolution of cosmic hydrogen also leave traces in the 21cm signal which have been investigated in \cite{Sethi:2009dd,Schleicher:2008gk}. It is found that
future radio surveys like SKA could probe magnetic fields with correlation scales of 0.1Mpc to several Mpc down to amplitudes of $10^{-10}$Gauss.

The modification of LSS formation in the presence on magnetic fields also affects gravitational weal lensing, as discussed in \cite{Pandey:2012kk}. Again constraints of a few nGauss are obtained for a spectral index $n_s \sim -2.9$ of the magnetic field spectrum. 

\subsubsection{Helical fields and the CMB}
If magnetic fields are helical, their induced gravitational fields are so as well which leads to non-vanishing cross correlation spectra of temperature and B-polarization and wall as E- and B-polarization \citep{Caprini:2003vc}. The non-detection of such cross-correlations can be used to place additional limits on helical magnetic fields. However, they are also of the order on nG as the other limits discussed above.

%%%%%%%%%%%%%%%%%%%%%%%%%%%%%%%%%%%%%%%%%%%%%%%%%%%%%%
\subsection{Limits from gravitational waves}
\label{s:GWs}
%%%%%%%%%%%%%%%%%%%%%%%%%%%%%%%%%%%%%%%%%%%%%%%%%%%%%%
The energy momentum of magnetic fields always gives raise to anisotropic stresses. Their transverse traceless component are sources gravitational waves. More precisely
\be\label{e5:gwevo}
\ddot h_{ij} + 2\HH \dot h_{ij} + k^2 h_{ij} = 16\pi Ga^{-2}\Pi^{(B)}_{ij} \,.
\ee
Here $h_{ij}$ is the gravitational wave amplitude and $\Pi^{(B)}_{ij}$ is the transverse traceless part of the anisotropic stress due to the magnetic field. On a scale $\la=2\pi/k\lsim \HH^{-1}$ within a Hubble time, gravitational waves with energy density~\citep{Caprini:2001nb,Caprini:2006jb}
\be\label{e5:gwOm}
\frac{\Om_{GW}(t,\la)}{\Om_{\rm rad}} \simeq  (\la\HH(t))^2 \left(\frac{\Om_{B}(t,\la)}{\Om_{\rm rad} }\right)^2
\ee
are generated. The dominant contribution to the gravitational waves on scale $\la$ are generated when this scale crosses the horizon so that we obtain
\be\label{e5:gwOm2}
\frac{\Om_{GW}}{\Om_{\rm rad}}(\la) \simeq   \left(\frac{\Om_{B}(t\simeq\la)}{\Om_{\rm rad} }\right)^2
\ee
Once generated, these gravitational waves propagate freely and do not interact anymore. Even when the magnetic field on these scales is damped away by fluid viscosity, the gravitational waves which it has produced remain. They contribute an additional relativistic component to the expansion of the Universe. The strongest limits on this come from combining nucleosynthesis constraints and CMB observations~\citep{Steigman:2008eb,Hinshaw:2012fq}, requiring $\Om_{GW}/\Om_{\rm rad} \lsim 0.1$ on all scales. As $\Om_{B}(t)/\Om_{\rm rad}$ decays during free turbulent evolution, the strongest constraints are obtained on scales of the order of the horizon scale at formation, $\la\sim t_*$ but
also there the above constraint just requires $\Om_{B}(t)/\Om_{\rm rad}\lsim 0.3$.

Of course for a specified process of magnetic field generation with a fixed spectrum, this can imply much 
stronger limits on much larger scales, for an overview see~\cite{Caprini:2009pr}. For example, if the formation is causal and evolution proceeds via incompressible MHD turbulence, the magnetic field energy on a given scale $k$ at late times is always smaller that its unprocessed value which behaves like
\be
\frac{d\rho_{B}(k)}{d\log k} \simeq k^3P_B(k) \equiv B^2(k) \propto k^5 \,,
\ee
so that on the scale $\la=2\pi/k$ we obtain the limit
\be\label{e5:lim1}
\Om_{\rm rad}^{-1}\frac{d\Om_{B}(k)}{d\log k} \lsim  (k/k_*)^5 \simeq (k/100\HH_*)^5\,.
\ee
Here we have assumed an initial correlation scale of 1\% of the initial Hubble scale. This is probably a reasonable  value for first order phase transitions which proceed vial bubble nucleation (see \citealt{Huber:2008hg}). 
For an arbitrary initial correlation scale $k_*$,   we have 
\be
B(k) \simeq B(k_*)\left(\frac{k}{k_*}\right)^{5/2} \,.
\ee
This provides a very strong limit on large scales. We use
\be
\frac{\Om_{B}}{\Om_{\rm rad} } \simeq \Om_{\rm rad}^{-1}P_B(k_*)k_*^3 = 5.6\left(\frac{B}{10^{-7}G}\right)^2 <0.3 \,.
\ee
If this limit is satisfied at the initial correlations scale $\la_*$
for magnetic fields generated at the electroweak phase transition, $H_* \simeq 10^{-2}$mHz$ \simeq (10^{-9}$Mpc$)^{-1}$, on the scale $\la\simeq 100$kpc a magnetic field of the order of
\be
B(\la\sim 100{\rm kpc}) \lsim 10^{-31}G \,. 
\ee

If we allow for compressible MHD evolution, with a scalar velocity component of similar amplitude as the vorticity, this constraint is significantly relaxed. Now the
correlation scale increases like $\bar\la=\la_*(t/t_*)^{2/5}$, and the magnetic power spectrum
remains roughly constant at the peak, $P_B(\bar\la)=P_B(\la_*)$, see Fig.~\ref{f4:Bevol}. Setting the endpoint of the free turbulent decay at recombination, 
$T_{\rm fin} \sim 0.3$eV, we obtain a correlation scale which is about 5 orders of magnitude larger,
$$
\bar\la_{\rm fin} \simeq (100{\rm GeV}/0.3{\rm eV})^{2/5}10^2{\rm sec} \simeq 1{\rm pc}. 
$$ 
The magnetic field power spectrum on this scale has the same amplitude as $P_B(\la_*)$ at the electroweak phase transition. At larger scales it decays like $\la^{-2}$. Inserting this in the 
condition~(\ref{e5:lim1}) yields
\be
B(\la\sim 100{\rm kpc}) \simeq \left[P_B(\bar\la)(\bar\la/\la)^2\la^{-3}\right]^{1/2}=
B(\la_*)(\la_*/\la)^{5/2}(\bar\la/\la_*) \,.
\ee
Hence the limit is relaxed by about 5 orders of magnitude with respect to the one coming from
incompressible turbulence. If the magnetic field is maximally helical, the limit is relaxed even more. These limits can also been obtained from the corresponding endpoints in Fig.~\ref{f4:evolP} by taking into account that on scales larger than the correlation scale, $\la>\la_B=\bar\la$ the magnetic field strength of causally generated fields decays like $(\la_B/\la)^{5/2}$.

%%%%%%%%%%%%%%%%%%%%%%%%%%%%%%%%%%%%%%%%%%%%%%%%%%%%%%
\subsection{Potential limits from ultra-high-energy cosmic ray observations}
\label{s:UHECR}
%%%%%%%%%%%%%%%%%%%%%%%%%%%%%%%%%%%%%%%%%%%%%%%%%%%%%%

Magnetic fields in IGM can be probed by measuring  their effect on trajectories of Ultra-High-Energy Cosmic Rays (UHECR), which are charged particles with energies  $E_{\rm UHECR}\sim 10^{20}$~eV penetrating in the Earth atmosphere. Such particles come to the Earth from yet unknown sources which are most probably situated outside the Milk Way. This means that UHECR traveling from their source to the Earth cross the IGM. A magnetic field in the IGM deflects UHECR trajectories from their straight line by an angle
\begin{eqnarray}
\label{UHECR_reg}
\theta_{\rm IGMF}= \frac{ZeB_\bot D}{E_{\rm UHECR}}\simeq 2.6^\circ Z\left(\frac{E_{\rm UHECR}}{10^{20}\mbox{ eV}}\right)^{-1}\left(\frac{B_\bot}{10^{-10}\mbox{ G}}\right)\left(\frac{D}{50\mbox{ Mpc}}\right),
\end{eqnarray}
where $Ze$ is the electric charge of UHECR particle, $D$ is the distance to the UHECR source and $B_\bot$ is the strength of IGMF component orthogonal to the line of sight.  In the above equation the IGMF is assumed to be coherent over the length scales larger than the distance $D$.  If the coherence length $\lambda_B$ is much shorter than $D$, UHECR experiences a sequence of deflections in different random directions during the passage through each distance interval $\lambda_B$, so that the overall deflection angle accumulated over the path $D$ is given by  
\begin{eqnarray}
\label{UHECR_turb}
\theta_{\rm IGMF}&\simeq& \frac{ZeB_\bot \sqrt{D \lambda_B}}{E_{\rm UHECR}}\simeq 0.4^\circ Z\left(\frac{E_{\rm UHECR}}{10^{20}\mbox{ eV}}\right)^{-1}\\ && \left(\frac{B_\bot}{10^{-10}\mbox{ G}}\right)\left(\frac{D}{50\mbox{ Mpc}}\right)^{1/2}\left(\frac{\lambda_B}{1\mbox{ Mpc}}\right)^{1/2}\,.
\nonumber
\end{eqnarray}

IGMF affect the arrival directions of UHECR particles by displacing them from the direction towards the source. This opens a possibility of the measurement of IGMF with the strength in the range $B\sim 10^{-9}$~G using UHECR observations \citep{Lee95,Lemoine97,Sigl98}. Indeed, if the sources of UHECR would be known, one would be able to measure the angular distribution of UHECR events around the source positions. In fact, the mere observation of isolated sources of UHECR on the sky would indicate that $\theta_{\rm IGMF}\ll 1$, so that IGMF can not be stronger than $\sim 10^{-9}$~G, provided that the distance to the UHECR sources is in the $D\sim 50$~Mpc range.

 Recent attempts to model the deflection of UHECR by the magnetic field of the intervening large scale structure, such as  galaxy clusters and/or filaments by two groups (see~\cite{Sigl04} and ~\cite{Dolag05})led to contradictory results, which reflect uncertainties of the structure of magnetic fields inside and around clusters and filaments.  In addition, if there is a single nearby UHECR source, the host galaxy or galaxy cluster of the source can span several degrees on the sky. Significant deflections of UHECR by magnetic fields in the host galaxy or galaxy cluster can then produce extensions of the UHECR emitting region of 1 up to10 degrees  \citep{Dolag09}. 
 
There are several obstacles for the measurement of IGMF with this method. The first is the uncertainty of the origin of UHECR and of their composition. The second is the presence of additional deflection of UHECR trajectories by the Galactic magnetic field.  Below we summarize the status of these uncertainties.

%%%%%%%%%%%%%%%%%%%%%%%%%%%%%%%%%%%%%%%%%%%%%%%%%%%%%%
\subsubsection{Sources and the composition of UHECR}
\label{s:UHECR_sources}
%%%%%%%%%%%%%%%%%%%%%%%%%%%%%%%%%%%%%%%%%%%%%%%%%%%%%%

The main problem for the localization and identification of the UHECR sources is the low statistics of the signal in the $10^{20}$~eV energy band. Starting from the earliest  days of UHECR observations and up to the present only about $\sim 10^2$ events have been collected by different experiments: Pierre Auger Observatory (PAO) \citep{Pao07_anisotropy,Pao08_GZK,Pao10}, HiRes \citep{HiRes08_GZK,HiRes10_composition,HiRes10_anisotropy}, AGASA \citep{AGASA_anisotropy}, Telescope Array (TA) \citep{TA_anisotropy,TA_GZK}.

\citet{HiRes10_composition} found evidence for a proton-dominated cosmic ray flux at the energies above $10^{18}$~eV, based on the analysis of the average depth of the maxima of Extensive Air Showers (EAS) of high-energy particles, produced by the UHECR penetrating in the Earth atmosphere. If the proton-dominated composition persists until the $10^{20}$~eV energy band \citep{HiRes_composition2}, a suppression of the UHECR flux due to the interactions of high-energy protons with the CMB photons, known as the Greisen-Zatsepin-Kuzmin (GZK) cutoff  \citep{Greisen66,Zatsepin66} is expected beyond this energy. This suppression was first observed in the UHECR flux \citep{HiRes08_GZK} in the HiRes data and later confirmed with better statistics by the PAO \citep{Pao08_GZK} and TA \citep{TA_GZK} experiments. 

An indication for clustering of UHECR arrival directions on small angular scales was first found in AGASA data \citep{AGASA_anisotropy}. A stronger small-scale anisotropy signal was subsequently found in Peirre Auger observatory data \citep{Pao07_anisotropy,Pao08_AGN}. The arrival directions of UHECR events were found to correlate with the sky positions of nearby Active Galactic Nuclei (AGN) situated within a distance $D<75$~Mpc. The correlation was found at angular scale $\theta\simeq 1^\circ$ equal to the angular resolution of the detector. In the initial analysis \citep{Pao07_anisotropy} a significant fraction of UHECR events was found to correlate with the AGN positions (see, however \citet{Gorbunov08}). However, the number of sources contributing to the signal and, as a consequence, the significance of the UHECR -- AGN correlation, has decreased with the accumulation of event statistics \citep{Pao10}.  No significant correlation with AGN was found in the analysis of the HiRes data with comparable event statistics \citep{HiRes_anisotropy} and in the Telescope Array data \citep{TA_anisotropy}.

Detection of small angular scale correlations of UHECR arrival directions with particular point sources on the sky would immediately imply that the UHECR trajectories are not strongly deflected by the IGMF. Thus, if the PAO result on the correlation with AGN holds, it immediately implies that the UHECR are protons ($Z=1$ in Eqs. \ref{UHECR_reg}, \ref{UHECR_turb}) and that a  IGMF can not be stronger than $\sim 10^{-10}$~G for $\lambda_B\sim D$ \citep{deAngelis07}. 

However,  the interpretation that most of the UHECR are protons appears to be in conflict with a recent finding of heavy composition of the highest energy events. \cite{Pao_composition} have concluded from the decrease of fluctuations of the depths of EAS maxima at energies above $10^{19}$~eV that these UHECR events are composed of heavier nuclei. A possible heavy composition of UHECR flux has to be verified with better statistics and using independent analyses from the HiRes / TA detector, which does not confirm this result at the moment \citep{HiRes_composition2}. 

The absence of firm source identifications and the ambiguities in the composition of UHECR at the highest energy end render the use of the UHECR techniques for the measurement of the IGMF very uncertain at the moment. A dramatic increase of the UHECR event statistics is required to remove the above mentioned uncertainties. This requires a dramatic increase (by at least an order of magnitude) of the collecting area of the UHECR detector. Existing ground-based detectors, like PAO and TA achieve collecting areas in the range of $(1-3)\times 10^3$~km$^2$. A further increase of the area is challenging because this would require homogeneous coverage of large surfaces with high-energy particle detectors, over the areas in the range comparable to the size of a small country like Switzerland. A particular UHECR detection technique, based on the measurement of ultraviolet fluorescence emission from the EAS, allows for a qualitatively different space-based approach: deployment of a single wide field-of-view downward looking telescope in space.  This approach allows to monitor a large part of the Earth surface and potentially reach a collecting area comparable to the Earth surface. A first attempt along this direction will be the planned next-generation space-based UHECR detector JEM-EUSO {\tt http://jemeuso.riken.jp}  \citep{jemeuso}. It will have a collecting area in the range of $10^5$~km$^2$ and an all-sky exposure capability. JEM-EUSO will collect some $\sim 10^3$~UHECR events in a $\sim 5$~yr exposure on board of the International Space Station. Possible usefulness of UHECR observations for constraining IGMF has to be re-assessed when the results of the all-sky small- and large-scale anisotropy analysis of JEM-EUSO UHECR data will be available.

%%%%%%%%%%%%%%%%%%%%%%%%%%%%%%%%%%%%%%%%%%%%%%%%%%%%%%
\subsubsection{UHECR deflections in the Galactic magnetic field}
\label{s:GMF}
%%%%%%%%%%%%%%%%%%%%%%%%%%%%%%%%%%%%%%%%%%%%%%%%%%%%%%

Apart from uncertainties of the sources and composition of UHECR, another major obstacle formeasuring IGMF with UHECR  is related to the uncertainties of the structure of the Galactic magnetic field (see also Section \ref{s:faraday}). The magnetic field of the Milky Way is conventionally modeled as the sum of a regular and a turbulent component of the field in the disk and in the halo of the Galaxy. Therefore, the deflections of UHECR by the Galactic magnetic field can be decomposed onto  four terms:
\begin{equation}
\label{UHECR_gal}
\theta_{Gal} =  \theta_{\rm Disk}^{\rm regular} + \theta_{\rm Disk}^{\rm turbulent} +\theta_{\rm Halo}^{\rm regular}+\theta_{\rm Halo}^{\rm turbulent}~.
\end{equation}
Deflection by the regular and turbulent components of Galactic disk and halo
% $\theta^{\rm Disk}_{\rm regular},\ \theta^{\rm Halo}_{\rm regular}$ and $\theta^{\rm Disk}_{\rm turbulent},\ \theta^{\rm Halo}_{\rm turbulent}$ , 
can be estimated by substituting the typical disk/halo size at the place of $D$ and the typical disk/halo field strength and correlation lengths at the place of $B$ and $\lambda_B$ in Eqs. (\ref{UHECR_reg}), (\ref{UHECR_turb}).  Deflections of UHECR by the regular field in the disk $\theta_{\rm Disk}^{\rm regular}$ were studied in many theoretical models  starting with a paper by \cite{Stanev97}. Typical values of parameters entering the analog of Eq.~(\ref {UHECR_reg}) imply  $D_{\rm Disk\; reg}\simeq 2$ kpc  (for sources located far from the Galactic Plane) and  $B_{\rm Disk\; reg}\simeq 2\ \mu$G, which give for   Eq.~(\ref{UHECR_reg}) $\theta_{\rm Disk}^{\rm regular} \simeq 4^\circ$. Turbulent fields are typically assumed to have a coherence  scale of $\sim 50$ pc and $B\simeq 4\ \mu$G, which, for  the same scale $D_{\rm Disk\; turn}\sim 2$ kpc gives  $\theta_{\rm Disk}^{\rm turbulent} \simeq 0.5^\circ$.   Contributions of the Halo fields are less certain, but result in deflections which are at least of the same order, see the recent discussions of all components by \cite{Sun08,Han09,Jansson09,Farrar12c}. 

Although the order of magnitude of the deflection angle of UHECR by the Galactic magnetic field can be readily estimated, uncertainties in the measurements of Galactic magnetic field and discrepancies between the existing measurements and existing theoretical models \citep{Sun08,Jansson09,Farrar12,Farrar12a} do not allow to predict the deflection angle and direction of deflection for particular lines of sight (toward UHECR sources).  This means that, most probably, the details of the structure of the Galactic magnetic field along the line of sight toward UHECR sources will have to be deduced from the UHECR data itself, rather than just taken into account in the UHECR data analysis~ \citep{Sigl98,Giacinti09b}. This, obviously, will introduce large uncertainties into the derivation of the properties of IGMF from the UHECR data.

Deflections of UHECR by the Galactic magnetic field are much stronger if UHECR particles are heavy nuclei, rather than protons, as indicated by the recent PAO results \citep{Pao_composition}. In this case the UHECR are distributed around their source on angular scales of $10^\circ-100^\circ$ \citep{Giacinti11}. Inhomogeneities of the exposure across the fields-of-view of  ground-based detectors is a significant obstacle for the recognition of such wide-angle patterns of UHECR distribution around their sources. Only all-sky detectors, like JEM-EUSO would be potentially able to localize the UHECR sources in this case.  It is clear that more sophisticated procedure for the identification of UHECR sources for heavy UHECR will automatically imply lower sensitivity of UHECR for the measurement of IGMF.

%%%%%%%%%%%%%%%%%%%%%%%%%%%%%%%%%%%%%%%%%%%%%%%%%%%%%%
\subsection{Constraints from gamma-ray observations}
\label{s:gamma}
%%%%%%%%%%%%%%%%%%%%%%%%%%%%%%%%%%%%%%%%%%%%%%%%%%%%%%

An alternative way to probe the weakest magnetic fields in the IGM is using high-energy (HE, photon energies in the 0.1-100~GeV range) and very-high-energy (VHE, photon energies in the 0.1-10~TeV range) $\gamma$-ray observations.  $\gamma$-rays propagating though the IGM occasionally interact with the abundant low energy photons and produce electron-positron pair. Electrons and positrons, being charged particles, are affected by the magnetic field. Therefore, secondary $\gamma$-rays produced by electrons and positrons interacting with CMB photons  carry information about the properties of IGMF. 

%%%%%%%%%%%%%%%%%%%%%%%%%%%%%%%%%%%%%%%%%%%%%%%%%%%%%%
\subsubsection{Absorption of  $\gamma$-rays in the IGM}
\label{s:tev_observations}
%%%%%%%%%%%%%%%%%%%%%%%%%%%%%%%%%%%%%%%%%%%%%%%%%%%%%%

To produce $e^+e^-$ pairs,  the energy of the primary $\gamma$-ray must be high-enough so that the center-of-mass energy in the photon-photon collision exceeds twice the electron mass. $\gamma$-rays with energy $E_\gamma$ propagating through a background of soft photons with energy $\epsilon$ can produce pairs if their energy is higher than the threshold \citep{Gould66}
\be
\label{e5:threshold}
E_\gamma\ge\frac{m_e^2}{\epsilon}\simeq 250\left(\frac{\epsilon}{1\mbox{ eV}}\right)^{-1}
\mbox{ GeV.}
\ee

\begin{figure}[h!]
\begin{center}
\includegraphics[width=\linewidth]{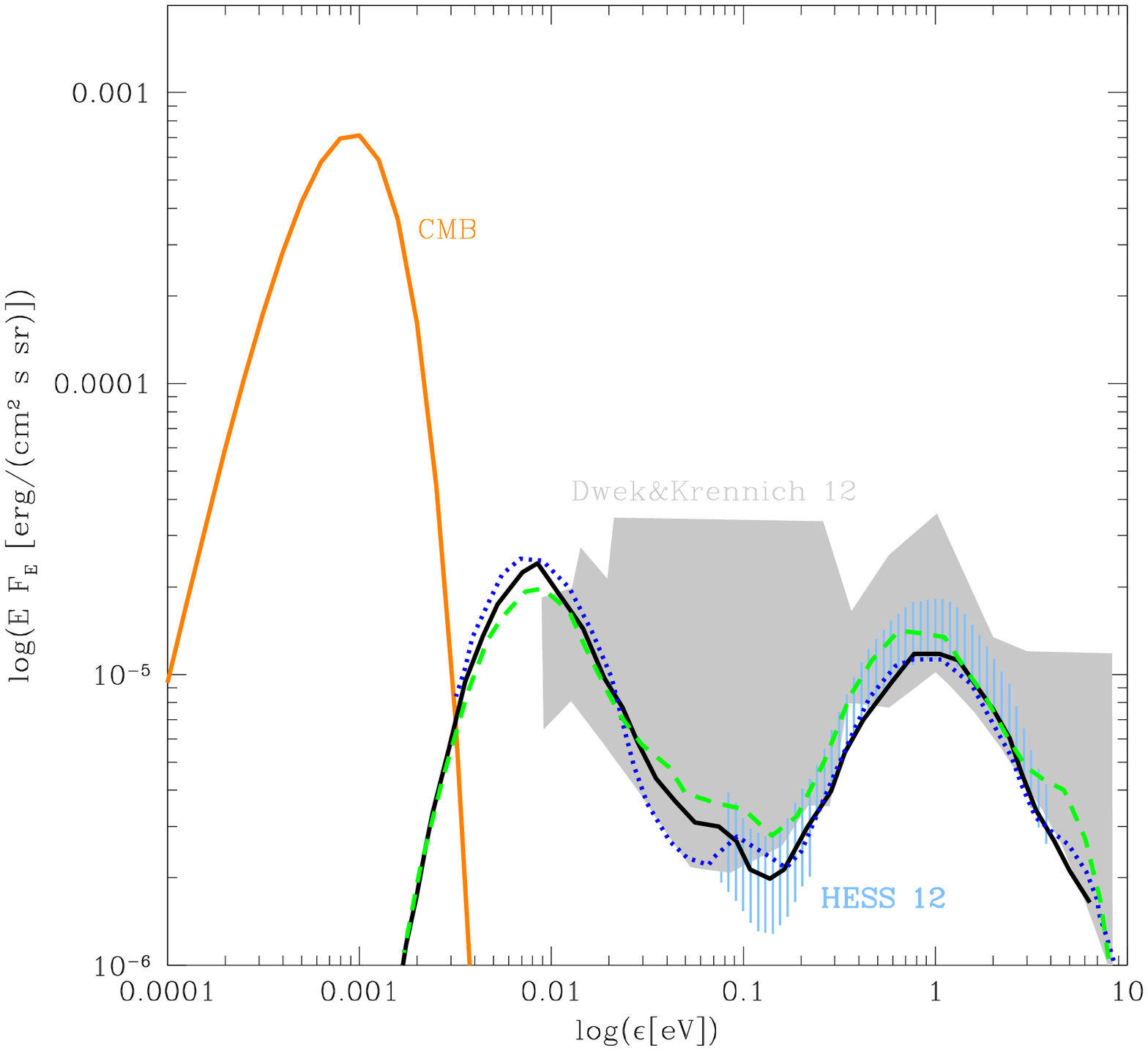}
\end{center}
\caption{Measurements, limits and model calculations of the spectral energy distribution of the EBL. The black solid curve shows the model calculation by \citet{Franceschini08}. The lue dotted curve shows the calculation of \citet{Dominguez11}, and the green dashed line shows the model by\citet{Kneiske08}. The grey shaded region  is the envelope of direct measurements summarized by \citet{Dwek12}. The blue shaded region is the measurement derived from $\gamma$-ray observations by \citet{HESS12_EBL}. The orange curve shows the CMB spectrum.}
\label{fig:EBL}
\end{figure}

Soft photons   with energies in the $0.1-10$~eV range are abundant in the Universe, because they are produced by  star formation in galaxies. The homogeneous and isotropic soft photon background in this energy range, known as Extragalactic Background Light (EBL) \citep{Mandau00,Dwek01,Franceschini08,Dwek12}, is accumulated throughout the history of the Universe, starting from the onset of star formation at  redshift $z\sim 10$. The spectral energy density of EBL has a characteristic two-bump shape with a near-infrared bump at $\sim 1$~eV being due to the direct starlight emission and the far-infrared bump at the energy $\sim 10^{-2}$~eV being produced by scattering of starlight with dust, see Fig. \ref{fig:EBL}. The energy density of the EBL is about $\rho_{\rm EBL}\sim 10^{-2.5}$~eV/cm$^3$ \citep{HESS12_EBL}, which is a factor of $\simeq 10^2$ lower than the energy density of the CMB. 

%%%%%%%%%%%%%%%%%%%%%%%%%%%%%%%%%%%%%%%%%%%%%%%%%%%%%%
%%%%%%%%%%%%%%%%%%%%%%%%%%%%%%%%%%%%%%%%%%%%%%%%%%%%%%

The precise value of $\rho_{\rm EBL}$ was somewhat uncertain (by a factor of $\simeq 2$) until recently, because it is not possible to measure it in the optical and infrared bands. The main obstacle for such a measurement is the much stronger foreground  of zodiacal light from the Solar system \citep{Puget96,EBL_DIRBE,Wright00,Gorjian00,Berstein02}. A proper subtraction of the zodiacal light foreground requires a precise knowledge of its spectral characteristics and angular distribution  in the sky. Uncertainties in these parameters introduce an uncertainty in estimates of the spectral characteristics of EBL based on the measurements of diffuse emission in the optical and infrared bands. This uncertainty  has affected early estimates of the EBL \citep{EBL_DIRBE}, which have turned out to be too high.

Alternatively, the amount of soft photons in the IGM can be estimated from source counts in deep observations by  infrared and optical telescopes. This approach provides lower bounds on the EBL density (since the contribution from undetected sources with fluxes below the telescope's sensitivity is not taken into account). Along these lines, tight lower bounds on the EBL density at different wavelengths were derived from the observations with telescopes in space and on the ground \citep{Hubble_EBL,EBL_IRTS,Spitzer_EBL1,Spitzer_EBL,Spitzer_EBL3,Herschel_EBL,Herschel_EBL2,EBL_Subaru}.
\citet{Dwek12} have summarized the constraints on the EBL spectrum obtained from the direct source count observations. These constraints are shown by the grey-shaded region in Fig. \ref{fig:EBL}.

\begin{figure}
\begin{center}
\includegraphics[width=\linewidth]{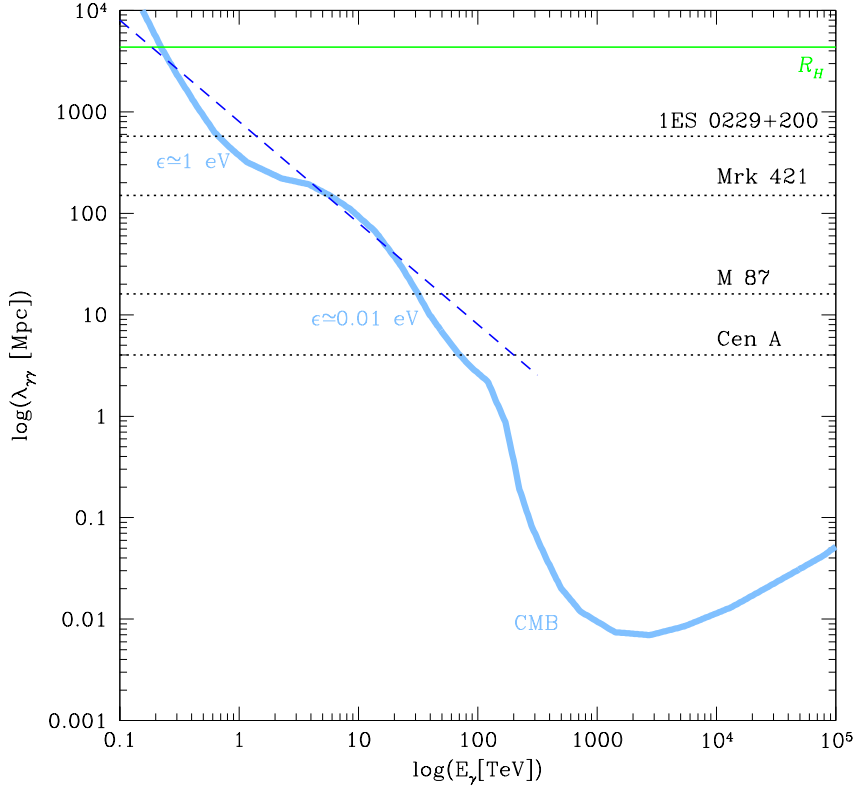}
\end{center}
\caption{The mean free path of VHE $\gamma$-rays as a function of energy. The horizontal green  line shows the Hubble radius. Dotted lines mark distances to some VHE $\gamma$-ray sources. The Inclined dashed blue line shows the estimate of Eq. \ref{eq:mfp}). Labels along the curve mark the energies of soft photons with which $\gamma$-rays interact.}
\label{fig:MFP}
\end{figure}

Interactions of Very-High-Energy (VHE) $\gamma$-rays (photon energies above 100~GeV) with the EBL photons can be used to constrain or measure the EBL density and spectrum. These interactions lead to an energy-dependent suppression of the $\gamma$-ray flux from extragalactic sources in the VHE band. Indeed, the mean free path of  $\gamma$-rays with energy above the threshold (\ref{e5:threshold}) is given by
\begin{eqnarray}
\label{eq:mfp}
\lambda_{\gamma\gamma}(E_\gamma)&=&\frac{1}{\int_{m_e^2/E_\gamma}^\infty\sigma_{\gamma\gamma}(E_\gamma,\epsilon)n_{\rm EBL}(\epsilon)d\epsilon}
\simeq \frac{1}{\sigma_{\gamma\gamma}n_{\rm EBL}}  \nonumber\\
&\sim& 0.8 \left[\frac{E_\gamma}{1\mbox{ TeV}}\right]^{-1}\left[\frac{\rho_{\rm EBL}}{10^{-2.5} \mbox{ eV/cm}^3}\right]^{-1}\mbox{ Gpc}\,.
\end{eqnarray}
Here $n_{\rm EBL}(\epsilon)$ is the number density of EBL photons, $n_{\rm EBL}\simeq \rho_{\rm EBL}/\epsilon$,   $\sigma_{\gamma\gamma}(\epsilon)$ is the pair production cross-section which reaches it maximum  $\sigma_{\gamma\gamma}\simeq 10^{-25}$~cm$^2$ at aout $4$ times the threshold energy \citep{Aharonian_book} and we have to insert the value of $\rho_{\rm EBL}$ at this maximum, $\ep_{\max} \simeq 4m_e^2/E_\ga$. The mean free path of $\gamma$-rays becomes shorter or comparable to  typical distances $D$ to extragalactic $\gamma$-ray sources in the TeV band, see Fig. \ref{fig:MFP}, so that the source fluxes are suppressed by a factor $\exp(-D/\lambda_{\gamma\gamma})$. If the intrinsic source spectrum is known, comparison of the observed and intrinsic spectrum of the source provides a measure of the suppression factor and, via Eq. (\ref{eq:mfp}), of the density of the EBL. 

The main class of extragalactic sources of VHE $\gamma$-rays are blazars. These are radio-loud AGN (active galactic nuclei) emitting jets closely aligned long the line of sight \citep{Aharonian_book}. Recent rapid development of the ground-based VHE $\gamma$-ray astronomy has resulted in the discovery of about $10^2$ extragalactic VHE $\gamma$-ray sources \citep{Aharonian08_review}. In a significant fraction of these sources, a high-energy suppression of the flux due to the interactions with EBL photons is clearly observed. 

In principle, if the quality of these data is sufficiently good, one can derive both, the intrinsic shape of the $\gamma$-ray spectrum and the energy-dependent suppression factor, directly from the $\gamma$-ray data. 
For this one has to use the fact that  the suppression factor $\exp(-D/\lambda_{\gamma\gamma})$ has a characteristic "two bump" shape (see Fig. \ref{fig:MFP}), related to the two components in the EBL spectrum. Fig. \ref{fig:MFP} shows the energy dependence of $\lambda_{\gamma\gamma}$ determined with Eq.~(\ref{eq:mfp}) from the EBL spectrum by \citet{Kneiske08}, which is consistent with all observational constraints (see Fig. \ref{fig:EBL}). The two enhanced absorption energy intervals at $E_\gamma\sim 1$~TeV (absorption on the near-infrared EBL)and $E_\gamma\sim 100$~TeV (absorption on far-infrared EBL)  are clearly seen as deviations  of the estimate (\ref{eq:mfp}) from a straight line. The shape of the enhancement of absorption around 1~TeV is determined by the shape of the EBL spectrum in the near-infrared band. Since $\lambda_{\gamma\gamma}^{-1}$ is proportional to the logarithm of the suppression factor $\exp(-D/\lambda_{\gamma\gamma})$. Fig. \ref{fig:MFP} shows in fact the shape of the EBL suppression factor, which can, in principle, be deduced from the $\gamma$-ray data. 

This approach has been followed recently using  HESS telescope date \citep{HESS12_EBL}. The spectrum of the EBL derived in this way from the $\gamma$-ray data is shown in Fig. \ref{fig:EBL} by the light-blue hatched band.  A similar analysis using data of the Fermi telescope at lower energies has allowed determine the level of the EBL at higher redshift $z\sim 1$ \citep{Fermi_EBL}.  Earlier calculations, based on lower quality data, had to rely on assumptions about a range of possible shapes for the intrinsic blazar spectra in the TeV energy band. This had then permitted to derive upper limits on the EBL density, rather than the full measurement of the EBL spectrum \citep{Biller95,Funk98,HESS06_0229,HESS_0229_2,HESS_0347,MAGIC_3C279,Mazin07,Finke09,HESS12_0414} .  

The direct measurements of the EBL and the determination using $\gamma$-ray data are supplemented by models deriving the EBL spectrum and its evolution with redshift from  models of cosmological evolution of optical and infrared luminosities and and of the luminosity function of galaxies \citep{Stecker06,Franceschini08,Kneiske08,Dominguez11,Finke10,Gilmore12}. These models have evolved with time, starting from early models which attempted to explain the relatively high level of the EBL derived from DIRBE data to the currently favored models which predict a EBL flux at the level close to that derived from the source count statistics, see Fig. \ref{fig:EBL} where example model spectra \citep{Franceschini08,Kneiske08,Dominguez11} are shown.

%%%%%%%%%%%%%%%%%%%%%%%%%%%%%%%%%%%%%%%%%%%%%%%%%%%%%%
%%%%%%%%%%%%%%%%%%%%%%%%%%%%%%%%%%%%%%%%%%%%%%%%%%%%%%

%%%%%%%%%%%%%%%%%%%%%%%%%%%%%%%%%%%%%%%%%%%%%%%%%%%%%%
\subsubsection{Secondary $\gamma$-rays from $e^+e^-$ pairs in the IGM. }
\label{s:cascade}
%%%%%%%%%%%%%%%%%%%%%%%%%%%%%%%%%%%%%%%%%%%%%%%%%%%%%%

The mean free path of VHE $\gamma$-rays (\ref{eq:mfp}) decreases monotonically with the energy $E_\gamma$. This is because the $\gamma$-rays of higher energies can interact with the more abundant lower energy photons from the EBL. From Fig. \ref{fig:MFP} we see that already for  relatively nearby blazars, such as Mrk 421 ($D\simeq 150$~Mpc), the mean free path of $\gamma$-rays with energy $E_\gamma\ge 10$~TeV  is shorter than the distance to the source. 

\begin{figure}
\begin{center}
\includegraphics[width=0.769\linewidth]{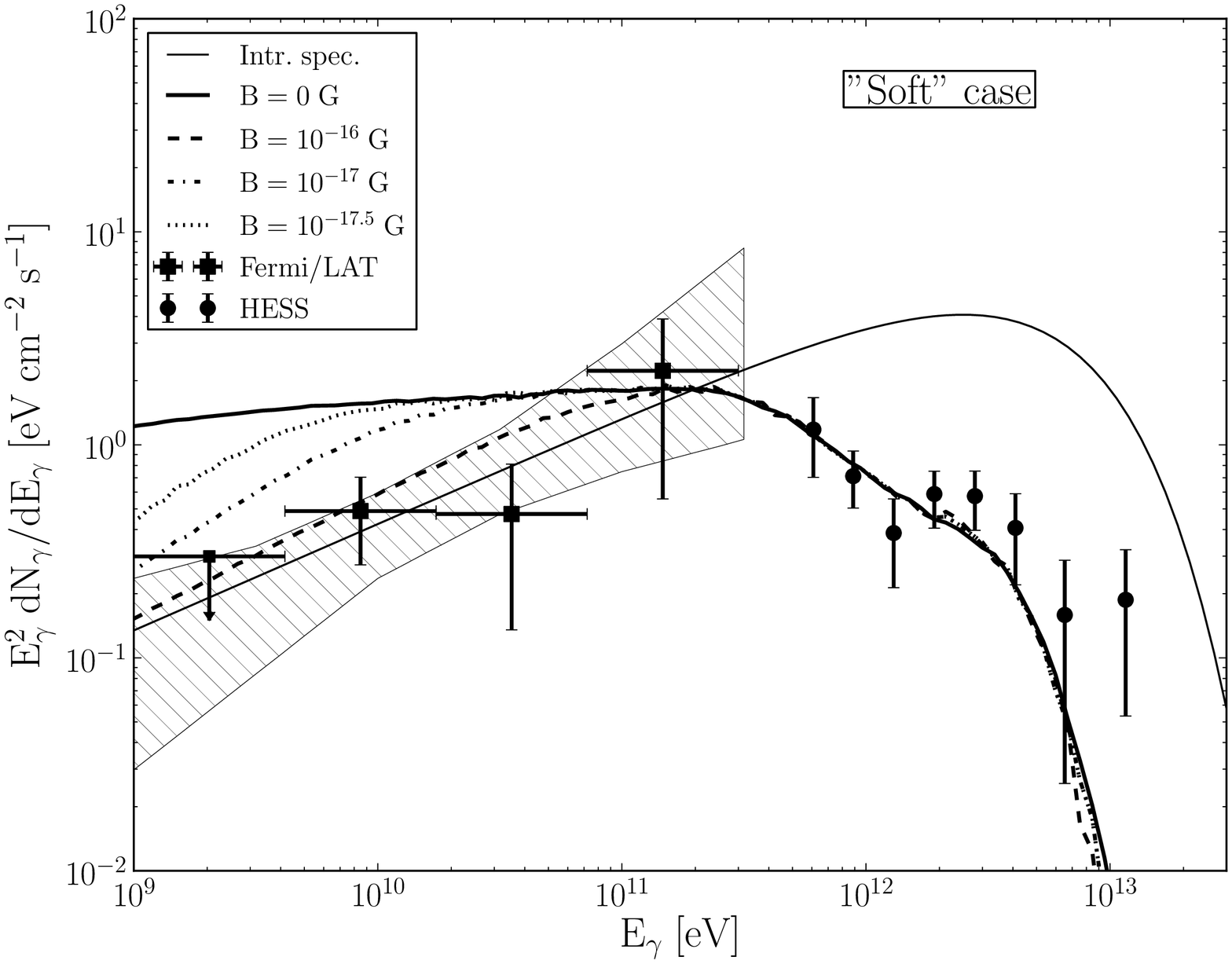}
\includegraphics[width=0.769\linewidth]{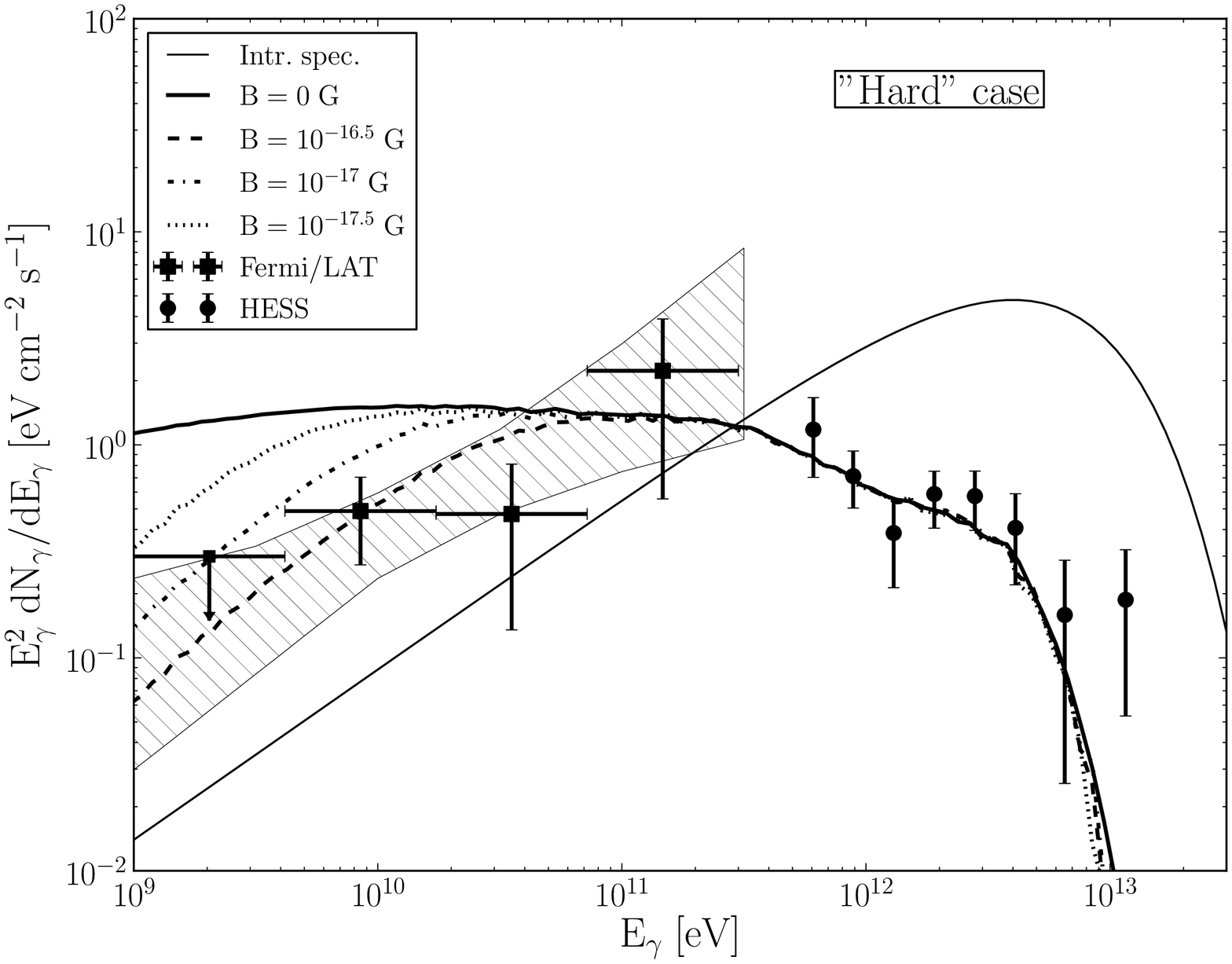}
\end{center}
\caption{The $\gamma$-ray spectrum of the blazar 1ES 0229+200, from \citet{Vovk12}. Thin solid lines show the intrinsic primary source spectrum. Thick solid lines shows the sum of the absorbed primary source spectrum and the secondary cascade spectrum after propagation of the $\gamma$-rays through the IGM from the source to the Earth. Dashed, dotted and dashed-dotted lines show modifications of the propagated spectrum in the presence of IGMF with the strength given  in the figure legends. }
\label{fig:1ES0229_spectrum}
\end{figure}

The absorption of the highest energy $\gamma$-rays which due to the pair production on EBL photons injects electron-positron pairs in the IGM. The pairs are highly relativistic and are injected along the $\gamma$-ray beam from the source, at a distance of about $\lambda_{\gamma\gamma}$. All the initial source power contained in the absorbed  VHE $\gamma$-rays is transferred to the $e^+e^-$ pairs. .  This power does not accumulate, because high-energy electrons and positrons rapidly  loose energy via inverse Compton scattering of CMB photons. The distance scale on which relativistic pairs of energy $E_e\simeq E_\gamma/2$ loose their energy is \citep{Blumenthal70}
\be
\label{e5:dic}
D_{\rm IC}=\frac{3m_e^2}{4\sigma_T\rho_{\rm CMB}E_e}\simeq  0.3\left(\frac{E_e}{1\mbox{ TeV}}\right)^{-1}\mbox{ Mpc}\,,
\ee
where $\rho_{\rm CMB}\simeq 0.25$~eV/cm$^{3}$ is the energy density of the CMB and $\sigma_T$ is the Thomson cross-section. This means that all power injected in the $e^+e^-$ pairs is finally converted into the power of inverse Compton $\gamma$-ray emission. The mean energy of the inverse Compton photons is
\be 
E_{\rm IC}=\frac{4\epsilon_{\rm CMB}E_e^2}{3m_e^2}\simeq 3\left(\frac{E_e}{1\mbox{ TeV}}\right)^2\mbox{ GeV,}
\ee
where $\epsilon_{\rm CMB}\simeq 3T_{CMB}$ is the mean energy of CMB photons. Thus, absorption of VHE $\gamma$-rays in the IGM finally results in the generation of secondary lower energy $\gamma$-ray emission from the IGM 
\citep{Aharonian94}. This situation is illustrated in Fig. \ref{fig:1ES0229_spectrum} where the spectrum of the secondary $\gamma$-ray emission generated along the $\gamma$-ray beam from a distant blazar 1ES 0229+200 is shown \citep{Vovk12}. One can see that absorption of the primary source flux in the TeV band leads to the production of the secondary (cascade) emission with spectral energy distribution stretching over several decades in energy, from $E<1$~GeV to $E>100$~GeV. The two panels of the Figure show calculations adopting different assumptions about the properties of the intrinsic $\gamma$-ray spectrum of the blazar. In both cases the intrinsic spectrum has the form of a cut-off power law 
\be 
\frac{dN_\gamma}{dE}\sim E^{-\Gamma}\exp\left(-\frac{E}{E_{\rm cut}}\right)
\ee
with slope $\Gamma$ and high-energy cut-off $E_{\rm cut}$. The model shown in the left panel assumes a softer slope, a power law with $\Gamma=1.5$, while the model in the right panel assumes a harder intrinsic spectrum, $\Gamma=1.2$. 

%%%%%%%%%%%%%%%%%%%%%%%%%%%%%%%%%%%%%%%%%%%%%%%%%%%%%%
%%%%%%%%%%%%%%%%%%%%%%%%%%%%%%%%%%%%%%%%%%%%%%%%%%%%%%

The secondary  $\gamma$-ray emission in the range 1-100~GeV is, in principle, detectable by telescopes sensitive to this energy band, such as  Fermi \citep{Atwood09} and AGILE \citep{Tavani09} which are currently in operation. 
Secondary emission at the energies above 100~GeV can also be detected by the ground-based $\gamma$-ray telescopes, HESS, MAGIC and VERITAS \citep{Aharonian08_review}.

The observed spectrum in the 1-100 GeV band shown by  the crosses in 
Fig.~\ref{fig:1ES0229_spectrum} contains two contributions: the intrinsic source spectrum in this band and the secondary cascade emission from $e^+e^-$ pairs generated by interactions of TeV photons with the EBL. If only the source spectrum is considered, we can in no way to distinguish between the direct and cascade contributions to the source flux in this energy band. However,  the two contributions can be distinguished based on their different imaging and timing properties. 

\begin{figure}
\begin{center}
\includegraphics[width=\linewidth]{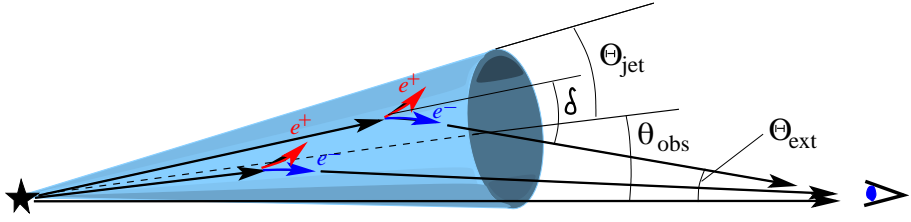}
\end{center}
\caption{Geometry of secondary $\gamma$-ray emission from electromagnetic cascade developing along the VHE $\gamma$-ray beam of the blazer. From \citet{GeV_jets}.}
\label{fig:geometry}
\end{figure}

Indeed, the secondary cascade emission is produced by an extended source in the IGM with the linear size given by the mean free path of the VHE $\gamma$-rays (\ref{eq:mfp}), while the primary emission is produced by a point source at the location of the blazar. A telescope with sufficient angular resolution can in principle resolve the overall emission in the 1-100~GeV band into a point source + extended emission from the IGM. A schematic representation of this system is shown in Fig. \ref{fig:geometry}.  A typical VHE $\gamma$-ray source, a blazar, emits most of the $\gamma$-ray flux into a narrow conical jet with an opening angle $\Theta_{\rm jet}\sim 0.1^o$, as inferred directly from the observations of blazar jets in the radio band and from indirect observations of relativistic bulk motions in blazar jets in radio-to-$\gamma$-rays \citep{Urry95}. Blazar jets are almost aligned with the line of sight, so that the direction toward the observer lies within the $\gamma$-ray emission cone. In genera, however, the jet axis is slightly misaligned by an angle $\theta_{obs}\le \Theta_{jet}$ with the direction toward the observer, as shown in Fig.~\ref{fig:geometry}. VHE $\gamma$-rays interacting in the IGM deposit electron positron pairs in the interior of the shaded cone tracing the $\gamma$-ray beam in Fig~ \ref{fig:geometry}. Secondary cascade emission comes from the locations within this cone, rather than directly from the primary point source  marked by the star in Fig. \ref{fig:geometry}.  

%%%%%%%%%%%%%%%%%%%%%%%%%%%%%%%%%%%%%%%%%%%%%%%%%%%%%%
%%%%%%%%%%%%%%%%%%%%%%%%%%%%%%%%%%%%%%%%%%%%%%%%%%%%%%

From Fig. \ref{fig:geometry} it is clear that only electrons and positrons moving in the direction of observer would contribute to the secondary $\gamma$-rays emission from the shaded conical region. A simple geometrical argument tells that if electrons/positrons would move exactly in the direction of the primary $\gamma$-ray, no extended emission from the shaded conical region in Fig. \ref{fig:geometry} would be observable. However, the pair trajectories are always misaligned with the primary $\gamma$-ray direction. The minimal possible misalignment angle $\delta$ is determined by the kinematics of the pair production collision, $\delta\ge m_e/E_\gamma\simeq 5\times 10^{-7}\left(E_\gamma/1\mbox{ TeV}\right)^{-1}$. 

Let us now consider am magnetic field which permeates the IGM.  In this case, electrons and positrons move along curved trajectories whose curvature radius $R_L$ is determined by the strength of magnetic field
\be
R_L=\frac{E_e}{eB}\simeq 10^2\left(\frac{E_e}{1\mbox{ TeV}}\right)\left(\frac{B}{10^{-17}\mbox{ G}}\right)^{-1}\mbox{ Mpc}\,.
\ee
Deflections of electron/positron trajectories by the magnetic fields lead to an additional misalignment of the particle trajectories with the primary $\gamma$-ray direction. After the propagation over an inverse Compton distance scale $D_{\rm IC}$ (\ref{e5:dic}), the misalignment angle is 
\be
\delta\simeq
\left\{
\begin{array}{ll}
\frac{\displaystyle D_{\rm IC}}{\displaystyle R_L}\simeq 0.2^\circ\left(\frac{\displaystyle E_e}{\displaystyle 1\mbox{ TeV}}\right)^{-2}\left(\frac{\displaystyle B}{\displaystyle 10^{-17}\mbox{ G}}\right), & D_{\rm IC}\gg \lambda_B\\
&\\
\frac{\sqrt{\displaystyle D_{\rm IC}\lambda_B}}{\displaystyle R_L}\simeq 0.2^\circ\left(\frac{\displaystyle E_e}{\displaystyle 1\mbox{ TeV}}\right)^{-1}\left(\frac{\displaystyle B}{\displaystyle 10^{-17}\mbox{ G}}\right)\left(\frac{\displaystyle \lambda_B}{\displaystyle 0.3\mbox{ Mpc}}\right)^{1/2}, & D_{\rm IC}\ll \lambda_B \,.\\
\end{array}
\right.
\label{e5:deflection}
\ee
For the second line above, we have taken into account many stochastic deflections which led the electron or positron perform a random walk.

A straightforward geometrical calculation based on Fig. \ref{fig:geometry} leads to the estimate of the angular size of extended emission from the IGM \citep{GeV_jets}
\begin{equation}
\label{eq:thetaext}
\Theta_{\rm ext}(B)=\mbox{min}\left[
\delta/\tau,\  \delta-\theta_{\rm obs}\right]<\Theta_{\rm ext,max}
\end{equation}
where $\tau=D/\lambda_{\gamma\gamma}$ is the optical depth for the primary $\gamma$-rays.

%%%%%%%%%%%%%%%%%%%%%%%%%%%%%%%%%%%%%%%%%%%%%%%%%%%%%%
\begin{figure}
\begin{center}
\includegraphics[width=\linewidth]{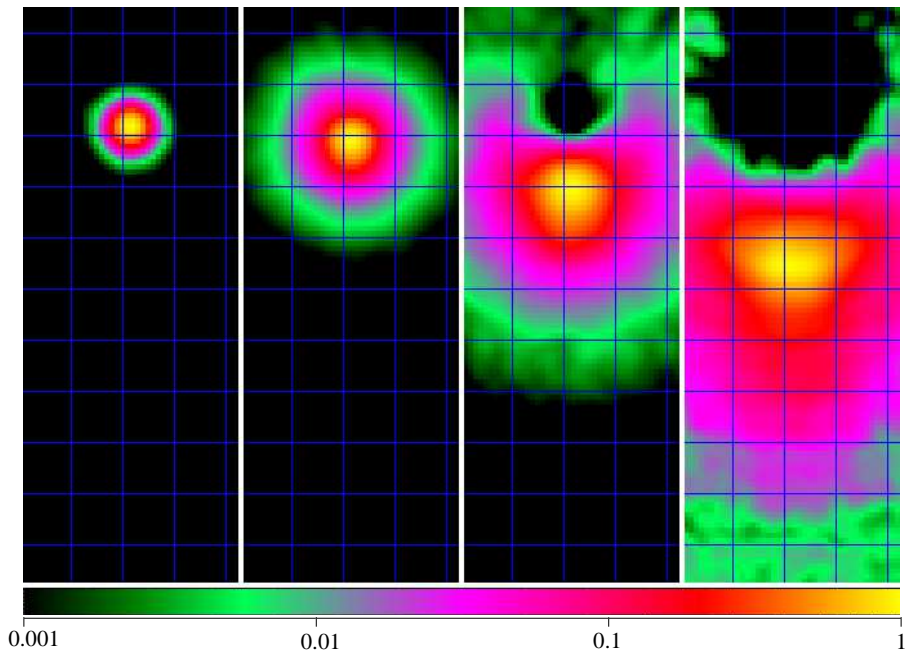}
\end{center}
\caption{Simulated 1-10~GeV band images of the sky region around a TeV blazar with a jet with $\Theta_{\rm jet}=\theta_{\rm obs}=3^\circ$ at different times after the instantaneous injection of 1~TeV \gr s at the source. The IGMF parameters are $B=10^{-16}$~G, $\lambda_B=1$~Mpc. From left to right: time-integrated emission; images after a delay time $t_{\rm del}$ $0<t_{\rm del}<10^5$~yr, $10^5$~yr$<t_{\rm del}<10^6$~yr, $10^6$~yr$<t_{\rm del}<3\times 10^6$~yr  and $3\times 10^6$~yr$<t_{\rm del}<10^7$~yr after an outburst. From \citet{GeV_jets}.  The grid spacing is $2^\circ$. The color code indicates the fraction of the total energy emitted in this region.}
\label{fig:delay}
\end{figure}
%%%%%%%%%%%%%%%%%%%%%%%%%%%%%%%%%%%%%%%%%%%%%%%%%%%%%%

The difference in the path between the direct $\gamma$-ray signal reaching the observer from  the primary source and the signal from the $\gamma$-rays which are converted to the $e^+e^-$ pairs in the IGM and then re-produced as inverse Compton emission leads not only to a displacement of the secondary emission from the primary source position but also to a time delay of the cascade signal. This is illustrated in Fig. \ref{fig:delay} in which the results of Monte-Carlo simulations of the  signal from the cascade developing in the IGM with a magnetic field of $B=10^{-16}$~G is shown \citep{GeV_jets}. The characteristic time delay scales with the off-source angle $\Theta_{\rm ext}$is given by \citep{GeV_jets}
\begin{eqnarray}
t_{\rm del}(\theta)&\sim&\frac{D}{c}\left(\frac{\sin\Theta_{\rm ext}+\sin(\theta_{\rm obs}+\Theta_{\rm jet})}{\sin(\Theta_{\rm ext}+\theta_{\rm obs}+\Theta_{\rm jet})}-1\right)\simeq \frac{D\Theta_{\rm ext}(\theta_{\rm obs}+\Theta_{\rm jet})}{2c}\nonumber\\
&\simeq&20\left(\frac{D}{150\mbox{ Mpc}}\right)\left(\frac{(\theta_{\rm obs}+\Theta_{\rm jet})}{1^\circ}\right)\left(\frac{\Theta_{\rm ext}}{1''}\right)
\mbox{ yr }\,.
\label{eq:tdelay}
\end{eqnarray}
The angular size of the extended emission from the IGM (\ref{eq:thetaext}) and, as a consequence, the time delay of the emission (\ref{eq:tdelay}) depend on the strength and correlation length of the IGMF. Thus the IGMF parameters can be deduced from imaging \citep{neronov07,Neronov09,Elyiv09} and/or timing \citep{Plaga95,Dai02,Neronov09,Ichiki08,Takahashi08} observations of the cascade signal by \gr\ telescopes sensitive in the 1-100~GeV band. 

If the gyro-radius of the high-energy $e^+e^-$ pairs is shorter than the inverse Compton cooling distance, pair trajectories are randomized by the IGMF before the pair energy is lost to the inverse Compton emission. In this case, the secondary emission is isotropic and its angular extent is determined only be the mean free path of the primary $\gamma$-rays. From Fig.~\ref{fig:geometry} one finds that in this case $\Theta_{\rm ext}=\lambda_{\gamma\gamma}\theta_{\rm obs}/D$. In this case $\Theta_{\rm ext}$ does not  depend on the IGMF. The condition $D_{\rm IC}<R_L$ gives the limiting IGMF strength measurable using $\gamma$-ray techniques 
\be
\label{e5:bmax}
B_{\max}=\frac{E_e}{eD_{\rm IC}}\simeq 3\times 10^{-15}\left(\frac{E_e}{1\mbox{ TeV}}\right)^2\mbox{ G}\,.
\ee
Properties of the extended \gr\ emission from IGM in the limit $B\ge B_{\max}$ were considered by \citet{Aharonian94}. The dependence of $\Theta_{\rm ext}$ on $\lambda_{\gamma\gamma}$ leads to a dependence  $\Theta_{\rm ext}\sim E^{-1/2}$ of the source size on the energy of the cascade photons. This characteristic energy dependence is, in principle, observable. Observation of this energy dependence would immediately imply a lower bound on the IGMF strength at the level given by 
Eq.~(\ref{e5:bmax}).

Searches for extended and/or time-delayed emission from the $\gamma\gamma$ pair cascade in the IGM were done with data from the Fermi telescope and from ground-based Cherenkov telescopes over the recent years. 

\citet{Aharonian01_HEGRA} searched for extended emission around the blazar Mrk 501 in the energy band above 500~GeV using  data from the HEGRA telescope. \citet{Aleksic10} have searched for extended emission around bright blazars Mrk 421 and Mrk 501 in the energy band above 300~GeV using data from the MAGIC telescope. No signal in excess of the instrument point-spread function (PSF) has been detected in both cases. This might impose some restrictions on IGMF in the range $B\sim 10^{-13}$~G, provided that the cascade signal in the energy band $E>300-500$~GeV is at a level higher than $\sim 1-10$\% of the primary source signal. However, this is not necessary so, as shown by \citet{Taylor:2011bn}. Thus the HEGRA and MAGIC data do not constrain the IGMF. 

\begin{figure}
\includegraphics[width=\linewidth]{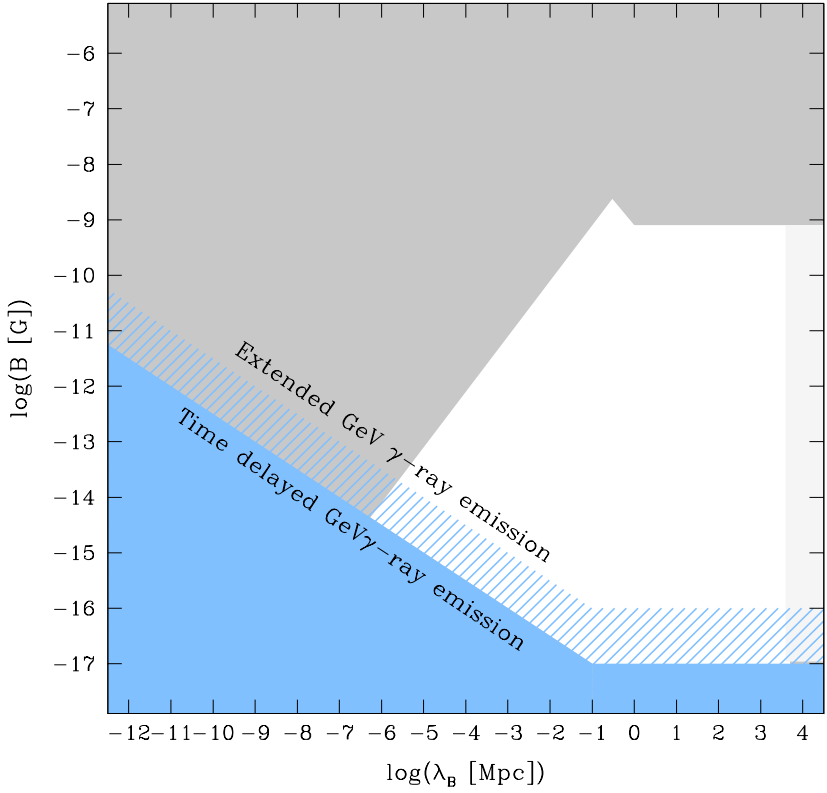}
\caption{Constraints on the  IGMF from the non-observation of $\gamma$-ray cascade emission.}
\label{f5:gamma}
\end{figure}

\citet{Neronov:2010,tavecchio-2010} have considered  data from the Fermi telescope and deduced a lower bound on the IGMF at the level of $B\gtrsim 10^{-16}$~G from the non-observation of the cascade signal in the GeV band. This limit is independent of the correlation length of the IGMF as long as $\lambda_B\gg D_{\rm IC}$. If, to the contrary, $\lambda_B\ll D_{\rm IC}$, the limit becomes more stringent with the decreasing $\lambda_B$. The necessary magnetic field increases  as $\lambda_B^{-1/2}$, because of the randomness of deflections of electron/positron trajectories, see Eq. (\ref{e5:deflection}). This limit is shown by the light-blue hatched region in Fig.~\ref{f5:gamma}.  Fermi observations are more sensitive to the cascade emission because a number of extragalactic VHE \gr\ sources like 1ES 0229+200, 1ES 0347-121 or 1ES 1101-232 observed with the Fermi telescope have hard intrinsic spectra  extending into TeV energy band. The intrinsic energy flux of these sources in the GeV band is much lower than that in the TeV band. Most of the intrinsic power of the sources in the TeV band is converted into cascade emission from the IGM, which is released in the GeV band. Thus, the expected energy flux of the cascade emission in the GeV band is much higher than the intrinsic flux. This makes the cascade emission more easily detectable. 

The limit found by \citet{Neronov:2010} and \citet{tavecchio-2010}  was derived assuming that the cascade signal is suppressed by the large angular extension of the cascade source. This is not necessarily so. An alternative way to suppress the cascade emission is via a sufficiently large time delay of the signal.  \citet{dermer11,Taylor:2011bn} used {\em simultaneous} observations of blazars in the GeV and TeV band by Fermi and ground-based Cherenkov telescopes to deduce the lower bound $B\gtrsim 10^{-18} - 10^{-17}$~G, under the assumption that the cascade emission is suppressed by the long time delay of the signal. The limit of \citet{dermer11} is lower than that of \citet{Taylor:2011bn} because of the difference in the modeling of the cascade signal. \citet{dermer11} used a semi-analytical model for the pair production and inverse Compton emission spectra, while \citet{Taylor:2011bn} used  Monte-Carlo simulations which take into account the detailed differential cross-sections for the two processes. The limit on IGMF derived by \citet{Taylor:2011bn}  is shown in Fig. \ref{f5:gamma} by the light-blue shaded region.  

%%%%%%%%%%%%%%%%%%%%%%%%%%%%%%%%%%%%%%%%%
%%%%%%%%%%%%%%%%%%%%%%%%%%%%%%%%%%%%%%%%%

\begin{figure}
\includegraphics[width=\linewidth]{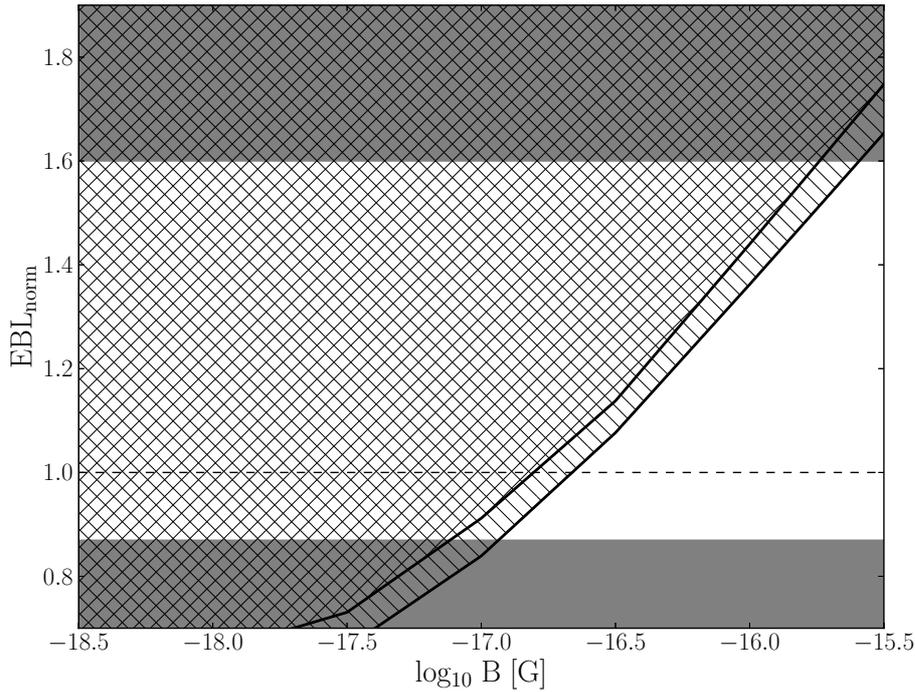}
\caption{Dependence of the lower bound for IGMF (at large correlation lengths) on the assumed level of the EBL density, in units of the model predictions by \citet{Franceschini08}. From \citet{Vovk12}.}
\label{f5:bound_with_EBL}
\end{figure}

An important uncertainty in the lower bounds on IGMF derived from $\gamma$-ray data stems from the uncertainty of the the EBL measurements. Indeed, the overall power of the cascade source is equal to the fraction of initial $\gamma$-ray power of the primary source, absorbed in the IGM. This fraction is proportional to the suppression factor $\exp(-D/\lambda_{\gamma\gamma})$, which, in turn, depends on the EBL density $n_{\rm EBL}$ via $\lambda_{\gamma\gamma}$, see Eq. (\ref{eq:mfp}). The dependence of the power of cascade source on $n_{\rm EBL}$ is exponential, so that even a moderate uncertainty of a factor $2$  in the EBL density induces an order-of-magnitude uncertainty in the cascade power. This uncertainty affects the prediction for the cascade flux in the GeV band, which is then compared to the data. \citet{Vovk12} have investigated the influence of this uncertainty on the lower bounds on IGMF and found that reducing to EBL density to the level of $0.8$ of the density assumed in the models of \citet{Franceschini08} and \citet{Dominguez11} reduces the lower bound on the IGMF  significantly, see Fig. \ref{f5:bound_with_EBL}. Similar conclusions have been reached by \citet{Arlen12} who found that uncertainties of the intrinsic source spectra combined with the uncertainties of the EBL measurement might even wash out the lower bound if the EBL  is significantly below the value found by \citet{Franceschini08}.

Note that recent measurement of the EBL spectrum  by HESS \citep{HESS12_EBL}  rules out  EBL spectra with normalizations lower than that of \citet{Franceschini08} and \citet{Dominguez11}, see Fig. \ref{fig:EBL}. This implies a lower bound on IGMF which is somewhat stronger than $10^{-17}$~G, see Fig. \ref{f5:bound_with_EBL}.  

%%%%%%%%%%%%%%%%%%%%%%%%%%%%%%%%%%%%%%%%%
%%%%%%%%%%%%%%%%%%%%%%%%%%%%%%%%%%%%%%%%%

To significantly suppress the cascade signal, the IGMF has to be present over a large fraction of the line of sight toward an extragalactic VHE \gr\ source. \citet{Dolag:2010ni} have found that this imposes a bound on the possible volume filling factors of the IGMF at the level of $\gtrsim 60\%$.

Detection of extended \gr\ emission around extragalactic VHE \gr\ sources, with an energy-dependent morphology, would imply a measurement, rather than just a bound on the IGMF. Such a measurement was claimed by \citet{Ando10}, based on  data from the Fermi telescope. However, the excess $\gamma$-ray signal coming from outside the Fermi PSF in the analysis of \citet{Ando10} turned out to be not due to a real signal, but due to the imperfect modeling of the telescope's PSF, as was shown by \citet{Neronov_PSF}. 

An alternative method to detect the cascade emission around blazars is to look for the signature of the gamma-ray induced cascades   in the anisotropy spectrum of the Extragalactic Gamma-Ray Background \citep{Venters12}.

A promising method for measuring an IGMF with strength close to the lower bounds derived from Fermi data is to look for an energy-dependent delayed emission following bright flares of TeV blazars. Indeed, from Eq. (\ref{eq:tdelay}) one can see that emission from very small off-source angles $\Theta_{\rm ext}\sim 1''$ might be detectable in this way. This range of the off-source angles can only be reached by imaging observations with \gr\ telescopes which have the PSF smaller than $\sim 0.1^\circ\simeq 3\times 10^2$~arcsec.  \citet{Takahashi12} have attempted a search for delayed cascade emission in Fermi, following a bright TeV flare of the blazar Mrk 501. Unfortunately, the flare flux and the spectrum of the flare in the TeV band were not high and hard enough to result in significant cascade emission which would be detectable on relatively short time scales (day-to-month) by Fermi, even if the IGMF would be negligibly small, see \citet{Neronov_Mrk501}. It is clear  that next strong TeV flare of a blazar with hard intrinsic spectrum extending into multi-TeV band, such as 1ES 0229+200, will provide strong constraints on IGMF or will allow us to  measure it, if the IGMF level is close to the currently existing lower bound. 

However, the known hard spectrum TeV blazars are surprisingly devoid of flaring activity. One possible explanation for this might be that the production mechanism of VHE \gr s in these sources is different from that operating in flaring sources. In particular, VHE \gr\ emission can be related to the emission of UHECR from these sources. A non-variable VHE \gr\ flux can be produced in interactions of a beam of protons with energies above $10^{18}$~eV during their propagation through the IGM \citep{Essey11a,Essey12}.

Observations of secondary VHE \gr\ emission from UHECR interactions in the IGM is possible only if the UHECR induced cascade develops in an anisotropic way along the UHECR beam, and is not isotropized due to the deflections of the cascade electrons and positrons by an IGMF. This implies that the UHECR cascade scenario for the VHE \gr\ emission works only if the IGMF strength is close to the lower bounds \citep{Essey11}.

The limits on IGMF derived from \gr\ observations stem from the non-observation of the cascade inverse Compton emission initiated by the absorption of the VHE \gr s in the IGM. A potential alternative possibility to suppress the cascade signal is to dissipate the absorbed \gr\ power not via inverse Compton scattering of CMB photons with $e^+e^-$ pairs, but through a different channel. Such a possibility was considered by \citet{Broderick12}, who considered the possible role of plasma instabilities on the geometry of the $e^+e^-$ beam in the IGM. In the absence of IGMF, electrons and positrons deposited along the primary VHE \gr\ beam, form a collimated beam of high-energy particles, with the density 
\be
n_{\rm beam}=\frac{F_{\rm VHE}D^2}{E_\gamma\lambda_{\gamma\gamma}^2}\simeq 10^{-23}\left(\frac{F_{\rm VHE}}{10^{-12}\mbox{ erg/(cm}^2\mbox{s)}}\right)\left(\frac{D}{600\mbox{ Mpc}}\right)^{2}\left(\frac{E_\gamma}{1\mbox{TeV}}\right)\mbox{ cm}^{-3}
\ee
where we have substituted the flux and the luminosity distance of 1ES 0229+200 as the reference values for $F_{\rm VHE}$ and $D$. This implies the mean distance between the beam particles $\ell_{\rm beam}\sim n_{\rm beam}^{-1/3}\simeq 10^8\mbox{ cm}$. 

The collimated electron-positron beam propagates through the ionized IGM with free charge density $n_{\rm IGM}=\Omega_b \rho\simeq 3\times 10^{-7}\left(\Omega_b/0.04\right)$~cm$^{-3}$. Plasma instabilities due to collective interaction of beam particles can occur if $\ell_{\rm beam}$ is shorter than the plasma skin depth $\ell_{\rm skin}\simeq 2\pi/\omega_p\simeq 6\times 10^9\left(n_{\rm IGM}/3\times 10^{-7}\mbox{ cm}^{-3}\right)^{-1/2}$~cm and/or the Debye length of the IGM plasma, 
\be
\ell_{\rm Debye}=\left(\frac{T_{\rm IGM}}{m_e}\right)^{1/2}\frac{1}{\omega_p}\simeq 10^6\left(\frac{\Omega_b}{0.04}\right)^{-1/2} \left(\frac{T_{\rm IGM}}{10^{4}\mbox{ K}}\right)^{1/2}\mbox{ cm}
\ee
where $\omega_p= \sqrt{4\pi e^2 n_{\rm IGM}/m_e}$ is the IGM plasma frequency and $T_{IGM}$ is the temperature of the IGM. \citet{Broderick12} have considered different possible instability modes of the electron-positron beam and found that the growth rate of an "oblique" mode of Laingmur waves in the linear regime is faster than inverse Compton cooling of electrons and positrons. Thus, plasma instabilities developing in the beam can potentially reduce the power of inverse Compton emission and instead inject this power into the collective plasma motions and the electromagnetic field along the beam path. The calculation of the growth rate of plasma instabilities by \citet{Broderick12} was performed in linear approximation, i.e. not taking into account  back reaction of the beam perturbations on the growth rate. However, back reaction is known to be important, especially for relativistic particle beams \citep{Kaplan73,Schlickeiser12,Miniati12}. In particular, the effect of non-linear Landau damping suppresses the growth of plasma instabilities and stabilizes the beam. The calculations by \citet{Kaplan73,Miniati12} and \citet{Schlickeiser12} agree on the fact that when nonlinear Landau damping is important, the growth of plasma instabilities is suppressed. However, they differ in the range of parameters of the beam for which nonlinear Landau damping is supposed to be important. Further investigations are required to asses the potential importance of plasma instabilities for the electron-positron beam.  

We also notice that development or suppression of plasma instabilities is highly sensitive to the angular and energy distribution of thze particles in the beam. The presence of cooled (rather than freshly injected) electrons in the beam with energies much below TeV and/or the presence of 
tiny IGMF with the strength much below the limits discussed above, can destroy the narrow collimation of the beam and suppress beam instabilities. This suggests that plasma effects might potentially be important only at the initial moment of the onset of blazar activity, on the time scales about the inverse Compton cooling time, $t_{\rm IC}\simeq D_{\rm IC}\simeq 10^6\left(E_e/1\mbox{ TeV}\right)^{-1}$~yr.

%%%%%%%%%%%%%%%%%%%%%%%%%%%%%%%%%%%%%%%%%%%%%%%%%%%%%%
\subsubsection{Prospects for IGMF measurement with next-generation $\gamma$-ray telescopes.}
\label{s:next_generation_gamma}
%%%%%%%%%%%%%%%%%%%%%%%%%%%%%%%%%%%%%%%%%%%%%%%%%%%%%%

If the IGMF strength is close to the lower bound derived from the GeV \gr\ data from the Fermi telescope, a measurement of the IGMF can be achieved using a combination of the Fermi telescope with one of the ground-based Cherenkov telescopes: HESS, MAGIC or VERITAS. Fields with the strength $\sim 10^{-17}$~G can be measured via the observation of the time delay of the cascade emission following an exceptionally bright and hard spectrum flares of TeV blazars, similar to the brightest flares of Mrk 421 and Mrk 501 in 1997 and 2000.  In the particular case of Mrk 501, the cascade emission in the GeV band might be detectable by Fermi only if the 1-10~TeV flux of the source is at the level of the historically brightest flare observed by HEGRA and CAT \citep{Neronov_Mrk501}. Otherwise, the quiescent source flux level of the source in the GeV band is high enough to hide the cascade emission. 

The duty cycle of exceptionally bright flares during which the TeV flux of a blazar can grow by up to two orders of magnitude is still unknown. After an all-sky monitoring of the \gr\ sky by the Fermi telescope of four years, no simultaneous GeV-TeV band observations of any of the exceptional flares is reported.  This suggests that either  flares are rare, with typical duty cycles much longer than a decade, or that the TeV band flares are not necessarily associated to the flaring activity in the GeV band \citep{Neronov_Mrk501}.  If this is so, the flares can be observed only by telescopes working in the TeV band. However, the narrow fields of views of ground-based Cherenkov telescopes do not allow for efficient monitoring of a significant fraction of the sky, similar to that done by the Fermi telescope in the GeV band. 

An all-sky monitoring in the TeV band is important for an efficient identification of  flaring episodes. The possibility of wide FoV telescopes in the TeV band is now demonstrated by the MILAGRO \citep{MILAGRO_results}, Tibet-AS$\gamma$  \citep{TIBET} and ARGO-ABJ \citep{ARGO} arrays. Next generation arrays, such as HAWC (\citet{HAWC}, {\tt http://hawc.physics.wisc.edu/}) and LHAASO  ({\tt http://english. ihep.cas.cn/ic/ip/LHAASO/ })  will have an order-of magnitude better sensitivity and somewhat lower energy threshold than MILAGRO, Tibet-AS$\gamma$ and ARGO-ABJ. 

A wide field-of-view monitoring of the sky will be also possible with the next-generation Cherenkov telescopes, like the Cherenkov Telescope Array (CTA) \citep{CTA}. The monitoring will be possible in the "sky survey" mode  in which some $\sim 20-30$ sub-arrays of Cherenkov telescopes with individual FoVs of $7^\circ-10^\circ$ will be pointing in slightly different directions, covering regions of ten(s) of degrees  on the sky.  

This will allow an efficient monitoring of blazar activity, rather than just source detection. If the Fermi telescope will still be in orbit at the time of full operation of HAWC and/or LHAASO, detection and detailed GeV-TeV monitoring of exceptional flares  will  strongly enhance the chance to observe the delayed cascade emission from the IGM and thereby to measure an IGMF in the range of $(10^{-17}-10^{-16})$G. 

The time delay of the cascade emission scales as \citep{Neronov09}
\begin{equation}
t_{\rm del}\simeq
\left\{
\begin{array}{ll}
0.3(1-\tau^{-1})(1+z)^{-3}&\\
\left[\frac{\displaystyle E}{\displaystyle 0.1\mbox{ TeV}}\right]^{-5/2}
\left[\displaystyle \frac{B}{\displaystyle 10^{-17}\mbox{ G}}\right]^2\mbox{ yr, }& \lambda_B\gg D_{IC}\\ \\
6\times 10^{-3}(1-\tau^{-1})
\left[\frac{\displaystyle E}{\displaystyle 0.1\mbox{ TeV}}\right]^{-2}&\\
\left[\frac{\displaystyle B}{\displaystyle 10^{-17}\mbox{ G}}\right]^2
\left[\frac{\displaystyle \lambda_{B}}{\displaystyle 1\mbox{ kpc}}\right]\mbox{ yr}, & \lambda_B\ll D_{IC}
\end{array}
\right.
\label{eq:Tdelay2}
\end{equation}
where $z$ is source redshift and $\tau$ is the optical depth for the primary absorbed $\gamma$-rays responsible for the secondary cascade photons at the energy $E$. 

\begin{figure}
\includegraphics[width=\linewidth]{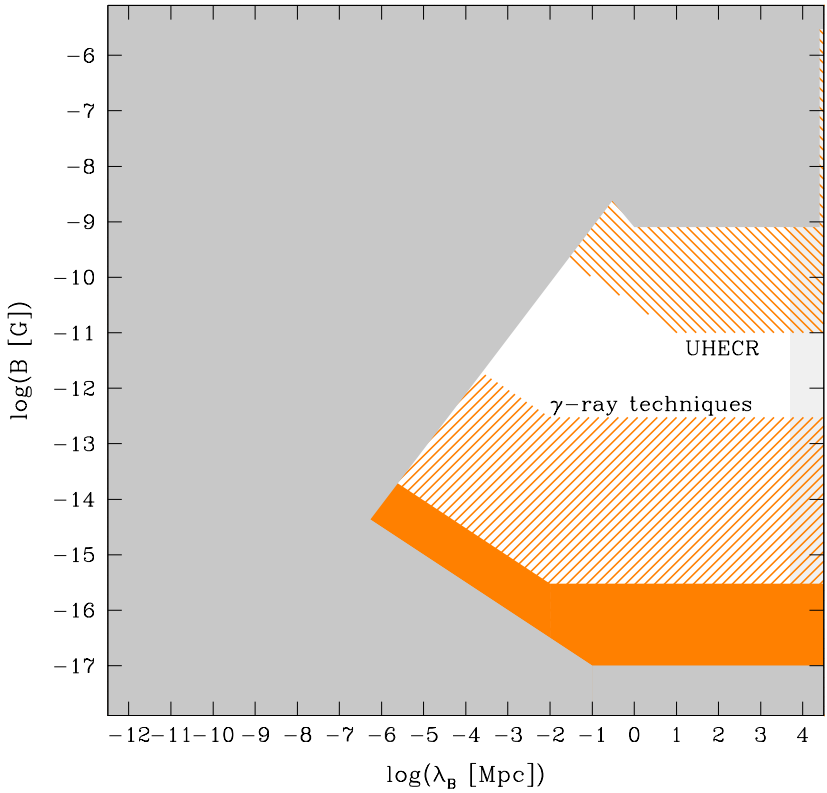}
\caption{The range of IGMF parameters accessible by next-generation \gr\ and UHECR telescopes. The orange shaded region shows the range of parameters for which delayed emission following bright flares is detectable at the energies below $\sim 1$~TeV. The lower orange hatched region shows the range of parameters for which extended emission is detectable below $\sim 1$~TeV. The upper orange hatched region shows the range of parameters which can be constrained by future  UHECR telescopes.}
\label{fig:future}
\end{figure}

A "natural" upper limit for the energy of \gr s produced by blazars is the energy at which the VHE \gr s can not escape from the blazar host galaxy because of pair production on the interstellar radiation field (ISRF) inside the host galaxy. Taking the nearby radio galaxy M 87 as a prototype host galaxy of TeV blazars, we estimate that photons with energies above $\sim 30$~TeV are efficiently absorbed while propagating through the ISRF \citep{NeronovAharonian07}. This implies that  VHE \gr\ blazars should have a high-energy cut-off in their intrinsic spectra at $\sim 30$~TeV energy. Taking into account the relation between the primary and cascade photon energies, 
\be
E=1\left(\frac{E_\gamma}{30\mbox{ TeV}}\right)^2\mbox{ TeV}\,,
\ee
we conclude that the delayed cascade emission is observable up to the 0.1-1~TeV energy band, by ground-based Cherenkov telescopes. Taking a time scale of $\sim 0.3-1$~yr as a reasonable span for an observation campaign following an exceptionally bright flare, we find that magnetic fields with strength up to $3\times 10^{-16}$~G are measurable in this way. Of course, it might be challenging to distinguish the delayed cascade emission signal  in the  0.1-1~TeV band from the intrinsic slowly decreasing flaring emission from the primary source in the same energy band. Only observations of multiple flares (from several sources) with characteristic energy dependent decay time given by Eq. (\ref{eq:Tdelay2}) will indicate that the delayed emission is due to the cascade, rather than being intrinsic to the sources. The range of IGMF measurable via observations of delayed emission is shown by the orange shading in Fig.~\ref{fig:future}. 

If IGMF is significantly stronger than $ 10^{-16}$G, the time delay of the cascade emission in any energy  band is too large to be directly measurable by \gr\ telescopes within a reasonable observation time span.  However, in this case the angular extent of the cascade emission becomes sufficiently large to be measurable by the \gr\ telescopes. For a source  with a jet aligned closely with the line of sight, the angular size of the extended source is \citep{Neronov09}
\begin{equation}
\Theta_{\rm ext}\simeq
\left\{
\begin{array}{ll}
8^\circ(1+z)^{-2}\frac{5}{\tau}
\left[\frac{\displaystyle E}{\displaystyle 0.1\mbox{ TeV}}\right]^{-1}
\left[\displaystyle \frac{B}{\displaystyle 10^{-13}\mbox{ G}}\right],& \lambda_B\gg D_{\rm IC}\\
&\\
1^\circ(1+z)^{-1/2}\frac{5}{\tau}
\left[\frac{\displaystyle E}{\displaystyle 0.1\mbox{ TeV}}\right]^{-3/4}\!
\left[\frac{\displaystyle B}{\displaystyle 10^{-13}\mbox{ G}}\right]\!
\left[\frac{\displaystyle \lambda_{B}}{\displaystyle 1\mbox{ kpc}}\right]^{1/2},&\lambda_B\ll D_{\rm IC}.
\end{array}
\right.
\label{eq:Thetaext2}
\end{equation}
Measurements of moderately extended (degree-scale) cascade emission at the highest energies $E\sim 1$~TeV  with CTA and/or HAWC and LHAASO can provide a measurement of IGMF, if its strength is below about $10^{-13}$G (for large correlation lengths $\lambda_B$). The range of IGMF parameters accessible via observations of extended emission around blazars in the 0.1-1~TeV band is shown as the lower orange hatched region in Fig.~\ref{fig:future}. 

%%%%%%%%%%%%%%%%%%%%%%%%%%%%%%%%%%%%%%%%%
%%%%%%%%%%%%%%%%%%%%%%%%%%%%%%%%%%%%%%%%%

As mentioned above, still stronger magnetic fields are not directly measurable using \gr\ techiques, because the properties of the extended emission from the IGM become independent of the IGMF. However, the presence of IGMF stronger than $\sim 10^{-12}$~G in the IGM can still be established from the \gr\ data, if the extended emission around blazars is detected \citep{Aharonian94,Neronov09}.

%%%%%%%%%%%%%%%%%%%%%%%%%%%%%%%%%%%%%%%%%%%%%%%%%%%%%%
\subsection{Constraints from initial seed fields for galactic dynamos}
\label{s:seeds}
%%%%%%%%%%%%%%%%%%%%%%%%%%%%%%%%%%%%%%%%%%%%%%%%%%%%%%

Cosmological magnetic fields might, in principle, play an important role in the cosmic magnetogenesis, because they can be the "seed" fields necessary for the action of dynamos in the galaxies \citep{Parker55,Ruzmaikin:1988,Kronberg:1993vk,Beck:1995zs,Kulsrud99,Grasso00,Widrow02,Brandenburg:2004jv}. 
Following the qualitative arguments used in Section \ref{s:evo} for the description of cosmological evolution of magnetic fields, one can estimate the growth time scale of Galactic dynamos $t_{\rm gal}$ based on the known  velocity scale $v\sim 10^7$cm/s ($v\sim 10^6$cm/s for turbulent motions on the scale $\lambda\le 100$pc in the Galaxy, and $v\sim 10^7$cm/s for the large scale motions on  distances $\lambda\sim (1-10)$kpc). The "eddy processing" time is estimated as 
\begin{equation}
t_{\rm gal}\sim \frac{\lambda}{v}\simeq 10^8\left(\frac{\lambda}{10\mbox{ kpc}}\right)\left(\frac{v}{10^7\mbox{ cm/s}}\right)^{-1}\mbox{ yr\,.}
\end{equation}
This eddy processing time is close to the typical  estimates of the growth rate (e-folding time) of the Galactic dynamo in the range of $\tau \sim 10^8-10^9$~yrs \citep{Kulsrud99}. This means that within some 20-30 e-foldings one would amplify the initially existing field by a factor of $\exp(t_0/\tau)\sim 10^9-10^{13}$. The observed amplitude of galactic magnetic fields of about $10\ \mu$G  can, therefore, be explained if the pre-existing seed magnetic field has the strength of at least $(10^{-21}-10^{-19})$G, taking into account additional amplification by about 3 orders of magnitude via compression during the collapse of primordial perturbations to galaxies,  see Eq~(\ref{e4:galflux}). Also galaxies at high redshift, $z\sim 1 - 2$ have similar magnetic fields. Since the age of the Universe at $z\sim 2$ is more than a factor of 2 less than $t_0$, the dynamo amplification in these galaxies is only about 10-15 e-foldings requiring correspondingly stronger seed fields of about $10^{-15}$G. The fact that galactic fields at redshifts $z=2$ are of the same order as this at $z=0$ hints to the fact that galactic dynamo amplification is saturated already at $z\sim 2$ and the amplitude of these fields is not determined by the strength of the seed fields.

If the mechanisms of the action of Galactic dynamos would be well constrained, we could  use observations  of the structure and strength of magnetic fields in different types of galaxies (and galaxy clusters) and at different redshift to gain information about the properties of initial seed fields for the dynamo action. However, the efficiency of dynamos in galaxies and galaxy clusters are uncertain, so that the existing estimates of the strength and spatial structure of the seed magnetic fields needed for the Galactic dynamos differ in a very broad range. 
 
The seed fields for the Galactic dynamos might, in fact, originate not from the pre-existing primordial fields, but occur locally in the forming galaxies. Possible mechanisms for the generation of magnetic fields during the gravitational collapse are the Weibel instability \citep{Schlickeiser03,Medvedev06} or battery effects \citep{Subramanian94,Kulsrud97,Gnedin00}. These mechanisms do not result in significant magnetic fields spread in the IGM. Therefore, in such scenarii, the seeding and dynamo amplification of magnetic fields in galaxies is not at all related to the properties of IGMF.

Besides, even if the seed fields for the Galactic dynamos stem form pre-existing magnetic fields produced in the Early Universe, the value of the seed fields might not  be closely related to the primordial field strength.  \citet{Ryu08,Schleicher10} argue that the small-scale turbulent dynamo operating in the IGM  during the collapse of proto-galaxies amplifies initial seed magnetic fields up to equipartition with the turbulent energy on time scales much shorter than the dynamical time scale, see also \cite{Beck:1994,Brandenburg:2004jv}. This implies that the initial conditions for Galactic dynamos are largely independent of the pre-existing seed field strength.

Thus, although cosmological fields produced before the onset of structure formation might serve as the initial seed fields for Galactic dynamos, no sensible constraints on the cosmological fields can be derived from the observed properties of Galactic magnetic fields. 

%%%%%%%%%%%%%%%%%%%%%%%%%%%%%%%%%%%%%%%%%%%%%%%%%%%%%%
\subsection{IGMF from Galactic winds}
\label{s:winds}
%%%%%%%%%%%%%%%%%%%%%%%%%%%%%%%%%%%%%%%%%%%%%%%%%%%%%%

Physical processes in the Early Universe are not the only possibility to generate magnetic fields in the IGM, in particular in the voids of the large scale structure. This means that potential measurement of the IGMF do not automatically imply the discovery of relic primordial fields. An alternative possibility is that the magnetic fields can spread through the IGM at late times by large-scale outflows from magnetized galaxies.

%%%%%%%%%%%%%%%%%%%%%%%%%%%%%%%%%%%%%%%%%
\begin{figure}
\includegraphics[width=\linewidth]{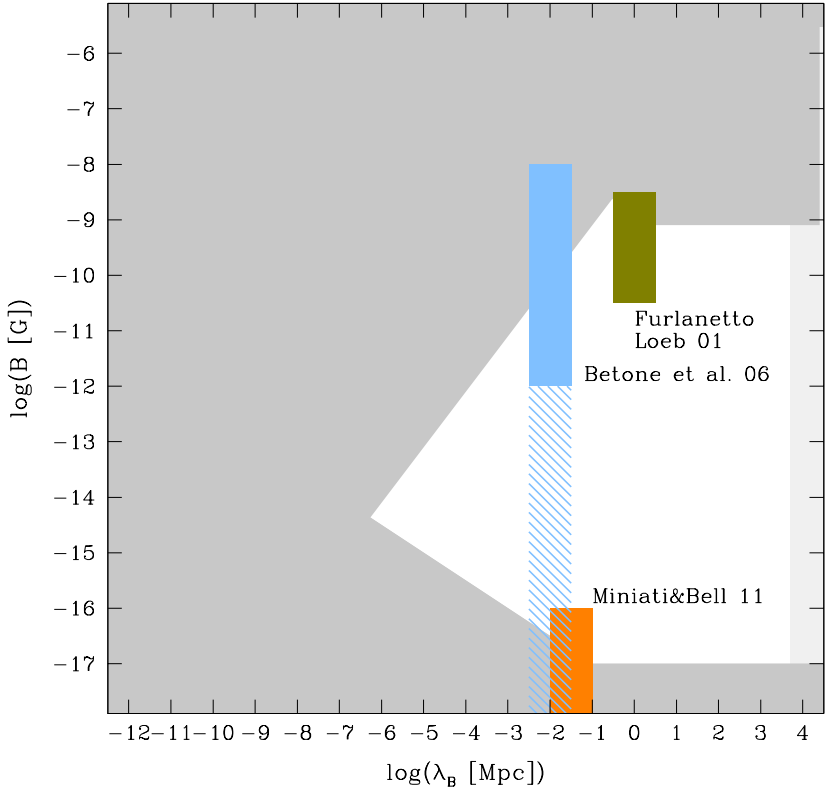}
\caption{Range of IGMF parameters expected in scenarii where IGMF is spread by supernova driven Galactic winds (light-blue shading and hatching) and/or cosmic rays (orange-shading).}
\label{fig:winds}
\end{figure}
%%%%%%%%%%%%%%%%%%%%%%%%%%%%%%%%%%%%%%%%%

The magnetic field might spread into the voids by the outflows from radio galaxies  \citep{Rees87,Daly90,Ensslin97,Kronberg01}. These outflows in the form of jets delivering high-energy particles into large scale radio lobes are a characteristic feature of  radio galaxies \citep{Urry95}. The observed size of the jets reach Mpc scales. \citet{Furlanetto01} discuss the possibility that magnetic fields spread by quasar outflows and find that by  redshift $z\sim 3$ (i.e. at the maximum of the quasar activity) some 5\%-20\% of the volume of the IGM might be "polluted" by  fields from quasar outflows. The strength of the field in the polluted regions would be at the level of $\sim 10\%$ of the thermal energy density of the IGM with temperature $T\sim 10^4$~K, which implies $B\sim 10^{-9}$G. The correlation length of these fields would be of the order of the size of the radio lobes, i.e. in the Mpc range. This range of parameters is shown as olive-shaded region in Fig.~\ref{fig:winds}. In general, the conclusions on the strength of the fields spread by radio galaxies and quasars strongly depend on assumptions about the generation of the outflows by quasars: the duration of the period of activity of the quasar central engine, its energetics, magnetization etc. 

Alternatively, magnetic fields might be spread into the IGM by the galactic winds driven by supernovae \citep{Kronberg99}. Supernova driven winds are most probably responsible for the metal enrichment of the IGM \citep{Aguirre01,Theuns02,Bertone05}. \citet{Bertone06} used a set of prescriptions for the magnetic field content of supernova driven winds, to estimate the volume filling factor of the resulting IGMF, based on numerical simulations of the wind spreading in the IGM \citep{Bertone05}. Their conclusion is that most of the IGM might be polluted with galactic wind fields with  strengths ranging from $10^{-12}$G to $10^{-8}$G. The correlation lengths of the wind-spread fields are expected to be about the size of their footprints in the galaxies, i.e. in the $\lambda_B\sim 1-10$~kpc. This estimates strongly depend not only on the "prescriptions" for the magnetization of the winds \citep{Bertone06}, but also on the "prescriptions" for galactic winds \citep{Bertone05} which are used in numerical simulations. In particular, varying model parameters, opposite results can be obtained: from relatively strong IGMF filling the voids with high volume filling factor  \citep{Bertone06} down to the small fraction of the volume (only about $10^{-2}$) occupied by the wind blown bubbles \citep{Bertone05}.  The range of parameters $B,\lambda_B$ of IGMF spread by galactic winds  is shown in Fig. \ref{fig:winds} by a light blue shaded/hatched region. \citet{Donnert09} further developed the model of \citet{Bertone06} and performed simulations aimed at testing the hypothesis of the origin of magnetic fields in galaxy clusters from galactic outflows. The results of the simulations show that the observed $\mu$G scale fields in the cluster cores can be fully explained by spreading of galactic winds. 

\citet{Miniati11} consider the possibility to generate IGMF by cosmic rays escaping from galaxies during the period of reionization at $z\sim 10$. In absence of strong primordial IGMF, cosmic rays generated by the first supernovae  at the onset of star formation freely stream out of the galaxies and produce an current $\bfj_{\rm CR}$  which is compensated by a return current of low-energy plasma from the IGM. Spatial variations of Ohmic resistivity of the IGM, $\eta\sim T_{\rm IGM}^{-3/2}$ caused by the variations of the temperature of the IGM, $T_{\rm IGM}$, provide a source term in the magnetic induction equation, $\partial \bB/\partial t\sim (\bnabla\eta)\wedge \bfj_{\rm CR}$ wherever the gradient of $\eta$ is misaligned with $\bfj_{\rm CR}$. This source term might lead to magnetic field generation in the IGM which stops as soon as the IGM gets heated by the UV radiation produced by the star formation activity. Soon after reionization, the resistivity drops and the source term in the induction equation disappears. This mechanism can lead to IGMF with strengths in the range $\sim 10^{-18}-10^{-16}$~G and with comoving correlation lengths in the range of typical fluctuations of the IGM temperature, $\lambda_{B}\sim 10-100$~kpc. We show this range of IGMF parameters by the orange shaded region in Fig.~\ref{fig:winds}.

% Constraints (s:constraints_testable)

%%%%%%%%%%%%%%%%%%%%%%%%%%%%%%%%%%%%%%%%%%%%%%%%%%%%%
\section{Constraints on the observationally testable mechanisms of cosmological magnetogenesis}
\label{s:constraints_testable}
%%%%%%%%%%%%%%%%%%%%%%%%%%%%%%%%%%%%%%%%%%%%%%%%%%%%%%

If the magnetic fields present in the IGM are of cosmological origin, understanding of the evolution of the fields from the moment of magnetogenesis until the present (see Section \ref{s:evo}) can be used to obtain information about the physical processes in the Early Universe, based on astronomical observations / limits  of the IGMF at $z=0$. This provides an attractive possibility of a new type of cosmological probe, potentially sensitive to the epochs preceding recombination (CMB decoupling) and Big Bang Nucleosynthesis. 

Processes leading to the generation of magnetic fields in the Early Universe typically produce magnetic fields on short distance scales (Section \ref{s:gen}). Turbulent decay and damping of the fields during subsequent evolution lead to the decrease of the field strength and increase of the correlation scale, where most of the magnetic field power is located. As a result, relic magnetic field which might be present in the IGM today may be relatively weak with  short correlation scale. 

Observational constraints on the IGMF restrict its strength and correlation length (see Section~\ref{s:obs}). If the strength  of the relic field expected in a cosmological magnetogenesis scenario is higher than of the observationally allowed IGMF $(B,\lambda_B)$ plane for a given correlation length, the scenario is not realized in the Universe. If it is lower than the lower limit on $B$ at the given correlation length, the dominant  magnetic field at this scale must come from some other mechanism and the model can not be directly tested by the observations. 
 The most interesting scenarios from the observational point of view are those which predict relic magnetic field parameters consistent with observational bounds on the IGMF. 
 
In this section we review the different models of production of magnetic fields in the Early Universe proposed in Section~\ref{s:gen} in view of their compatibility with observational constraints on IGMF.
  
%%%%%%%%%%%%%%%%%%%%%%%%%%%%%%%%%%%%%%%%%%%%%%%%%%%%%%
\subsection{Inflation}
%%%%%%%%%%%%%%%%%%%%%%%%%%%%%%%%%%%%%%%%%%%%%%%%%%%%%%

As it is discussed in Section~\ref{ssec:infF}, the power spectra of the fields generated during inflation by coupling the electromagnetic field either to curvature or to the inflaton are expected to be blue, with most of the power concentrated on short scales. The correlation length of the fields at the end of inflation is of the order of the size of the cosmological horizon. This means that the power of the fields generated by inflation happening at very high energy scales (in principle, up to the Planck scale) is initially concentrated at very short scales. This power is subject to fast dissipation, so that the final strength of the relic magnetic field at $z=0$ is extremely small.

%%%%%%%%%%%%%%%%%%%%%%%%%%%%%%%%%%%%%%%%%%%%%%%%%%%%%%
\begin{figure}
\includegraphics[width=\linewidth]{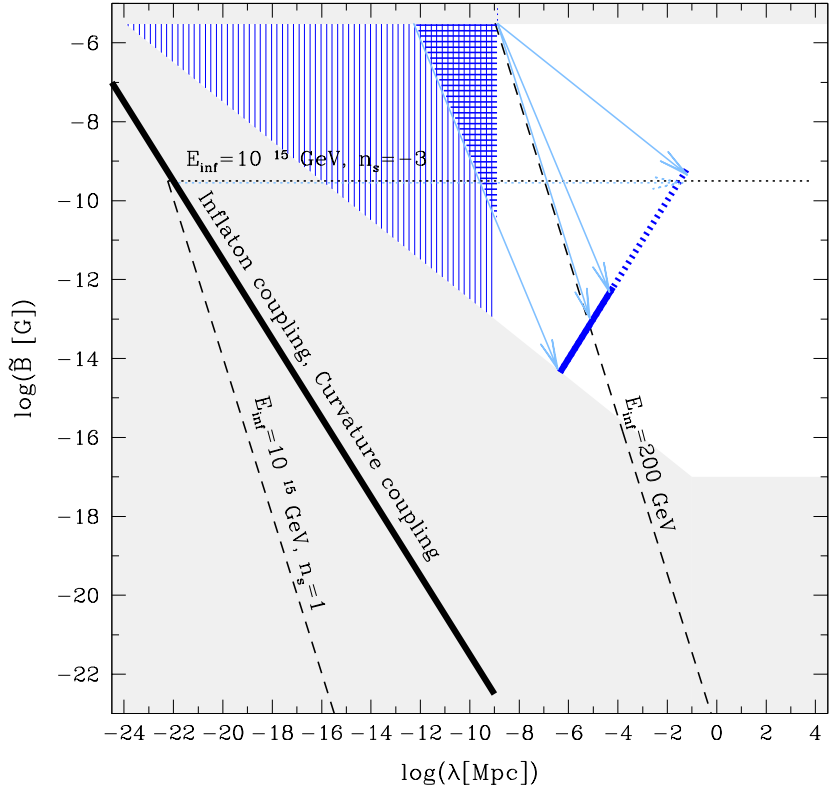}
\caption{Observationally testable region for inflation generated magnetic fields in scenarios with inflaton coupling to the electromagnetic field. The shaded region shows the range of magnetic fields  and comoving correlation lengths for which are excluded by observational bounds on IGMF. Arrows show evolutionary tracks of the field correlation length and field strength (see Section \ref{s:evo}). Dashed lines show the power spectrum of the maximally possible field generated at inflation occurring at different energy scales. }
\label{fig:exclusion_inflation}
\end{figure}
%%%%%%%%%%%%%%%%%%%%%%%%%%%%%%%%%%%%%%%%%%%%%%%%%%%%%%

As discussed in Section \ref{s:gen}, the fields generated during inflation have typically a blue spectrum peaking at the wavenumbers $k_*\sim H_{\rm inf}$ where $H_{\rm inf}$ is the expansion rate. The energy density of magnetic field is $\rho_B\sim (H_{\rm inf}/M_P)^2\rho$, so that the magnetic field is largely sub-equipartition, unless $H_{\rm inf}\sim M_P$. The generically expected relation between $B_*\sim \sqrt{\rho_B}$ and $\lambda_*\sim k_*^{-1}$ is $B_*\sim \lambda_*^{-1}$. This range of expected initial values of magnetic field parameters is shown by a thick black solid line in Fig. \ref{fig:exclusion_inflation}. The spectrum of the field is generically expected to be blue, with the slope $n_s=1$ in the most optimistic case. As an example, we show by the dashed line in Fig. \ref{fig:exclusion_inflation} the initial field spectrum for the case of inflation happening at the Grand Unification energy scale $E_{\rm inf}\sim 10^{15}$~eV.

Evolution of the field strength and correlation length from the inflation till the present day is governed by the turbulent decay and proceeds along one of the tracks shown in Fig. \ref{f4:evolP}. The choice between the three possible tracks depends on the helicity of the field and on the type of turbulence governing the evolution. Evolution of the fields with initial parameters lying on the thick solid line in Fig. \ref{fig:exclusion_inflation} and a blue spectrum could not produce a final field configuration in the observable range (unshaded area in Fig. \ref{fig:exclusion_inflation}) even in the most optimistic case of evolution of maximally helical field via inverse cascade. The only hypothetical possibility to have an observable  field strength today is to generate a field with the scale invariant spectrum $n_s\simeq -3$. In this case, the possible field parameters $\tilde B, \lambda_B$ today 
satisfy the relation (\ref{e4:limit0}) with the field strength reaching $\sim 10^{-9}$~G for the field generated at the Grand Unification scale, see Fig. \ref{fig:exclusion_inflation}. However, as it is discussed in Section \ref{s:gen}, up to now there have been no self-consistent models for the generation of the scale-invariant magnetic field at Inflation. 

Another possibility for an observable field generated at inflation is to assume that the field is significantly amplified  e.g. by dynamo action during reheating in order to become observable in the present Universe. In this case the resulting field energy density could increase up to the equipartition with the plasma energy density $\rho_B\sim \rho$. If this is the case, the initial field strength might be high enough to result in an observable field in the present day Universe. For this to happen, the initial field parameters $\tilde B_*, \lambda_*$ should be in the blue-shaded region in Fig. \ref{fig:exclusion_inflation}. The denser-shaded region shows the allowed range of parameters of the non-helical fields, while the light-shaded region shows the range of possible parameters of the helical fields. 

The lowest possible energy scale for inflation is just above the Electroweak energy scale, see e.g. \cite{German:2001tz}. The correlation length of magnetic fields which might be generated at this low-energy inflation can reach $\lambda\sim 0.01$~pc (see Section \ref{ssec:infF}).  The maximal possible strength of the initial magnetic field,  $B(k)\sim \sqrt{d\rho_B/d\log k} <\sqrt{\rho}$, from inflation for a typical spectrum, $n_s =1$ for lowest possible scale $E_{\rm inf}\sim 200$GeV  is shown in Fig.~\ref{fig:exclusion_inflation} as dashed line. 

The decay of these magnetic fields in the course of cosmological evolution can proceed in three possible ways. First, for the case of incompressible turbulence, no amplification of the field at large scales occurs, the decay just removes power from the scales smaller than the scale of the largest processed eddies. The evolutionary track of the field then follows the initial power spectrum (dashed line in Fig. \ref{fig:exclusion_inflation} up to the point where the time scale of the processing of the eddies is comparable to the Hubble time. The locus of the present-day strengths and correlation lengths of magnetic fields is along the blue thick line in Fig.~\ref{fig:exclusion_inflation}. 

Alternatively, the field excites compressible turbulence right after generation. In this case, the field strength at large scales is continuously driven into equipartition with fluid turbulence. This leads to a mild amplification of the field on large scales so that the evolutionary track of the field strength -- correlation length follows the $B\sim \lambda^{-1.5}$ path discussed in Section \ref{s:evo}. This results in somewhat stronger relic field and somewhat larger final correlation length. The locus of the possible present day parameters of the relic field is still along the inclined thick blue line in Fig.~\ref{fig:exclusion_inflation}, but the evolutionary track is along the two less steep arrows (not parallel to the black dashed lines).

Finally, if the field is helical, as suggested e.g. by~\cite{Durrer:2010mq}, helicity conservation forces it to evolve along the significantly flatter path, $B\sim \lambda^{-0.5}$.  Again, evolution ends on the thick blue line fixed by the relation $\la_B/v_A =t_0$.

The explanation of these different evolution paths is given in Section~\ref{s:evo}.

Tracing back the evolution of $B,\lambda_B$ from the observationally allowed range at $z=0$ (shown by the thick blue line in the diagram \ref{fig:exclusion_inflation}), we find the  range initial  values $B,\lambda_B$  for observationally testable inflationary magnetogenesis models. This range is shown by the blue hatched region in 
Fig.~\ref{fig:exclusion_inflation}. Non-helical magnetic fields from inflation fall in the observationally allowed range only if the energy scale of inflation is below $E_{\rm inf}\sim 10^5$~GeV. This limit is removed for helical fields.

%%%%%%%%%%%%%%%%%%%%%%%%%%%%%%%%%%%%%%%%%%%%%%%%%%%%%%
\subsection{The electroweak phase transition}
%%%%%%%%%%%%%%%%%%%%%%%%%%%%%%%%%%%%%%%%%%%%%%%%%%%%%%

\begin{figure}
\includegraphics[width=\linewidth]{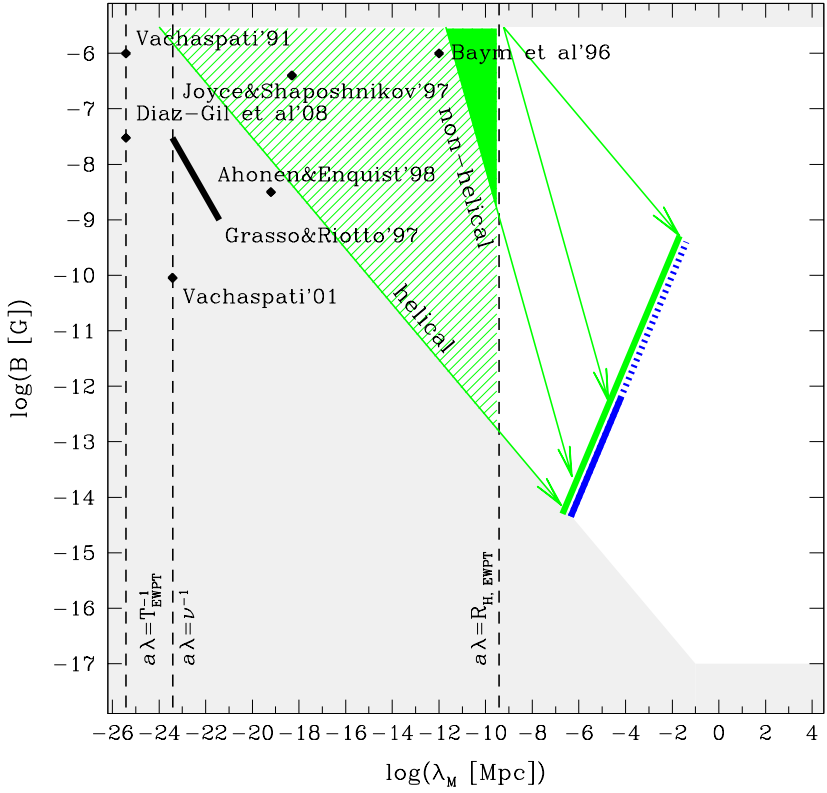}
\caption{Observationally testable region for the models of non-helical (green colored region) and helical (green hatched region) magnetic field generation during the electroweak phase transition. Arrows show the evolution paths of non-helical (compressible turbulence) and helical magnetic fields. Points with labels show various model predictions. For comparison, the blue thick line shows the range of relic magnetic field strengths and correlation length from testable inflationary magnetogenesis scenarios.}
\label{fig:exclusion_EWPT}
\end{figure}

In a similar way one can delimit the range of magnetic field parameters $(B,\la_B)$ for observationally allowed models of electroweak magnetogenesis. This range can be found by back-tracing the evolutionary tracks of magnetic fields in the $B,\lambda_B$ plane as it is shown in Fig.~\ref{fig:exclusion_EWPT}. The main difference with the inflationary scanarios is that many models discuss the production of helical fields for which the field strength and correlation length evolve along the line $B\sim \lambda^{-1}$ (see Section~\ref{s:evo}). This results in slower decay and potentially larger correlation length at $z=0$, provided that the initial correlation length of the fields reaches a scale of the order of the cosmological horizon at the electroweak phase transition. 

%%%%%%%%%%%%%%%%%%%%%%%%%%%%%%%%%%%%%%%%%%%%%%%%%%%%%%
%%%%%%%%%%%%%%%%%%%%%%%%%%%%%%%%%%%%%%%%%%%%%%%%%%%%%%

The scenarios considered by \citet{vachaspati91} and \citet{enquist93} result in fields outside the observationally testable region. Order-of magnitude estimates for the field strength which can be produced via such process is $B\sim m_{W}^{2}$, with typical correlation length $\lambda_B\le m_{W}^{-1}$, where $m_{W}$ is the W-boson mass. These initial values of $B,\lambda_B$ are shown in the diagram of Fig. \ref{fig:exclusion_EWPT}.  However, taking into account the order-of-magnitude nature of the estimate of the boundaries of the region of observationally testable models once can still suggest that the scenario of \citet{vachaspati91} might result in observable magnetic fields in the IGM, roughly at the level of the currently existing lower bound on the field strength and correlation length. This requires that the fields are produced immediately in the maximally helical configuration, otherwise, initial faster field decay toward the maximally helical configuration reduces the field strength to the range much below the level consistent with the observational lower bounds.

The scenario discussed by \citet{vachaspati91} and \citet{enquist93}  has a problem in that the initial distance scale on which the magnetic field is generated is shorter than  the scales of both viscous and Ohmic dissipation in the primordial plasma soon after the epoch of Electroweak phase transition. Indeed, the mean free path of particles at the moment of the Electroweak phase transition is $a\lambda_{\rm mfp}\simeq g^{-4}T^{-1}$ where $g\sim 0.1- 0.3$ is the coupling constant. The effective viscosity in the gas of relativistic is related to the mean free path of the least coupled particle is $\tilde\nu=\lambda_{\rm mfp}/5$,  see \cite{Weinberg71} and Appendix~\ref{a:visc}, so that the viscous dissipation scale $\lambda_d\simeq \tilde\nu/v$ is close to the particle mean free path for the characteristic plasma/fluid velocities $v$. Assuming that the characteristic velocity scale is $v \sim v_A$, one finds that in the case of magnetic field with strength $B\sim m_W^2$, $v\sim v_A\sim 1$. Furthermore, at these energies the magnetic diffusivity is $\si^{-1}\simeq  g^{-2}T^{-1}$, is of the same order of magnitude as the viscosity and the field is also damped by magnetic diffusion, see Section~\ref{s:evo}. However, on these small scales the MHD equations might be not applicable right after magnetic field generation. Instead, the evolution of the coupled magnetic field-plasma system on the distance scales comparable to particle mean free path has to be modeled using the Boltzmann equation for the particle distributions.

The scenario  of \cite{Diaz-Gil08a,diaz-gil08} considers  topologically non-trivial configurations of the inflaton and Higgs fields at the end of the post-inflation preheating as a source of magnetic fields. This operates in the temperature range of the electroweak phase transition. The estimates for the characteristic correlation scale and the magnetic field strength in \cite{Diaz-Gil08a,diaz-gil08} are similar to those of  \citet{vachaspati91}, with a correction for the fact that more elaborated numerical calculations (rather than qualitative arguments) result in magnetic field energy density which is two orders below equipartition with the energy density of the Universe at the temperature $T_*\sim 100$~GeV. The expected initial parameters for the magnetic field in this scenario are also shown in 
Fig.~\ref{fig:exclusion_EWPT}. With its much lower initial field strength, the fields generated in this model are clearly outside the observationally testable range.

A larger initial correlation length of magnetic field is expected in the models of  \cite{grasso97} and \cite{Ahonen98} in which the electroweak phase transition is assumed to be first order and the generation of magnetic field  is related to the  collision of bubbles of new phase which leads to macroscopic classical vortex-like configurations of the gauge fields forming at the Electroweak phase transition. \cite{grasso97}  found that the initial correlation length of magnetic field might be larger than $\lambda\sim T_*^{-1}$, while the field strength can still be somewhat below the equipartition. This larger correlation length (a numerical estimate $\lambda\sim 10^4T_*^{-1}$ in the case of a first order phase transition and $\lambda\sim 10T_*^{-1}$ for the second order phase transition is given in the paper) relaxes the problem of immediate dissipation of the magnetic field energy via plasma viscosity, which operates at a smaller scale. From Fig.~\ref{fig:exclusion_EWPT} we see that the strength and initial correlation length of magnetic field estimated by \cite{grasso97} are also outside the region of directly testable models. 

\citet{Vachaspati:2001nb} has considered the generation of helical magnetic fields during the electroweak phase transition, see also \cite{Semikoz:2009ye,Semikoz:2012ka}. A generation of non-zero helicity is related to the changes of Chern-Simons number for the SU(2)$\times$U(1) gauge field which accompanies the generation of non-zero baryon number in electroweak baryogenesis. In this setting the helicity of the magnetic field generated simultaneously with baryons is directly related to the baryon number, $\mathpzc{h}\simeq 10^2n_B$. Taking into account that the field is generated with a characteristic scale $L\sim (e^2 T_*)^{-1}$, one finds that the field strength is roughly in equipartition with the energy density of baryons, which is ${\eta_b} \rho_{\rm rad}$, where ${\eta_b}\simeq 10^{-8.5}$ is the baryon number of the Universe. This means that the magnetic field strength at generation is by a factor of $\sqrt{\eta_b}\sim 10^{-4}$ below equipartition at the moment of generation. The initial values of the magnetic field strength and correlation length in the scenario of \citet{Vachaspati:2001nb}, shown in Fig.~\ref{fig:exclusion_EWPT} lie also outside the testable range. 

Similarly to the scenario of \citet{vachaspati91} and \citet{enquist93}, fields generated in the scenario of \citet{Vachaspati:2001nb} are initially on the distance scales close to the viscous damping scale. A potentially significant difference between the two scenarios might be in the fact that the field strength in the model of \citet{Vachaspati:2001nb} is strongly sub-equipartition. The Alfv\'en velocity is, therefore strongly non-relativistic $v_A\sim \sqrt{B}\sim 10^{-4}$. Comparing the viscous damping scale $\lambda_d\simeq v\nu\simeq v\lambda_{\rm mfp}\sim 10^2VT_*^{-1}$ with the field correlation length $\lambda_B\sim 10^2T_*^{-1}$ one can find that if the velocity scale of the dissipating modes is $v\sim v_A$, the field energy is not directly damped by the viscosity. However, it can be damped my magnetic diffusion. 

A much larger initial correlation length of the magnetic field $\lambda\sim 10^7/T_*$ is possible in the model of \citet{Joyce:1997uy}, who consider the production of helical magnetic field due to an imbalance of right- and left-handed electrons and positrons. This imbalance leads to a non-zero chemical potential $\mu$ for right-handed positrons. The chemical potential results in a dynamo-like term in the induction equation for the magnetic field. This term is responsible for the field amplification at the  scale $\lambda\sim \mu^{-1}\gg T_*^{-1}$. The dynamics of the field generation and evolution in this model in the temperature range $T\lesssim T_*$ and at higher temperatures $T\lesssim 80$~TeV has recently been considered by \citet{Boyarsky12,Boyarsky12a}  and \citet{Dvornikov12}. The presence of relatively strong hypermagnetic fields at the moment of the electroweak phase transition can modify the dynamics of the phase transition itself making it first order \citep{Elmfors98} and can further modifying the magnetic induction equation describing the evolution of the magnetic field \citep{Semikoz08}.

\citet{Baym96} consider much larger bubbles of new phase in a first-order electroweak phase transition, with sizes reaching $\sim (10^{-3}-10^{-2})\ell_{H}$. Developed MHD turbulence with  largest eddies of the size of these bubbles leads to equipartition between the energy density of magnetic fields and the energy density of fluid motions. Taking into account that the bubble walls can reach mildly relativistic speeds, the energy density of fluid motions can, in fact, be comparable to the energy density of the Universe. The magnetic fields are, therefore, also amplified to an  energy density comparable to the energy density of the Universe. The initial magnetic field parameters in the scenario of \citet{Baym96}, shown in Fig.~\ref{fig:exclusion_EWPT} are clearly well inside the observationally testable region. In fact, if a measurement of IGMF today with parameters $(B,\lambda_B)$ which, after back tracing to the Early Universe will imply initial magnetic field strength close to equipartition with the rest of matter in the Universe at scales close to the Electroweak horizon scale would provide a strong argument in favor of electroweak magnetogenesis and in favor of a first order electroweak phase transition.  

%%%%%%%%%%%%%%%%%%%%%%%%%%%%%%%%%%%%%%%%%%%%%%%%%%%%%%
\subsection{QCD phase transition}
%%%%%%%%%%%%%%%%%%%%%%%%%%%%%%%%%%%%%%%%%%%%%%%%%%%%%%

\begin{figure}
\includegraphics[width=\linewidth]{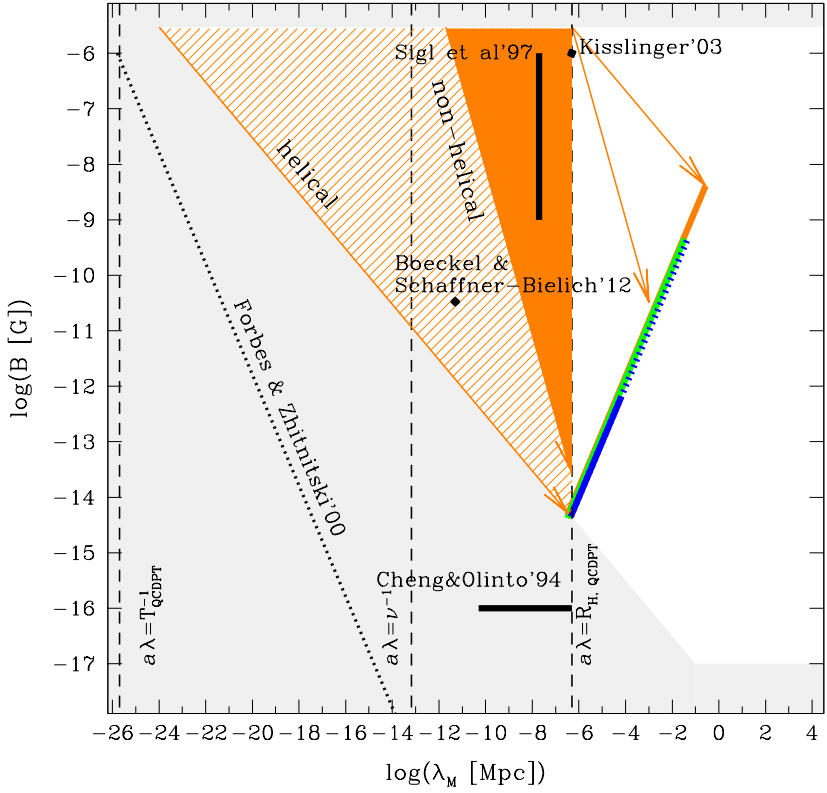}
\caption{The observationally testable region for models of non-helical (orange colored region) and helical (orange hatched region) magnetic fields generated during the QCD phase transition. Arrows show the evolution paths of non-helical and helical magnetic fields. Points with labels show various model predictions. }
\label{fig:exclusion_QCDPT}
\end{figure}

For the QCD phase transition, the range of possible initial correlation lengths extends to larger values up to the Hubble radius at the temperature $T_*\sim 100$~MeV. This somewhat broadens the range of possible final field strengths and correlation lengths.  However, similar to the case of the electroweak phase transition, most of the models consider field generation at very short scales, which lead to initial field parameters outside the observationally testable range, see Fig. \ref{fig:exclusion_QCDPT}. 

%%%%%%%%%%%%%%%%%%%%%%%%%%%%%%%%%%%%%%%%%%%%%%%%%%%%%%
%%%%%%%%%%%%%%%%%%%%%%%%%%%%%%%%%%%%%%%%%%%%%%%%%%%%%%

In the model of \cite{forbes00} the field is generated at the domain walls of the bubbles of a first-order QCD phase transition. The initial correlation length of the field is $a\lambda_* \sim T_*^{-1}$ and its strength is close to equipartition, $B\sim T_*^2$. The domain walls subsequently coalesce increasing the correlation length and decreasing the field strength as $\tilde B\sim \lambda^{-1}$. Since the exact value of the final correlation length is not known, the estimates of the field strength and correlation length in this model can only be represented in the form of a line in the diagram of Fig.~\ref{fig:exclusion_QCDPT}. This line lies outside the region in which the magnetic field might leave a detectable imprint in the IGM today. 

\cite{kissliger03} has proposed that, contrary to the assumption of \cite{forbes00}, the magnetic field strength for the large domain walls might not be suppressed by a factor $\lambda^{-1}T_*$. He assumes that a field strength $B\sim T_*^2$ right at the scale of cosmological horizon at the moment of QCD phase transition (no special motivation for this assumption is provided though, although amplification of the field up to equipartition by turbulence, similar to the scenario discussed by \citet{Baym96} in the context of Electroweak phase transition is, in principle, possible).This is clearly the most optimistic scenario which is logically possible. In this  setting, the field can produce measurable effects on the polarization of the CMB and also induce a gravitational wave background. The model predictions for the field strength and correlation length for the fields obtained by  \cite{kissliger03} fall  within the range of observationally testable models, see Fig.~\ref{fig:exclusion_QCDPT}. The effects of helicity on the evolution of magnetic field in this scenario was discussed by \citet{Tevzadze:2012kk}.

In the scenario of \cite{boeckel12} the magnetic field is generated together with the baryon asymmetry of the Universe in result of the period of a "little inflation" accompanying a first order QCD phase transition. The energy density of the magnetic field generated in this way can reach the equipartition with the baryon energy density, so that $B$ can be estimated as $B\sim \sqrt{2 {\eta_b}\rho_{\rm rad}}$. The characteristic correlation length is determined by the diffusion  of baryons across the walls of bubbles of new phase, occurring at a first order QCD phase transition. It is estimated to be in the range of $\sim 10$~cm physical length which is about $10^{-5}\ell_{H,{\rm QCD}}$ at the moment of the QCD phase transition. These reference values of the field strength and correlation length are shown in Fig. \ref{fig:exclusion_QCDPT}. The model prediction lies within the observationally testable region, if the field is helical.

\citet{Cheng94} also consider the generation of magnetic fields by the propagating domain walls separating the two phases. The magnetic field is produced by an electrical current arising as a result of uncompensated charges on different sides of the wall. The field strength produced via this mechanism is significantly below equipartition with either baryon or total energy density and falls below the observationally testable range, see 
Fig.~\ref{fig:exclusion_QCDPT}. Subsequent amplification of the fields by MHD instabilities and turbulence was considered by \citet{Sigl97} who found that the fields can, in principle be increased up to a strength approaching equipartiton with the thermal energy of the plasma. This moves  the initial field parameters in the range of observationally testable models, see Fig.~\ref{fig:exclusion_QCDPT}.

%Conclusion (s:con)
%%%%%%%%%%%%%%%%%%%%%%%%%%%%%%%%%%%%%%%%%%%%%%%%%%%%%%
\section{Conclusions}
\label{s:con}
%%%%%%%%%%%%%%%%%%%%%%%%%%%%%%%%%%%%%%%%%%%%%%%%%%%%%%

In this work we have discussed the generation and evolution of cosmological magnetic fields (IGFM) and we have discussed their possible observation. 

Inflation only generates magnetic fields if conformal invariance is broken. But also then it must either happen at relatively low energy, $E_{\rm inf}\lsim 10^5$GeV or generate helical fields in order to yield relic fields which are observable today. Otherwise the correlation length is too short and the fields are damped by viscosity during the subsequent MHD evolution below the amplitude required by observational lower bounds. If the magnetic field generated
during inflation is not amplified subsequently to near equipartition, we expect a very low initial magnetic field amplitude determined by $\rho_B/\rho \simeq (H_{\rm inf}/M_P)^2$ which is well below the observable range for all inflation scales.

We have also considered magnetic field generation by cosmological phase transitions, in particular we have discussed the electroweak and the QCD transition.
There we have seen that if the transition is of first order and proceeds via bubble nucleation, the correlation length can be sufficiently large and the amplitude can be sufficiently high for the fields to be observable today. 

This is very interesting as standard model physics predicts a cross-over for both, the QCD and the electroweak transition. Therefore, relic magnetic fields from the electroweak phase transition would be a signal of physics beyond the standard model. A first order QCD phase transition is possible only if the chemical potential of the neutrinos is large, see \cite{Schwarz:2009ii}. In this case, we expect strong helicity in the leptonic sector which modifies the evolution of magnetic fields via a significant contribution from the electroweak anomaly. This is expected to  induce magnetic field helicity as outlined by~\cite{Boyarsky12a}.

Observationally accessible magnetic fields could be generated in the standard model electroweak phase transition if the correlation length of the field is much larger than the inverse temperature scale, as proposed by \citet{Joyce:1997uy}.

%%%%%%%%%%%%%%%%%%%%%%%%%%%%%%%%%%%%%%%%%%%%%%%%%%%%%%
\begin{figure}
\includegraphics[width=\linewidth]{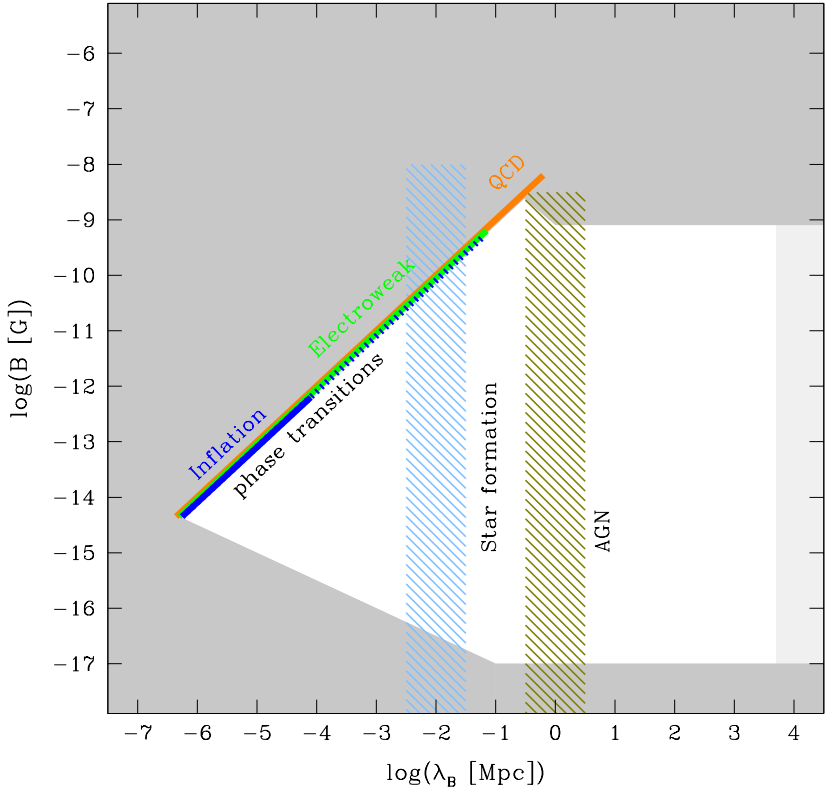}
\caption{A summary of observational constraints and model predictions for IGMF. Grey shaded region shows the range of parameters excluded by observations and theroetical arguments presented in Section \ref{s:obs}. Solid lines mark the locus of possible present day strength and correlation length of relic magnetic fields produced at the phase transition in the Early Universe, as discussed in Seciton \ref{s:evo}. Green and blue hatched regions show possible ranges of IGMF spread by the outflows from galaxies (Section \ref{s:winds}). }
\label{fig:summary}
\end{figure}
%%%%%%%%%%%%%%%%%%%%%%%%%%%%%%%%%%%%%%%%%%%%%%%%%%%%%%

A summary of the existing observational constraints and different model predictions for the  IGMF is shown in 
Fig.~\ref{fig:summary}. This figure allows us to assess possible future measurements of the IGMF. Unfortunately cosmologically produced fields and field ejected by the galaxies span the same range of  field strengths, from the existing lower bounds from $\gamma$-ray  observations ($\sim 10^{-17}$~G at large distance scales) up to the upper bounds from radio observations ($\sim 10^{-9}$~G  at large distance scales). Measurement of the field strength alone cannot provide a clue on the origin of the IGMF. In order to distinguish between early (cosmological) and late (galaxy formation) origin of IGMF, a combined measurement of the field strength and the correlation length is necessary. The detection of fields with correlation length shorter than $\sim$ kpc  favors a cosmological origin of the IGMF. If the IGMF correlation length is larger than $\sim$ kpc, measurement of its strength significantly below $\sim 10^{-9}$~G seems to imply that the field is produced by  galactic outflows, rather than by the processes in the Early Universe. 

It is, therefore, crucially important to develop observational methods which allow a measurement of not only field strength (or of a particular combination of field strength and correlation length), but also separate measurement of the field strength and the correlation length. 

Another possibility to distinguish cosmological IGMF from the fields spread by galactic winds is to search for turbulence in the voids of LSS. Indeed, cosmologically produced IGMF decays by transferring its power to turbulent motions of the plasma in the Universe. The latest episode of the magnetic field driven turbulence might have happened recently, after reionization of the Universe by the star formation activity. The strength and correlation length of the relic cosmological magnetic fields today is such that  the time scale of turnover of the eddies of the size comparable to the correlation length is just about the Hubble time,  $\la_B/v_A =t_0$ (this is the relation which defines the thick colored line in Fig.~\ref{fig:summary}).  Therefore,  the IGM in the voids of the LSS might be turbulent today if cosmological magnetic fields are present.  A search of turbulent plasma motions in voids may therefore provide an alternative way to the discovery of primordial magnetic fields.

\begin{acknowledgements}
We thank Chiara Caprini, Alexey Boyarski, Karsten Jedamzik, Oleg Ruchayskiy, Misha Shaposhnikov and Kandu Subramanian for discussions and useful suggestions.
This work is supported by the Swiss National Science Foundation.
\end{acknowledgements}

%Appendix (a:vosc)

\vspace{1cm}

\noindent{\Large\bf Appendix} \vspace{0.2cm}\\

\appendix
\section{Viscosities and Reynolds numbers}\label{a:visc}

In this work the Reynolds number of a given scale which is inversely propositional to the viscosity plays an important role since it defines the time when the velocity field and with it the magnetic field on the given scale is damped into heat. We closely follow the treatment 
of~\cite{Caprini:2009yp}.

\subsection{Kinematic viscosity}
The kinematic viscosity is given by
\be
\nu=\frac{\eta}{\rho+p},
\ee
where 
$\eta$ is the shear viscosity. The kinematic viscosity 
characterizes the diffusion of transverse momentum due to collisions, and is given roughly by the mean free path $\ell_{\rm mfp}$ of the particles. In this Appendix $\ell_{\rm mfp}$ is the physical mean free path, while $\la_{\rm mfp}$ appearing in the main text is the comoving mean free path. The relation is simply $\ell_{\rm mfp} =a\la_{\rm mfp}$.

A more precise expression for the shear viscosity is, see~\cite{Weinberg71}
\be
\eta=\frac{4}{15} \frac{\pi^2}{30} g_* T^4 \ \ell_{\rm mfp} \qquad \mbox{so~that}\qquad \nu=\frac{\ell_{\rm mfp}}{5}\,.
\ee
The largest viscosity comes from the weakest interactions. However, non-interacting particles do not contribute to the viscosity. For this reason simple analytical approximations to the viscosity have unphysical jumps whenever a species decouples from the plasma.

Estimates from kinetic theory show that 
the shear viscosity of highly relativistic particles, $T\gg m$, behaves as (to leading-log accuracy): 
\be
{\eta}= C \frac{T^3}{g^4 \log g^{-1}},
 \ee
where  $g$ is the appropriate coupling constant (depending on the temperature and the length scale at which one wants to compute the Reynolds number) and 
$C$ is a numerical coefficient that can only be obtained from a detailed analysis.

At temperatures larger than the electroweak phase transition, neutrino interactions are not  suppressed. The shear viscosity is dominated by right handed lepton transport and is given by \citep{Arnold00}
\be
{\eta} \approx \left(\frac{5}{2}\right)^3\zeta(5)^2\left(\frac{12}{\pi}\right)^5\frac{3/2}{9\pi^2+224 (5+1/2)} \frac{T^3}{g'^4\log g'^{-1}}
\ee
where $g'$ is the hypercharge coupling. This leads to
\be \label{e:nuHi}
\nu ( T\gtrsim 100 \mbox{ GeV}) \approx  \frac{21.6}{T}\,.
\ee

After electroweak symmetry breaking, neutrino interactions are suppressed by a factor $(T/M_W)^4$. In this regime, neutrinos have the longest mean free path and dominate the viscosity. We use \cite{Heckler93}
\be
\ell_{\rm mfp}\approx (3 G_F^2 T^5)^{-1}\,,
\ee
leading to 
\be\label{e:nu100+}
\nu ( T\lesssim 100 \mbox{ GeV}) \approx 4.9 \times 10^8 \ \frac{\mbox{GeV}^4}{T^5}\,.
\ee
At temperatures smaller than $100$ MeV, after the QCD phase transition, the 
remaining relativistic particles in the cosmic plasma are electron/positrons, neutrinos and photons and  the neutrino mean free path increases to
\be
\ell_{\rm mfp}\approx \frac{10}{9} (G_F^2 T^5)^{-1}
\ee
such that
\be\label{e:nu100-}
\nu ( T\lesssim 100 \mbox{ MeV}) \approx 1.6 \times 10^9 \ \frac{\mbox{GeV}^4}{T^5}\,.
\ee
The neutrino mean free path determines the viscosity until neutrinos
decouple at $T\sim 1.4$ MeV, after which photons take over. Below 1MeV, when electrons and positrons annihilate and the remaining electrons become non-relativistic, the viscosity can be approximated by
\be
\nu(T<0.5 \mbox{MeV}) \approx (\si_T n_e)^{-1} \simeq 0.5\times 10^{-22} \mbox{ GeV}^{-1}
\left(\frac{ \mbox{ GeV}}{T}\right)^3 \,.
\ee

After neutrino decoupling, the viscosity drops by about 30 orders of magnitude and the Reynolds number increases correspondingly. Therefore, all scales on which turbulence is maintained until $T\sim 1$MeV will the remain turbulent until decoupling, $T\sim 1$eV.
  
  But even if turbulence is lost before neutrino decoupling, as long as the magnetic field survives, it will become turbulent again after $T\sim 1$MeV and we expect equipartition between the magnetic field and the velocity field to be re-established.

Summing up all the results we find
\be
\nu(T) \approx \left\{ \begin{array}{ll}
21.6\mbox{GeV}^{-1}\left(\frac{ \mbox{ GeV}}{T}\right)& \mbox{if } ~  T\gtrsim 100 \mbox{GeV}\\
4.9 \times 10^8 \mbox{GeV}^{-1} \left(\frac{ \mbox{ GeV}}{T}\right)^5  & \mbox{if } ~  100 \mbox{GeV}> T\gtrsim 100 \mbox{ MeV}\\
1.6 \times 10^9 \mbox{GeV}^{-1} \left(\frac{ \mbox{ GeV}}{T}\right)^5 & \mbox{if } ~  100 \mbox{MeV}> T\gtrsim 1 \mbox{MeV}\\
0.5\times 10^{-22} \mbox{GeV}^{-1}
\left(\frac{ \mbox{GeV}}{T}\right)^3 & \mbox{if } ~  0.5\mbox{ MeV}> T\gtrsim 1 \mbox{eV}
 \,. \end{array} \right.
\ee
 The unphysical jumps come from regions where our approximations are invalid. Nevertheless, at neutrino decoupling viscosity is significantly reduced and turbulence resumes on the relevant scales. This is different after matter and radiation equality since then the Alfv\'en speed decays
and the coupling of the magnetic field to the velocity field soon becomes negligible.

\begin{figure}[htb!]
\begin{center}
\includegraphics[width=10cm]{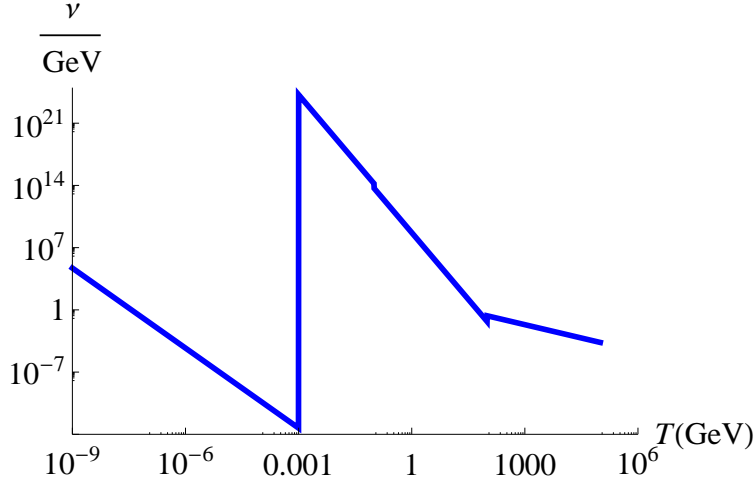}
\caption{\label{f:visc} \small Evolution of the kinematic viscosity $\nu(T)$ as a function of temperature for $T>1$eV. The unphysical discontinuities and kinks come from our crude approximation.}
\end{center}
\end{figure}

The evolution of $\nu$ with temperature is plotted in Fig.~\ref{f:visc}.

%%%%%%%%%%%%%%%%%%%%%%%%%%%%%%%%%%%%%%%%%%%%%%%%%%%%%%%%%%
\subsection{Magnetic diffusivity}
%%%%%%%%%%%%%%%%%%%%%%%%%%%%%%%%%%%%%%%%%%%%%%%%%%%%%%%%%%

Here we derive expressions for the
magnetic diffusivity also called resistivity for relativistic electrons
in the cosmic plasma with temperatures $1$ MeV $< T< 100$ GeV. Again, we follow the treatment of~\cite{Caprini:2009yp}.

To determine the magnetic diffusivity, we derive an expression for the 
conductivity $\si(T)$, which is the inverse of the diffusivity.  The Lorentz force acting on an electron is
$$m_e\frac{du^\mu}{d\tau} = eF^{\mu\nu}u_\nu \,.$$ 
If we average this equation over a fluid element containing many electrons, the magnetic field
term is sub-dominant. Even though the electrons are highly relativistic, the average fluid velocity
is small. Furthermore $\gamma =1/\sqrt{1-v_e^2} \simeq T/m_e$ is nearly constant and we may
neglect the contribution $d\ga/d\tau$ from $du^i/d\tau = d(\ga v^i)/d\tau$ above. 
With $d\tau = \ga^{-1}dt = (m_e/T)dt $, this yields the following equation for
the mean velocity of the electron fluid: 
$$ \frac{d\bv}{dt} = \frac{e}{T}\,{\bf E} \, .$$
If we denote the collision time for the electrons by $t_c$, they can 
acquire velocities of the order $\bv \simeq \frac{e}{T}\,{\bf E}\,t_c$ 
between successive collisions. Hence the current is 
$${\bf J} \simeq en_e{\bf v} \simeq t_c\frac{e^2n_e}{T}{\bf E}\equiv 
\si {\bf E}$$ 
so that the conductivity  becomes
$$ \si =  t_c\frac{e^2n_e}{T} \,. $$
We now derive an estimate for $t_c$ from Coulomb interactions.
For a strong collision between the electron and another charged particle we
need an impact parameter $b$ such that $e^2/b > E_e \simeq T$. Hence the
cross section becomes $\si_t \sim \pi b^2 \simeq \pi e^4/ T^2$ (this simple 
argumentation neglects the Coulomb logarithms which enhance the cross section
by $\ln(1/\al_{\min})$ where $\al_{\min}$ is the minimal deflection 
angle~(see \citealt{LL6})). With $v_e=1$ the time between collisions is therefore
$ t_c= 1/(\si_tn_e) \simeq T^2/(\pi  e^4n_e )$ and
\be\label{e:si}
\si \simeq \frac{T}{\pi e^2} \,. 
\ee
Note that this result is independent of the electron density. This is 
physically sensible as $n_e$ enhances the current on the one hand but it 
reduces in the same way the collision time.

With (\ref{e:si}) we obtain for the magnetic diffusivity
\be\label{e:mu}
\frac{1}{\si} \simeq \frac{e^2(T)}{4T} \simeq 
\frac{10^{-1}}{T}  - \frac{10^{-2}}{T} \, .
 \ee
 The first value applies close to $T\sim 100$GeV, where $\al=e^2/4\pi \sim 0.1$, while the second value corresponds to low energies, $T\sim 1$MeV. For non-relativistic electrons we obtain the standard result for the conductivity by simply replacing $T$ by the electron mass and multiplication by $v^3 \simeq (m_e/T)^{3/2}$ so that~\citep{Spitzer:1978}
\be\label{e:mulow}
\frac{1}{\sigma} \simeq \frac{e^2m_e^{1/2}}{T^{3/2}} \,.  
 \ee
   
 \subsection{Reynolds numbers and Prandl number}
The kinematic Reynolds number  is given by
\be
{\rm R_k}(T) =  \frac{v_K\la_K}{\tilde\nu(T)} \,,
\ee
where $\nu$ is the  kinematic viscosity, $v_K = \sqrt{k_K^3P_v(k_K)}/(2\pi^2)$ is the mean velocity which is roughly the velocity at the integral scale $\la_K =2\pi/k_K$, and $\tilde\nu = \nu/a$, see Section~\ref{s:evo}.

Correspondingly, the magnetic reynolds number is defined by
\be
{\rm R_m}(T) =  \frac{v_A\la_B}{1/\tilde\si(T)} \,,
\ee

Inserting the resistivity from Eq.~(\ref{e:mu}) and the kinematic viscosity from Eqs.~(\ref{e:nu100+}) or (\ref{e:nu100-}), assuming equipartition so that $v_A=v_K$ and $\la_B=\la_K$,
we obtain for the Prandl number
\be\label{e:Papp}
{\rm P_m}  \equiv \frac{{\rm R_m}(k,T)}{{\rm R_k}(k,T)} = \nu(T)\sigma(T)\simeq 10^{12}\left(\frac{\rm GeV}{T}\right)^4 \, .
\ee
This number is  larger than $1$ for all temperatures $1$ MeV$<T\ll 100$ GeV 
where the derivation applies. 

The non-linearities in the Euler and induction equation are stronger than the damping term whenever the Reynolds numbers are larger than unity. In this  regime  MHD turbulence develops.

\subsection{The Prandl number at very high energy}

Finally let us consider the situation at very high temperature assuming that all particle interactions are
 given by the same coupling strength $g^2$ and all particles are relativistic and in thermal equilibrium. This approximation is roughly valid above the electroweak scale.
 (We neglect strong interactions in this picture.)  The cross section then is of the order of $\si_c \simeq g^4T^{-2}$ and
 \be
  t_c=\la_{\rm mfp} = (\si_c n)^{-1} \simeq \frac{1}{g^4 T} \simeq \nu\,, \qquad  T> 100\mbox{GeV}\,.
\ee
This qualitatively reproduces Eq.~(\ref{e:nuHi}). For the conductivity we have with the same approximations
\be
\si = t_c\frac{g^2n}{T} \simeq \frac{T}{g^2} \,,  \qquad  T> 100\mbox{GeV}\,.
\ee
In this case the Prandl number becomes
\be
P \simeq g^{-2}  \simeq 10 \,, \qquad  T> 100\mbox{GeV}\,.
\ee

 It will be important for our discussions that at the electroweak phase transition $T\sim 100$GeV, both, 
 the kinetic viscosity and  the magnetic diffusivity are actually of the same order. At significantly
 lower temperatures, the magnetic diffusivity is always much smaller than the kinetic viscosity.
 This is due to the fact that the kinetic viscosity is governed by the most weakly interacting particles, the neutrinos while the conductivity is of course determined by the stronger electromagnetic interactions.

\section{Maxwell's equation in curved space times}\label{a:max}
We consider the 4-velocity $u^\mu$ with $u_\mu u^\mu = -1$ in an arbitrary curved spacetime
and define the electric and magnetic fields as in Section~\ref{s:mhd}, eqs.~\ref{e2:EB},
\be\label{a2:EB}
E_\mu = F_{\mu\nu}u^\nu\,, \qquad B_\mu = \ep_{\mu\nu\ga}F^{\nu\ga}/2\,,
\ee
such that
\be
 F_{\mu\nu} = u_\mu E_\nu -u_\nu E_\mu + \ep_{\mu\nu\ga}B^\ga\,.
\ee
We define the expansion rate $\theta$, the shear $\si_{\mu\nu}$, the vorticity $\om_{\nu\mu}$
and the acceleration $a_\mu$ of the 4-velocity $u^\mu$ by
\bea
\theta &=& u^\mu_{;\mu}\, , \quad \si_{\mu\nu} =\frac{1}{2}\left(u_{\mu;\nu} + u_{\nu;\mu} -\frac{1}{3} \theta h_{\mu\nu}\right) \;,  \\
\om_{\mu\nu} &=& \frac{1}{2}\left(u_{\mu;\nu} - u_{\nu;\mu} \right) \;,  \quad  a_\mu = u^\nu u_{\mu;\nu}
\,.
\eea
Here $h_{\mu\nu} = g_{\mu\nu} +u_\mu u_\nu$ is the projector to the tangent space normal to $u$.

In terms of these quantities the homogeneous Maxwell equations, $F_{(\mu\nu,\al)} =0$ 
become (see ~\citet{Barrow:2006ch})
\bea\label{a2:maxh}
h_{\mu\nu}u^\al {B^{\nu}}_{;\al} + \ep_{\mu\nu\ga}E^{\nu;\ga} &=& \left( \si_{\mu\nu} + \om_{\mu\nu}
+\frac{1}{3} \theta h_{\mu\nu}\right)B^{\nu} -\ep_{\mu\nu\al}a^\nu E^\al  \\
 {B^{\nu}}_{;\nu} &=&  \ep_{\mu\nu\al}\om^{\nu\al}E^\mu \,.
\eea
The 3-component $ \ep$--tensor is given in terms of the totally antisymmetric tensor  $\eta_{\beta\mu\nu\al}$ by $\ep_{\mu\nu\al} = u^\beta \eta_{\beta\mu\nu\al}$, see Section~\ref{s:mhd}.
 
Introducing the 4-current $j_\mu$, the charge density $\rho_e = -u^\mu j_\mu$ and the 3-current
$J_\mu = j_\mu - \rho_e u_\mu$ we obtain for the inhomogeneous Maxwell equations,
$ {F^{\mu\nu}}_{;\nu} = j^\mu$,
\bea\label{a2:maxi}
{E^{\nu}}_{;\nu} &=& \rho_e - \ep_{\mu\nu\al}\om^{\nu\al}B^\mu \,, \\
-h_{\mu\nu}u^\al {E^{\nu}}_{;\al} + \ep_{\mu\nu\ga}B^{\nu;\ga} &=&-\left( \si_{\mu\nu} + \om_{\mu\nu}
+\frac{1}{3} \theta h_{\mu\nu}\right)E^{\nu} -\ep_{\mu\nu\al}a^\nu B^\al   +J^\mu \; .
\eea

The energy momentum tensor of the Maxwell field in terms of $E$ $B$ and $u$ is
\bea
T_{\mu\nu}^{\rm (em)} &=& -F_{\mu\al}{F^\al}_\nu -  \frac{1}{4}F_{\al\beta}F^{\al\beta}g_{\mu\nu}
\nonumber \\   \label{a2:Tmunu}
 &=& \frac{1}{2}(E^2 + B^2)u_\mu u_\nu + \frac{1}{2}(E^2 + B^2)h_{\mu\nu}
-E_\mu E_\nu -B_\mu B_\nu + P_\mu u_\nu + u_\mu P_\nu \,,
\eea
where $P_\mu = \ep_{\mu\al\beta}E^\al B^\beta$ is the Poynting vector, i.e. the energy flux seen
 by an observer with 4-velocity $u$. The energy density is $\rho^{\rm (em)} = T_{\mu\nu}^{\rm (em)}u^\mu u^\nu = (E^2 + B^2)/2$ and $T_{\mu}^{\rm (em) \; \mu} =0$ as we expect it from the electromagnetic energy momentum tensor.

In a Friedmann universe, for a co-moving observer with $u = a^{-1}\dd_t$, these equations simplify considerably. Since $\om_{\mu\nu} = \si_{\mu\nu} = a_\mu =0$ and $\theta = 3\dot a/a^{2} = 3H$,
we obtain with $(B_\mu) = a(0,\bB)$ and $(E_\mu) = a(0,\bE)$.
\bea\label{a2:maxF1}
\dd_t(a^2 \bB)+ a^2\bnabla\wedge \bE &=& 0 \\
 \bnabla\cdot\bB &=& 0 \,, \\
 \bnabla\cdot\bE &=& a\rho_e  \,, \\
-\dd_t (a^2\bE) +a^2\bnabla\wedge \bB &=&  a^3\bJ \; . \label{a2:maxF4}
\eea
Note that $\bB$ and $\bE$ scale like $1/a^2$. In term if the re-scaled quantities  $a^2\bB$, $a^2\bE$ and $a^3\bJ$,  the Maxwell equations assume the same form as in Minkowski space, the expansion factor can be 'scaled out'.

% BibTeX users please use one of
\bibliographystyle{spbasic}      % basic style, author-year citations
%\bibliographystyle{spmpsci}      % mathematics and physical sciences
%\bibliographystyle{spphys}       % APS-like style for physics
%\bibliography{}   % name your BibTeX data base

%%%%%%%%%%%%%%%%%%%%%%%%%%%%%%%%%%%%%%%%%%%%%%%%%%%%%%
%\bibliographystyle{AARev}
\bibliography{MF-refs}

% Non-BibTeX users please use
%\begin{thebibliography}{}
%
% and use \bibitem to create references. Consult the Instructions
% for authors for reference list style.
%
%\bibitem{RefJ}
% Format for Journal Reference
%Author, Article title, Journal, Volume, page numbers (year)
% Format for books
%\bibitem{RefB}
%Author, Book title, page numbers. Publisher, place (year)
% etc
%\end{thebibliography}

\end{document}